\documentclass[aoas]{imsart}

\RequirePackage{amsthm,amsmath,amsfonts,amssymb}
\RequirePackage[authoryear]{natbib}
\RequirePackage[colorlinks,citecolor=blue,urlcolor=blue]{hyperref}
\RequirePackage{graphicx}

\usepackage{mathtools}
\mathtoolsset{showonlyrefs}
\usepackage{dsfont,bm}
\usepackage[inline]{enumitem}
\usepackage[frozencache,cachedir=minted-cache]{minted}
\newcommand{\mycode}[1]{\mintinline{shell}{#1}}

\usepackage{subcaption}
\usepackage[figurename=Fig, font=footnotesize, labelfont=sc, textfont=it]{caption}
\usepackage{numprint}
\npthousandsep{,}\npthousandthpartsep{}\npdecimalsign{.}

\renewcommand{\vec}[1]{\boldsymbol{#1}}

\newcommand{\abs}[1]{\vert{#1}\vert}

\startlocaldefs

\endlocaldefs

\usepackage{multibib}
\newcites{SM}{Supplementary references}

\begin{document}

\begin{frontmatter}
\title{Nested Dirichlet models for unsupervised attack pattern detection in honeypot data}
\runtitle{Nested Dirichlet models for attack pattern detection in honeypot data}

\begin{aug}
\author[A]{\fnms{Francesco}~\snm{Sanna Passino}\ead[label=e1]{f.sannapassino@imperial.ac.uk}\orcid{0000-0002-4571-6681}},
\author[B]{\fnms{Anastasia}~\snm{Mantziou}\orcid{0000-0002-2762-023X}},
\author[A]{\fnms{Daniyar}~\snm{Ghani}\orcid{0000-0001-8611-9966}},
\author[A]{\fnms{Philip}~\snm{Thiede}},
\author[C]{\fnms{Ross}~\snm{Bevington}}, \\
\and
\author[A]{\fnms{Nicholas A.}~\snm{Heard}\orcid{0000-0002-8767-0810}}
\address[A]{Department of Mathematics, Imperial College London, London (United Kingdom)\printead[presep={,\ }]{e1}}
\address[B]{Department of Statistics, University of Warwick, Coventry (United Kingdom)}
\address[C]{Microsoft Threat Intelligence Center (MSTIC), Cheltenham (United Kingdom)}
\end{aug}

\begin{abstract}
Cyber-systems are under near-constant threat from intrusion attempts. Attacks types vary, but each attempt typically has a specific underlying intent, and the perpetrators are typically groups of individuals with similar objectives. Clustering attacks appearing to share a common intent is very valuable to threat-hunting experts. This article explores Dirichlet distribution topic models for clustering terminal session commands collected from honeypots, which are special network hosts designed to entice malicious attackers. The main practical implications of clustering the sessions are two-fold: finding similar groups of attacks, and identifying outliers. A range of statistical models are considered, adapted to the structures of command-line syntax. In particular, concepts of primary and secondary topics, and then session-level and command-level topics, are introduced into the models to improve interpretability. The proposed methods are further extended in a Bayesian nonparametric fashion to allow unboundedness in the vocabulary size and the number of latent intents. The methods are shown to discover an unusual MIRAI variant which attempts to take over existing cryptocurrency coin-mining infrastructure, not detected by traditional topic-modelling approaches.
\end{abstract}


\end{frontmatter}


\section{Introduction} \label{sec:intro}

The increasing reliance of enterprises on information technologies, such as cloud services, gives rise to new challenges for protecting customer data and computer systems from intrusions. To tackle these cyber threats, enterprises increasingly resort to quantitative methods for the development of the next-generation intrusion detection techniques \citep{Hero23}.
\textit{Honeypots} \citep[see, for example,][]{Mokube07} play an important role in the detection and understanding of attacker behaviours. A honeypot is a host located within a computer network designed to entice malicious attackers. Security teams use 
the commands issued by attackers during interactive sessions with the honeypot, as well as other meta-data 
such as 
the source IP address,
in order to understand the attack and the attacker's intent to better protect their networks from compromise \citep[see, for example,][]{Wang10}. 
Honeypots therefore provide cyber analysts with \textit{session data}, where each session is comprised of multiple commands issued by the user; each command can be interpreted as a sequence of \textit{instructions} in \textit{command language} (for example, \textit{shell} programming languages), similar to \textit{words} in natural language \citep[see, for example,][]{Hanif22}.

Honeypot session data provide a rare insight into the operational techniques of cyber attackers, such as their automated or interactive nature, the individual scripting styles and their overall objectives. This makes honeypot tracking systems particularly attractive for developing robust quantitative methods for cyber-security \citep{Highnam21}. The volume of traffic passing through a honeypot can be surprisingly high, and so automating the understanding of these sessions, classifying them and detecting new emerging patterns provides a challenging research problem which is addressed in this article.

Typically, attackers have one main objective after gaining access to a network host. For example, an intruder might want to 
infect the machine with ransomware, build a cryptocurrency miner, take over existing infrastructure, 
copy information for data leakage or sale, or collect intel about the organisation. 
Recently, the most common 
malicious behaviours have described and classified in the MITRE ATT\&CK\textregistered\footnote{For more details, see \url{https://attack.mitre.org/}.} \citep[for example, see][]{Yevo22} knowledge base for enterprise attacks, extensively used in cyber-security research \cite[see the survey of][]{Roy23}. 
Therefore, each observed session could be thought to have an underlying \textit{latent intent}. Importantly, such intents evolve and change over time, creating new threats for the security of cyber-systems. From a statistical perspective, the problem of estimating latent intents from a collection of attempted attacks can be framed as a \textit{clustering} task. Hence, the main objective of this work is to develop clustering models for command line data observed in cyber-security applications. Such clustering models could then be used for automated online classification of network intrusions, providing a valuable tool for threat experts and enterprises to discover underlying patterns that would have not been easily detectable otherwise \citep{Adams18}. Automated threat detection can be viewed as complementary to deterministic classification frameworks, such as MITRE ATT\&CK\textregistered, providing a further level of sophistication to attack pattern detection.

In the present work, ideas borrowed from the literature on topic modelling in text analysis are used to detect attack patterns, with sessions playing the role of documents and commands playing the role of sentences. 
Command line instructions are modelled under a \textit{bag-of-words} assumption, leading to a generative model for the instructions dependent on the latent intent characterising the corresponding session. This fundamentally differentiates our approach from mixed membership strategies to language modelling, such as Latent Dirichlet Allocation \citep[LDA,][]{Blei03}.
The drawbacks of LDA for the scope of modelling command lines are threefold:
\begin{enumerate*}[label=(\roman*)]
\item attackers usually have mainly \textit{one} intent per session;
\item for the purposes of attack pattern detection, analysts and threat experts would typically prefer to have one label per session, rather than interpreting the results of a mixed-membership model which might contain surplus information for their requirements;
\item models based on LDA often present unidentifiability and convergence difficulties, making reproducibility of results problematic.
\end{enumerate*}
Such difficulties are addressed in this work, presenting an approach that assigns a \textit{single topic}, or intent, to each session, 
which is then easy to interpret for threat experts through statistical summaries of sessions assigned to the same group. Furthermore, one class of proposed models incorporate the additional idea of command-level intents, establishing a two-level clustering structure. 

Later models and inferential procedures discussed in this work admit the possibility of an unknown and unbounded number of latent intents and an unbounded vocabulary size. These extensions are particularly important for practical deployment in computer network security, since attack vectors frequently evolve and new command line instructions appear. The number of topics in LDA models is usually chosen using scree-plot criteria using the perplexities calculated from a holdout dataset \citep{Teh06}. However, optimising for perplexity might not yield interpretable topics \citep{Ding18}. In this work, an alternative strategy based on Bayesian hierarchical nonparametric Griffiths-Engen-McCloskey priors \citep[GEM,][]{Pitman06} is used, admitting the possibility of previously unobserved intents and instructions.

The rest of the article is structured as follows: in the remainder of this section, the data sources used in this work are described along with a review of the related literature. Section~\ref{sec:models} describes models for session data, and Section~\ref{sec:inference} presents inferential procedures. The methodology is then extended to the cases of unbounded numbers of topics and vocabulary size in Sections~\ref{sec:dp_model}~and~\ref{sec:dp_model_inference}. Finally, the proposed methods are applied to real-world session data from honeypots in Section~\ref{sec:honey}, and the practical implications of the results are discussed.

\subsection{Honeypot session data} \label{sec:datadesc}

When a user connects to a honeypot through certain protocols, a \textit{session} starts, and every action the user performs on this host is recorded until logout, when the session ends. A user will run a sequence of \textit{commands}, which are strings of code which perform actions on the host. Each command comprises a sequence of \textit{words} drawn from the syntax of the chosen protocol.
In the following example session, the intruder first attempts to access a convenient directory (through multiples uses of the \mycode{cd} command), then tries three methods of downloading a \textit{Bash} script from the web (\mycode{wget}, \mycode{curl} and \mycode{tftp get}, representing different commands having the same underlying intention), before attempting to execute and delete the script. The real IP address which was used in the attack is masked using the string \mycode{abc.def.ghi.jkl}. 
\begin{minted}{shell}
cd /tmp || cd /var/run || cd /mnt || cd /root || cd /
wget http://abc.def.ghi.jkl/Zerow.sh
curl -O http://abc.def.ghi.jkl/Zerow.sh
chmod 777 Zerow.sh
sh Zerow.sh
tftp abc.def.ghi.jkl -c get tZerow.sh
chmod 777 tZerow.sh
sh tZerow.sh
rm -rf Zerow.sh tZerow.sh
\end{minted}

Transforming commands into sequences of words, known as \textit{tokenisation} in the literature, is not a trivial task in cyber-security \citep[see, for example,][]{Hanif22}. 
Consider the web address \mycode{http://abc.def.ghi.jkl/Zerow.sh}, which appeared in the second command of the session. One might consider the \textit{entire string} as a word, or split it into different words, such as \mycode{http}, \mycode{abc.def.ghi.jkl} and \mycode{Zerow.sh}. Furthermore, the entire Internet Protocol (IP) address \mycode{abc.def.ghi.jkl} could be considered as a word, or just its subnet \mycode{abc.def}. Similarly, \mycode{Zerow} could be considered as an individual word, excluding the file extension \mycode{.sh}. More details on the preprocessing 
will be given in Section~\ref{sec:preprocessing}.

\subsection{Related literature} \label{sec:relit}

In topic models, each document is usually considered as a bag-of-words, and the words are assumed to be exchangeable. Under this assumption, the information carried by paragraphs and sentences in natural language is lost. In cyber-security, documents correspond to sessions and sentences correspond to commands, which are expected to have a specific intent. For attack pattern detection, it would be informative to also capture such latent intents at the command-level. In the literature, document and sentence clustering have been considered as two independent problems.

The problem of clustering documents has been extensively studied in the natural language processing, computer science and information retrieval communities \citep[for a survey, see][and references therein]{Aggarwal12}.
Common approaches include matrix factorisation techniques \citep{Xu03} and spectral clustering \citep{Cai11}. Furthermore, \cite{Wallach08} proposed a cluster-based topic model extending LDA, where each group is assigned a cluster-specific Dirichlet prior on the document-specific topic distribution. \cite{Xie13} also propose a multi-grain topic model with clustering where documents are assigned global and group-specific topics.

Sentence-level structure within topic models has been largely overlooked in the literature. \cite{Balikas16} propose to extend LDA by sampling words from sentence-specific topic distributions. Furthermore, \cite{Jiang19} propose to model the sentence-specific topic distribution as a mixture between the topic distributions of adjacent sentences, weighted by a topic association matrix. In the present article, a new framework is proposed which permits joint inference of the latent structure for both sessions and commands.



Usually, one of the main difficulties for practitioners with LDA models is the interpretation of the output of a mixed membership model for each document. For example, in cyber-security, analysts would need a single label assigned to each document, not a mixture of topics, even if such topics seem interpretable \citep{Chang09}. In the literature, sparse topic models \citep{Williamson10,Archambeau15,Zhang20} might alleviate this potential issue by enforcing sparsity in the topic-specific word distributions. \cite{Doshi15} proposed Graph-Sparse LDA, that used relationships between words to improve interpretability.

Also, the performance of LDA methods heavily relies on suitable preprocessing of the data. For example, high-frequency words are often removed before \citep{Fan19} or after \citep{Schofield17} inference, under the assumption that such words make limited contributions to the meaning of the documents. In practical application, an issue with this approach is that models would need to be retrained multiple times before finding the optimal tradeoff in terms of number of words removed. In large-scale applications, this may not be possible. 
In order to avoid data pruning, alternative term-weighting schemes have also been proposed 
\citep{Wilson10}. Here, a further possible solution is proposed: a \textit{secondary} topic, shared across all documents, can be used to capture high-frequency words and lead to more interpretable \textit{primary} topics characterising individual documents. 

Another possible explanation of the issues of LDA with high-frequency terms is that, in natural language, word counts have a power-law distribution \citep{Sato10}. Therefore, \cite{Sato10} proposed a Pitman-Yor LDA model, which admits power-laws by construction. Another approach to the problem of modelling power-laws is the latent IBP compound Dirichlet allocation model \citep{Archambeau15}. It is unclear whether power-laws apply to the word counts in command line data and cyber-security applications. Such structures could be easily accounted for in the methodology proposed in this paper via two-parameter GEM prior distributions, corresponding to stick-breaking proportions of a Pitman-Yor process \citep{Pitman06}. 

Cyber-security applications require the number of latent intents and vocabulary to be unbounded for practical deployment. Hierarchical Dirichlet Processes \citep{Teh06} have been successfully used within the context of LDA models to admit an unbounded number of topics. Furthermore, \cite{Zhai13} developed an online LDA algorithm with unbounded vocabulary size, proposing a multinomial and $n$-gram prior distribution for a conventional character language. However, $n$-grams tend to suffer from data sparsity issues \citep{Allison06}. In this work, the words are simply interpreted as tokens, and therefore a GEM prior is employed instead, corresponding to a prior distribution over the natural numbers. This strategy avoids the difficulty of specifying a prior distribution on the command line syntax, which would be an undesirable additional task.
Furthermore, GEM priors are also assigned to the number of latent topics. 

The approach proposed in this work is closely related to existing methodologies in the Bayesian nonparametric literature aimed at clustering observations divided into groups, such as the Hierarchical Dirichlet Processes proposed in \cite{Teh06} and \cite{Muller04}. In particular, the proposed method is most linked to the nested Dirichlet process \citep[NDP,][]{Rodriguez08}. Note that the terms \textit{nested} and \textit{hierarchical} are also used in the literature for topic models with a tree structure \citep[see, for example,][]{Blei03_2,Blei10,Paisley14}. \cite{Camerlenghi19} prove that the NDP can degenerate to a fully exchangeable model when ties across samples are observed under a continuous base measure, and proposes a class of latent nested nonparametric priors that overcome the issue. This is not an issue for the model proposed in this work, as the vocabulary is considered either finite or countably infinite, implying an underlying discrete base measure. This is further confirmed via simulations in the Supplementary Material, which did not show issues of posterior degeneracy for recovering the correct groups. 

It must be remarked that the task of estimating the number of components in a finite mixture model also presents issues with consistency both under finite parametric \citep{Cai21} and nonparametric priors \citep{Miller13,Miller14} under model misspecification, unless a prior distribution on the concentration parameter of the Dirichlet process is appropriately specified \citep{Ascolani22}. Therefore, in practice, it is not expected to \textit{always} recover the exact number in the data generating process. Simulations under the model proposed in this work did not show difficulties in estimating the correct number of components in a number of scenarios (\textit{cf.} Supplementary Material). 

In cyber-security, some attempts have been made at analysing command line data via statistical and machine learning methods. 
For example, \cite{Sadique21} aim to predict the next command of the attacker by using an edit distance training model on the sequence of commands input. In a similar setup, \cite{Crespi21} aim to identify attacker behaviours from command logs using supervised NLP methods. 
Also, \cite{Shrivastava19} focus on classifying types of attacks from commands using a series of machine learning techniques such as naive Bayes, random forests and support vector machines. 
Lastly, 
attacker behaviour from session data has been analysed using Hidden Markov Models, as seen in the studies of \cite{Rade18} and \cite{Deshmukh19}. However, none of  
these studies consider topic modelling approaches for the analysis of sessions.

\section{Models for clustering session data} \label{sec:models}

Command line data are observed in \textit{sessions}, where each session is divided into \textit{commands}, and each command is composed of different \textit{words} drawn from a vocabulary $\mathcal{V}$ of size $V=\abs{\mathcal{V}}$. For $D\in\mathds N$ observed sessions, let $N_d,\ d=1,\dots,D$ denote the number of commands in each session, and $M_{d,j},\ j=1,\dots,N_d$
represent the number of words in the command $j$ of the session $d$. Following the standard LDA model \citep{Blei03}, 
the number of commands $N_d$ in session $d$ and the number $M_{d,j}$ of words in each command $j$ within that session are assumed to be Poisson distributed:
\begin{align}
N_d &\sim \text{Poisson}(\zeta),\ d=1,\dots,D, \\
M_{d,j} &\sim \text{Poisson}(\omega),\ j=1,\dots,N_d,
\end{align}
where $\zeta,\omega\in\mathds R_+$. 
The $i$-th word in the $j$-th command of the $d$-th session is denoted $w_{d,j,i}$, which has a corresponding probability mass function $\bm\xi_{d,j,i}\in\mathds R_+^{ V}$ over the vocabulary $\mathcal{V}$.

The stated aim is to develop clustering algorithms for sessions, where the clusters represent shared \textit{intents} of the intruders, or \textit{groups} of attackers with similar behaviour. To achieve this aim, a 
variety of structures are considered, establishing shared distributions $\bm\xi_{d,j,i}$ across groups of sessions and commands to identify clusters. In particular, this work focuses on two approaches:
\begin{enumerate*}[label=(\roman*)]
\item \textit{Constrained}: Each session has a primary topic and a global secondary topic;
\item \textit{Nested}: Each session topic is a distribution on command-level topics, which introduces two layers of latent topics.
\end{enumerate*}
These approaches are discussed in detail in the next sections.

\subsection{Constrained Bayesian clustering with primary and secondary topics}

As a most basic approach, each document $d$ could have a latent assignment to one of $K$ possible topics, where each topic $k\in\{1,\dots,K \}$ is characterised by a probability mass function $\bm\phi_{k}$ on the vocabulary $\mathcal{V}$. Let $t_d$ denote the topic assignment for document $d$, and let $\bm \lambda = (\lambda_1,\ldots,\lambda_K)$ where $\lambda_k$ denotes the probability that $t_d=k$. It could then be assumed $\bm\xi_{d,j,i}=\bm\phi_{t_d}$, such that conditional on the session-specific topic $t_d$, all the words $w_{d,j,i}$ in session $d$ are sampled from the same distribution $\bm\phi_{t_d}$. Assuming conjugate Dirichlet prior distributions for the probability distributions $\bm\lambda$ and $\{\bm\phi_k\}$ implies the following model:
\begin{align}
\bm\lambda &\sim \text{Dirichlet}_K(\bm\gamma), \\
\bm\phi_k &\sim \text{Dirichlet}_V(\bm\eta),\ k=1,2,\dots,K , \\
t_d \mid \bm\lambda &\sim \text{Categorical}_K(\bm\lambda),\ d=1,\dots,D, \\
w_{d,j,i} \mid t_d, \{\bm\phi_k\} &\sim \text{Categorical}_V(\bm\phi_{t_d}),\ i=1,\dots,M_{d,j},\ j=1,\dots,N_d,
\label{mod1}
\end{align}
where $\bm\gamma\in\mathds R_+^{K }, \bm\eta\in\mathds R_+^{ V}$. In the cyber-attack context, these topics correspond to 
\textit{intents}. 
This model imposes a constraint on LDA, by assuming that each document contains only one topic. Therefore, the approach is denoted constrained Bayesian clustering (CBC).  

The simple CBC model above can be used as a starting point for exploring more complex clustering structures. For example, to better identify differences between topic-specific distributions, it could be assumed that words in a document are sampled either from a topic-specific probability distribution, or from a baseline probability distribution shared across all documents. The shared distribution represents words that are commonly used in \textit{all} sessions, but are not key for characterising the intent of a session. For example, in natural language, such a baseline distribution could give probability mass to conjunctions (e.g. \textit{but}, \textit{and}, \textit{if}), articles (e.g. \textit{a}, \textit{an}, \textit{the}), or pronouns (e.g. \textit{she}, \textit{he}, \textit{they}). Similarly, for command lines in a cyber-security context, the shared distribution might give weight to common \textit{Bash} commands such as \mycode{ls} (list contents), \mycode{ps} (list the running processes), or \mycode{cd} (change directory). The shared distribution will be used to make the session-specific topics more representative of the attacker's intents, 
reducing the probability of words commonly used across topics. 

In particular, an extended model assumes each topic has an associated probability $\theta_k\in[0,1],\ k=1,\dots,K $, representing the mixing proportion between the topic-specific word distribution $\bm\phi_k$ and the shared distribution $\bm\phi_0$. Each word is then sampled with probability $\theta_{t_d}$ from $\bm\phi_{t_d}$, or from $\bm\phi_0$ with probability $1-\theta_{t_d}$, implying the following revised model:
\begin{align}
\bm\phi_k &\sim \text{Dirichlet}_V(\bm\eta),\ k=0,1,2,\dots,K , \\
\theta_k &\sim \text{Beta}(\alpha_k , \alpha_0),\ k=1,\dots,K , \\
z_{d,j,i} \mid t_d, \{\theta_k\} &\sim \text{Bernoulli}(\theta_{t_d}),i=1,\dots,M_{d,j},\ j=1,\dots,N_d,\\
w_{d,j,i} \mid z_{d,j,i}, t_d, \{\bm\phi_k\} &\sim \text{Categorical}_V(\bm\phi_{t_dz_{d,j,i}}),\ i=1,\dots,M_{d,j},\ j=1,\dots,N_d.
\label{mod2}
\end{align}
This model essentially imposes a sparsity constraint on LDA, by assuming that each document contains only two topics:
\begin{enumerate*}[label=\normalfont(\roman*), ]
\item a \textit{primary} topic $t_d$, chosen from $K $ primary topics, and
\item a \textit{secondary} topic shared across \textit{all} documents, denoted ``topic $0$'' for notational convenience.
\end{enumerate*}

It must be remarked that the proposed approach is different from simply removing words that occur frequently in the documents \cite[see, for example,][]{Schofield17,Fan19}: the weight $\theta_k\in[0,1]$ given to the shared topic represented by $\bm\phi_0$ is considered here to be part of the characterisation of the $k$-th topic. 

\subsection{Nested constrained Bayesian clustering with session-level and command-level topics} \label{sec:hctm}

The models in \eqref{mod1} and \eqref{mod2} assume that words in each command within a given session are sampled from the \textit{same} topic-specific distribution, or from a distribution shared across documents. The information about the structure of a session as a sequence of commands is therefore ignored, which might be limiting in practical settings. Instead, it would be reasonable to assume that session-specific intents share similar commands for specific tasks. Such tasks could be interpreted as \textit{command-level intents}, 
and the distribution of the tasks characterises the \textit{session-level} topic. Let $H$ be the assumed number of command-level topics. It could be assumed that each session-level topic has an associated $H$-dimensional probability distribution $\bm\psi_k$ across command-level intents, with each command within a given session being assigned a command-specific topic $s_{d,j}\in\{1,\dots,H \}$ sampled from $\bm\psi_{t_d}$. Conditional on $s_{d,j}$, the words in the command are then sampled independently from a $ V$-dimensional distribution $\bm\phi_{s_{d,j}}$, specific to the command-level topic. Therefore, the model in \eqref{mod1} becomes: 
\begin{align}
\bm\psi_k &\sim \text{Dirichlet}_H(\bm\tau),\ k=1,2,\dots,K , \\
\bm\phi_h &\sim \text{Dirichlet}_V(\bm\eta),\ h=1,2,\dots,H , \\
s_{d,j} \mid t_d, \{\bm\psi_k\} &\sim \text{Categorical}_H(\bm\psi_{t_d}),\ j=1,\dots,N_d, \label{mod3} \\
w_{d,j,i} \mid s_{d,j}, \{\bm\phi_h\} &\sim \text{Categorical}_V(\bm\phi_{s_{d,j}}),\ i=1,\dots,M_{d,j},\ j=1,\dots,N_d,
\end{align}
where $\bm\tau\in\mathds R_+^{H }$. In this model, there are two layers of topics, and corresponding indices:
\begin{enumerate*}[label=(\roman*)]
\item Command topic indices, $s_{d,j}$, used to match the words in the corresponding command to distributions $\bm\phi_1,\dots,\bm\phi_{H }$ over $\mathcal{V}$;
\item Document topic indices, $t_d$, used to match the commands in the corresponding session to distributions $\bm\psi_1,\dots,\bm\psi_{K }$ over the command-level topics.
\end{enumerate*}
Letting $\vec\Phi$ be the $H \times V$ matrix with $j$-th row $\bm\phi_j$, and letting $\vec\Psi$ be the $K  \times H $ matrix with $k$-th row $\bm\psi_k$, then marginally $\bm\xi_{d,j,i} =\bm\lambda^\intercal\cdot\vec\Psi\cdot\vec\Phi$, whereas $\bm\xi_{d,j,i}=\bm\phi_{s_{j,d}}$ conditionally. Since the model imposes a nesting structure in addition to the constraints discussed in the previous section, the model is denoted nested constrained Bayesian clustering (NCBC). 

\subsection{Combining the two approaches: NCBC with secondary topics} 

\begin{figure}[t]
\centering
\includegraphics[height=6cm]{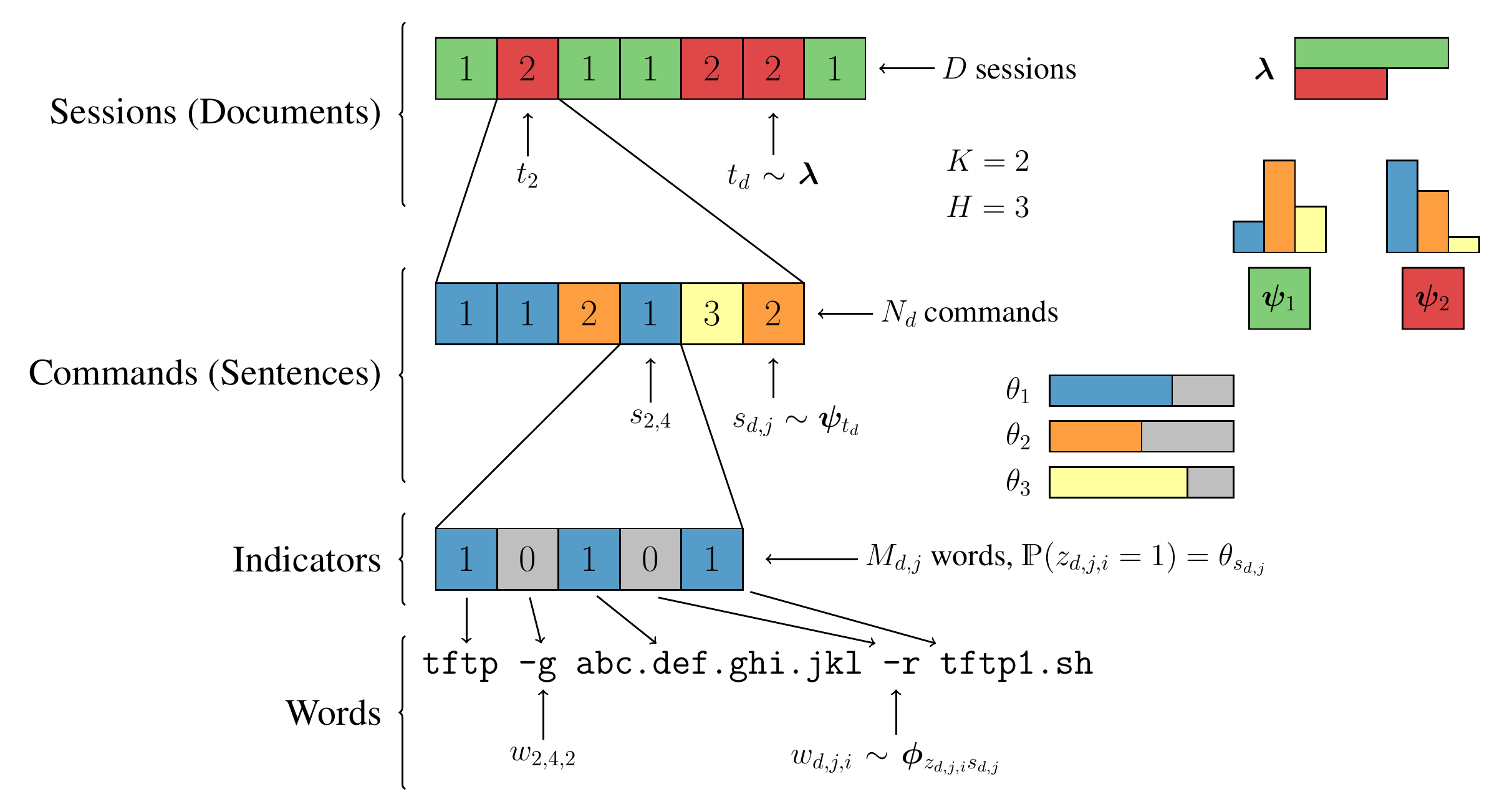}
\caption{Cartoon representation of the full Nested Constrained Bayesian Clustering (NCBC) model.}
\label{fig:model}
\end{figure}

To aid interpretability of the command-level topics, it is possible to use the same constraint from model \eqref{mod2}: it could be assumed that words are sampled either from $\bm\phi_{s_{d,j}}$, where $s_{d,j}$ is the command-level topic, or from a distribution $\bm\phi_0$ shared across all commands and sessions. As in \eqref{mod2}, each command-level topic $h\in\{1,\dots,H\}$ has an associated probability $\theta_h\in[0,1]$ for sampling words from $\bm\phi_h$. Therefore, the full model, which combines \eqref{mod1}, \eqref{mod2} and \eqref{mod3}, takes the form: 
\begingroup 
\allowdisplaybreaks
\begin{align}
\bm\lambda &\sim \text{Dirichlet}_K(\bm\gamma), \\
\bm\psi_k &\sim \text{Dirichlet}_H(\bm\tau),\ k=1,2,\dots,K , \\
\bm\phi_h &\sim \text{Dirichlet}_V(\bm\eta),\ h=0,1,\dots,H , \\
\theta_h &\sim \text{Beta}(\alpha_h , \alpha_0),\ h=1,2,\dots,H , \\
t_d \mid \bm\lambda &\sim \text{Categorical}_K(\bm\lambda),\ d=1,\dots,D, \\
s_{d,j} \mid t_d, \{\bm\psi_k\} &\sim \text{Categorical}_H(\bm\psi_{t_d}),\ j=1,\dots,N_d, \\
z_{d,j,i} \mid s_{d,j}, \{\theta_h\} &\sim \text{Bernoulli}(\theta_{s_{d,j}}), i=1,\dots,M_{d,j},\\
w_{d,j,i} \mid z_{d,j,i}, s_{d,j}, \{\bm\phi_h\} &\sim \text{Categorical}_V(\bm\phi_{s_{d,j}z_{d,j,i}}),\ i=1,\dots,M_{d,j}.
\label{mod4}
\end{align}
\endgroup
A representation of model \eqref{mod4} is given in Figure~\ref{fig:model}. It is also possible to consider variations of \eqref{mod4} through 
changes to the specification of the prior distributions on the hyperparameters. For example, document-specific mixing topic proportions $\theta_d\sim\text{Beta}(\alpha,\alpha_0)$ could be used.

Section~\ref{sec:inference} will describe inferential methods for model \eqref{mod4}. Deriving the inferential procedures for \eqref{mod1}, \eqref{mod2} and \eqref{mod3} follows similar guidelines, with minor modifications required. 

\subsection{Parent-Child Nested Bayesian Clustering} \label{sec:pcnbc}

\begin{figure}[t]
\centering
\includegraphics[height=6cm]{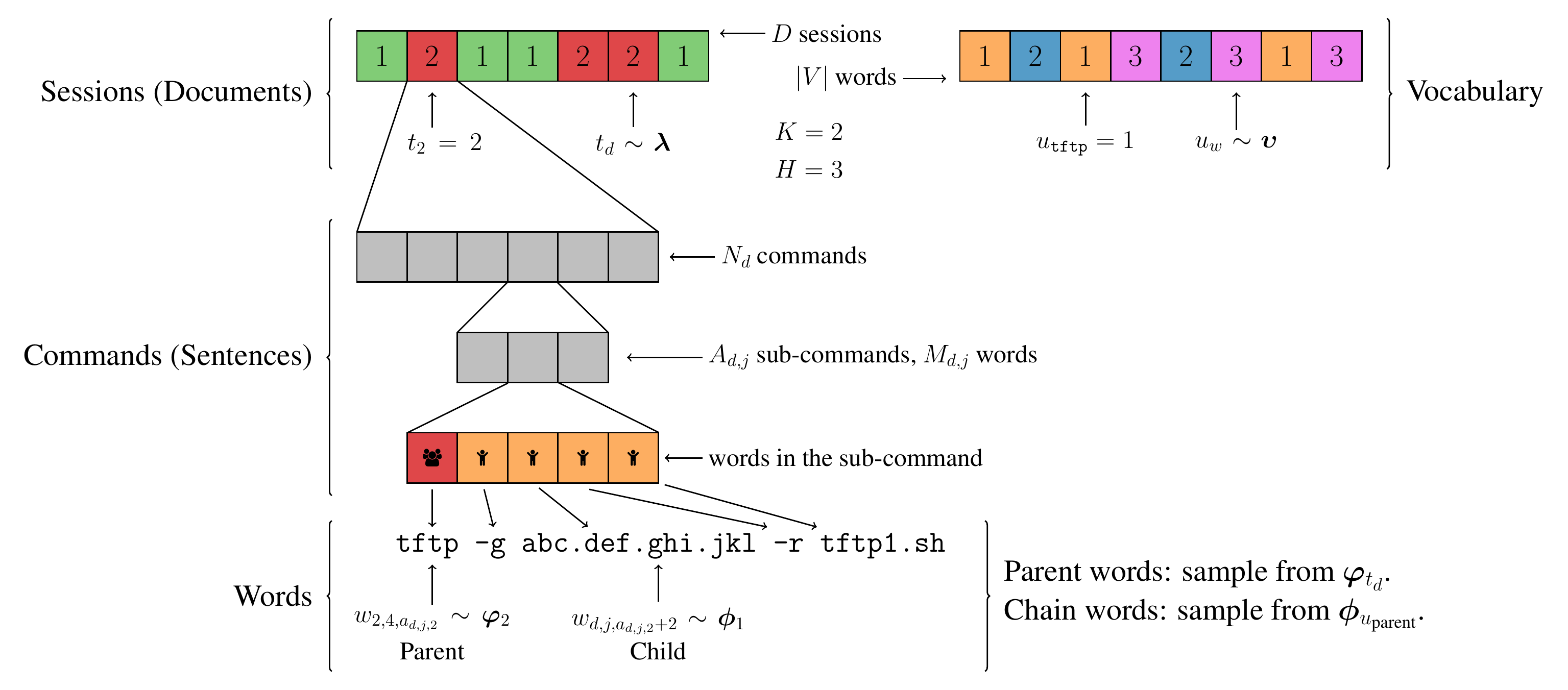}
\caption{Cartoon representation of the Parent-Child Nested Bayesian Clustering (PCNBC) model.}
\label{fig:model_parent_child}
\end{figure}

The models proposed in the previous sections are generally applicable to any corpus of text, not only limited to data collected on honeypots. 
In this work, this general approach is also compared to a model specifically tailored towards session data. 
In particular, in command lines, the first word of each command could be highly indicative of the intent of the entire command. 
For example, consider \mycode{chmod 777 Zerow.sh} from the code block in Section~\ref{sec:datadesc}: \mycode{chmod} indicates that the intent of the command is to change permissions for a file or directory. 
Furthermore, individual commands in \textit{Bash} scripting are often concatenated and piped together by symbols such as "\mycode{;}", "\mycode{|}", "\mycode{>}" or "\mycode{<}", forming a \textit{unique} command. Therefore, 
each command could be deterministically divided into sub-commands via the separators "\mycode{;}", "\mycode{|}", "\mycode{>}" or "\mycode{<}", and the first word of each sub-command could be interpreted as a \textit{parent} word, which determines the distribution of the remaining words in the sub-command, called \textit{child} words. Assuming $A_{d,j}\in\mathds N$ parent words for the $j$-th command in the $d$-th document, with $A_{d,j}\leq M_{d,j}$, let $a_{d,j,1},\dots,a_{d,j,{A_{d,j}}}\in\{1,\dots,M_{d,j}\}$ denote the parent word indices, where $a_{d,j,1}=1$ by definition of a parent word, and $a_{d,j,h}<a_{d,j,\ell}$ for $h<\ell$. Additionally, let $\mathcal A_{d,j}=\{a_{d,j,1},\dots,a_{d,j,A_{d,j}}\}$ be the set of parent word indices for the $j$-th command in the $d$-th document, $a_{d,j,\ell}^\ast=\max\{h\in\mathcal A_{d,j}:h\leq\ell\}$ the index of the parent word preceding the word $w_{d,j,\ell}$, and $w^\ast_{d,j,\ell}=w_{d,j,a_{d,j,\ell}^\ast}$ the corresponding parent word. 
Sessions are assigned a topic indicator $t_d\in\{1,\dots,K\}$ used to determine the distribution $\bm\varphi_{t_d}$ of the parent words in the session. 
Furthermore, every word $w$ in the vocabulary index set $\{1,\dots, V\}$ is assigned a cluster indicator $u_w\in\{1,\dots,H\}$; this determines the topic-specific distribution $\bm\phi_{u_w}$ used to sample child words within each sub-command, after observing parent word $w$. This results in the following model: 
\begin{align}
\bm\lambda &\sim \text{Dirichlet}_K(\bm\gamma), \\
\bm\upsilon &\sim \text{Dirichlet}_H(\bm\chi),\\
\bm\varphi_k &\sim \text{Dirichlet}_V(\bm\tau),\ k=1,2,\dots,K , \\
\bm\phi_h &\sim \text{Dirichlet}_V(\bm\eta),\ h=1,\dots,H, \\
t_d \mid \bm\lambda &\sim \text{Categorical}_K(\bm\lambda),\ d=1,\dots,D, \\
u_w \mid \bm \upsilon &\sim \text{Categorical}_H(\bm\upsilon),\ w=1,\dots, V,\ \\
w_{d,j,i} \mid t_d, \{u_w\}, \{\bm\varphi_k\},\{\bm\phi_h\} &\sim \left\{\begin{array}{lll} 
\text{Categorical}_V(\bm\varphi_{t_d})\ &&\ \text{if}\ i \in\mathcal A_{d,j} \\
\text{Categorical}_V(\bm\phi_{u_{w^\ast_{d,j,i}}})\ &&\ \text{if}\ i \notin\mathcal A_{d,j}
\end{array}\right., \label{eq:pcnbc}
\end{align}
where $\bm\chi\in\mathds R_+^H$, $\bm\tau\in\mathds R_+^V$, and $i=1,\dots,M_{d,j},\ j=1,\dots,N_d,\ d=1,\dots,D$. Figure~\ref{fig:model_parent_child} shows a cartoon representation of this \textit{Parent-Child Nested Bayesian Clustering} (PCNBC) approach.

\subsection{Unbounded number of topics and vocabulary} \label{sec:dp_model}

\begin{figure}[!t]
\centering
\begin{subfigure}[t]{.475\textwidth}
\centering
\caption{Before preprocessing}
\label{fig:pre_pre}
\includegraphics[width=\textwidth]{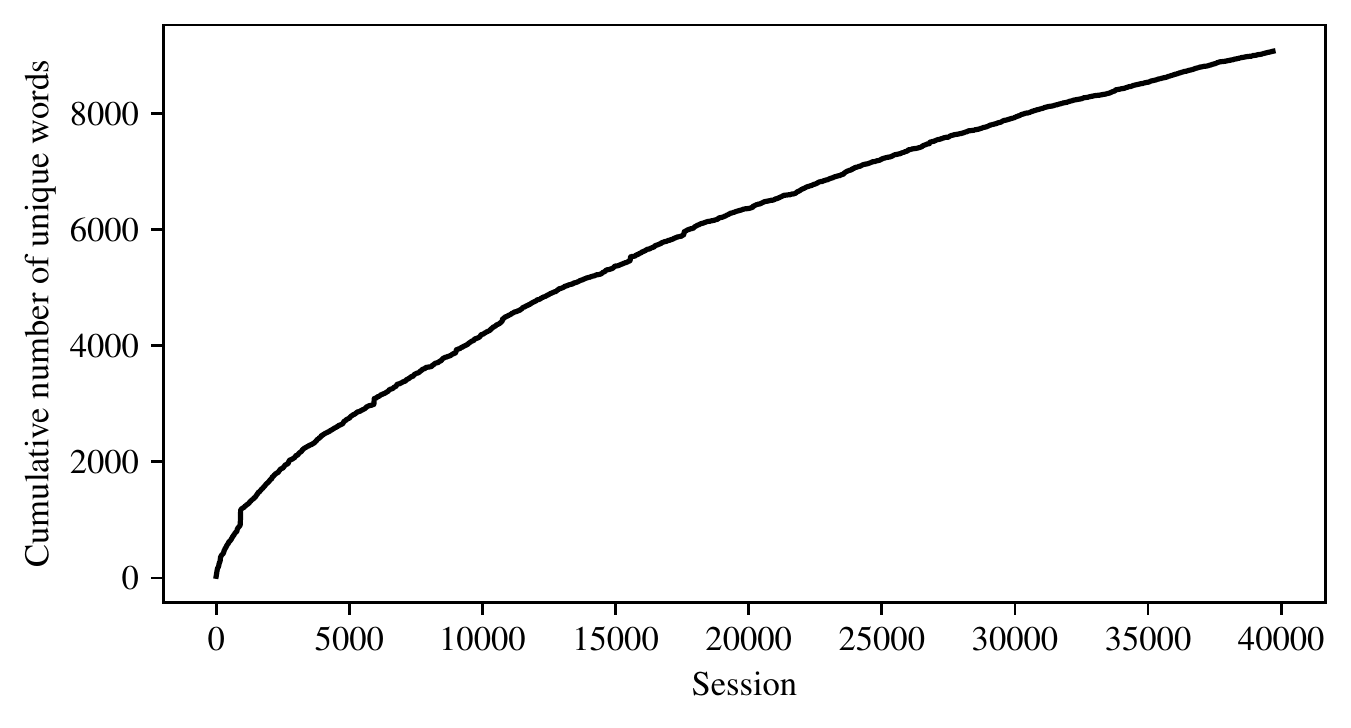}
\end{subfigure}
\hspace*{0.025\textwidth}
\begin{subfigure}[t]{.475\textwidth}
\centering
\caption{After preprocessing}
\label{fig:post_pre}
\includegraphics[width=\textwidth]{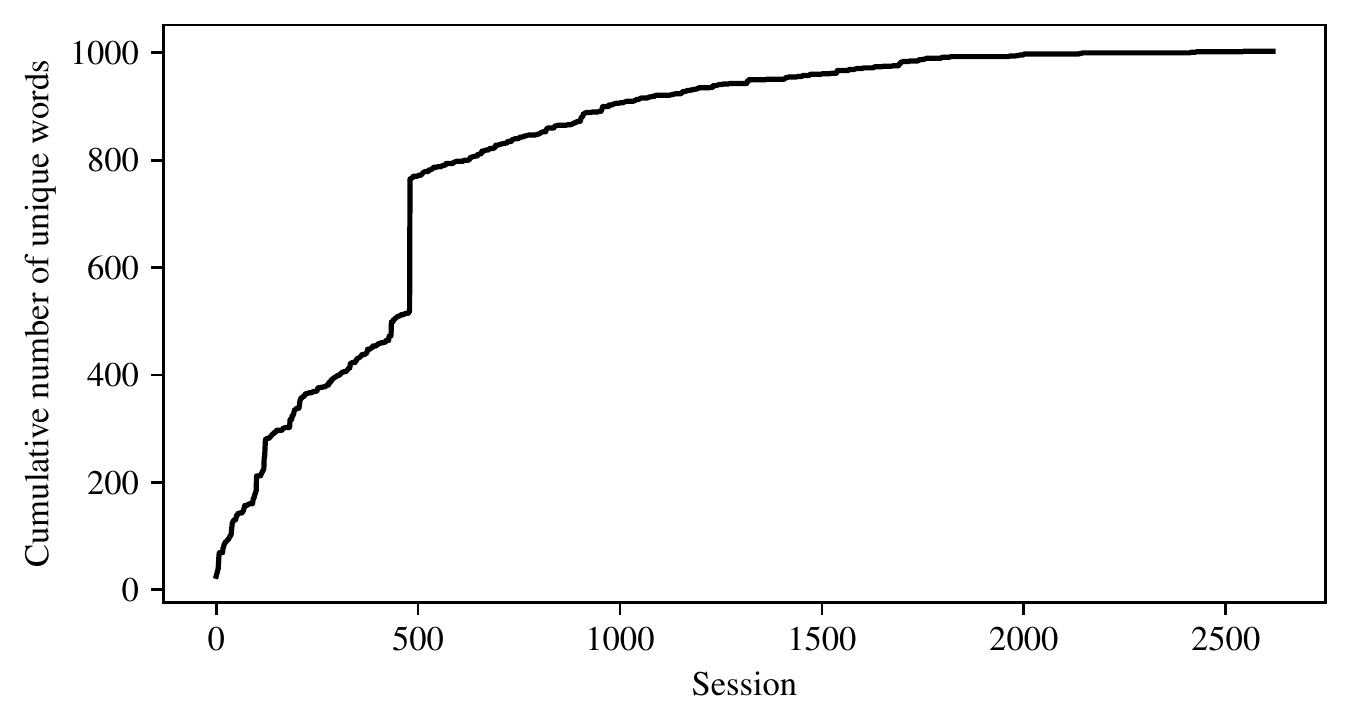}
\end{subfigure}
\caption{Cumulative number of unique words in the vocabulary $\mathcal V$ in the Imperial College London honeypot data, before and after preprocessing. The preprocessing steps are described in Section~\ref{sec:preprocessing}.}
\label{fig:cum_words}
\end{figure}

All models discussed in previous sections assume a fixed size $ V$ of the vocabulary, and a fixed number of session-level and command-level topics, $K $ and $H $ respectively.  
Such assumptions might be problematic if the model is used for clustering 
held-out sessions, 
since it would not be possible to cluster 
\textit{new} commands, composed of previously unobserved words. Therefore, 
a potentially \textit{infinite} vocabulary must be considered \citep[see][]{Zhai13,Waters14}, admitting a probability of observing \textit{new words} in future sessions. 
This occurrence is common in honeypot data, as demonstrated in Figure~\ref{fig:cum_words}, which shows that the cumulative number of unique words 
increases rapidly, especially if no preprocessing is performed (\textit{cf.} Figure~\ref{fig:post_pre}).
Also, the behaviour of attackers is expected to evolve and change over time, and it is possible that \textit{new} attack patterns or intents arise. Therefore, for real-world attack pattern detection, it is beneficial to assume an unbounded number of session-level and command-level topics. These allowances require a modification to the Dirichlet distributions used in the previous sections, instead assuming:
\begin{align}
\bm\lambda \sim \text{GEM}(\gamma), & &
\bm\psi_k \sim \text{GEM}(\tau),\ k=0,1,2,\dots, & & 
\bm\phi_\ell \sim \text{GEM}(\eta),\ \ell=0,1,2,\dots,
\end{align}
where $\tau,\eta,\gamma\in\mathds R_+$.
The GEM (Griffiths-Engen-McCloskey) distribution \citep{Pitman06} corresponds to the proportions calculated using the stick-breaking representations of the Dirichlet process \citep{Sethuraman94}, 
and to the limit for $K \to\infty$, $H \to\infty$ and $ V\to\infty$ of the Dirichlet distributions in Section~\ref{sec:models} with $\bm\gamma=\gamma\bm 1_{K }/K $, $\bm\tau=\tau\bm 1_{H }/H $, and $\bm\eta=\eta\bm 1_{ V}/ V$. For the NCBC model in Section~\ref{sec:hctm}, the construction with GEM priors corresponds to a hierarchical version of a nested Dirichlet process \citep{Rodriguez08}. 

\section{Bayesian inference via Markov Chain Monte Carlo} \label{sec:inference}

This section describes inferential procedures for the 
models discussed in Section~\ref{sec:models}. The full NCBC model is considered, with primary-secondary topics and session-level and command-level clustering. 
The posterior distribution of the parameters is only available up to a normalising constant, therefore inference must be performed using Markov Chain Monte Carlo (MCMC) methods. 
Because of the Categorical-Dirichlet conjugacy properties, sampling from the posterior distribution $p(\bm z, \bm s,\bm t, \bm\theta,\bm\lambda,\{\bm\phi_h\}, \{\bm\psi_k\} \mid \bm w)$ of the model parameters via Gibbs sampling is immediate.
The main objective of the inferential procedure is estimating $\bm t$, the session-level clusters. Hence, the remaining parameters could be interpreted as nuisance, and integrated out when possible. The parameters $\bm\theta,\bm\lambda,\{\bm\phi_h\}$ and $\{\bm\psi_k\}$ can be analytically marginalised, 
resulting in the marginal posterior density
\begin{equation}
p(\bm z, \bm s,\bm t \mid \bm w)
\propto p(\bm w, \bm z, \bm s,\bm t) 
= p(\bm t) 
\times p(\bm s \mid \bm t)
\times p(\bm z \mid \bm s)
\times p(\bm w \mid \bm z,\bm s). 
\label{eqn}
\end{equation}
Each term in the right-hand side of the marginal posterior \eqref{eqn} can be calculated explicitly by conjugacy of the Categorical-Dirichlet and Beta-Bernoulli distributions. The marginal distribution for the session-level intents is:  
\begin{equation}
p(\bm t)
=\frac{B(\bm\gamma+\bm T)}{B(\bm\gamma)}, \label{marg_t}
\end{equation}
where $B(\bm x)=\prod_{i}\Gamma(x_i)/\Gamma(\sum_i x_i)$ is the multivariate beta function, 
and $\bm T=(T_1,\dots,T_{K })$, where $T_k=\sum_d\mathds{I}_{\{k\}}(t_d)$ is the number of sessions assigned to topic $k$.
Similar calculations lead to the marginal distribution for the command-level topics, given the session-level intents:
\begin{equation}
p(\bm s\mid \bm t)
=\prod_{k=1}^{K }\frac{B(\bm\tau+\bm S_k)}{B(\bm\tau)}, \label{marg_s}
\end{equation}
where $\bm S_k=(S_{1,k},\dots,S_{H ,k})$, and $S_{k,h}=\sum_{d}\mathds{I}_{\{k\}}(t_d)\sum_{j}\mathds{I}_{\{h\}}(s_{d,j})$ denotes the number of commands assigned to the command-level topic $h$, only from the subset of sessions with session-level topic $k$.
Similarly, the marginal distribution of the primary-secondary topic indicators $\bm z$ has a closed form expression from the Beta-Bernoulli conjugacy:
\begin{equation}
p(\bm z\mid \bm s)
=\prod_{h=1}^{H }\frac{B(Z_h+\alpha_h,M_h^\ast-Z_h+\alpha_0)}{B(\alpha_h,\alpha_0)}, \label{marg_z}
\end{equation}
where $Z_h=\sum_{d,j} \mathds{I}_{\{h\}}(s_{d,j}) \sum_{i=1}^{M_{d,j}}z_{d,j,i}$ denotes the number of words assigned to the primary topic, only from commands with command-level topic $h$, and 
$M_h^\ast=\sum_{d,j}\mathds{I}_{\{h\}}(s_{d,j}) M_{d,j}$ 
denotes the total number of words in commands with topic $h$, across all documents.
The final component of the marginal posterior \eqref{eqn} is the marginal likelihood for the observed words $\bm w$, conditional on the indicators $\bm z$ and topic-level allocations $\bm s$:
\begin{equation}
p(\bm w \mid \bm z, \bm s)
=\prod_{h=0}^{H }\frac{B(\bm\eta+\bm W_h)}{B(\bm\eta)}, \label{marg_w}
\end{equation}
where $\bm W_h=(W_{h,1},\dots,W_{h, V})$, and $W_{h,v}=\sum_{d,j,i} \mathds{I}_{\{h\}}(z_{d,j,i} s_{d,j})\mathds{I}_{\{v\}}(w_{d,j,i})$ denotes the number of times word $v$ is assigned to the command-level topic $h$.

The marginal distributions \eqref{marg_t}, \eqref{marg_s}, \eqref{marg_z} and \eqref{marg_w} are the building blocks for the collapsed Gibbs sampler \citep{Liu94} used for inference on the model parameters. Collapsed Gibbs samplers are commonly used for inference in LDA models \citep[see, for example,][]{Griffiths04}. The Gibbs sampler consists of three basic moves: resample the session-level topic allocations $\bm t$, resample the command-level topic allocations $\bm s$, and resample the primary-secondary topic indicators $\bm z$. Also,  convergence of Gibbs sampling algorithms for clustering usually benefits from split-merge proposals, which are evaluated using a Metropolis-Hastings acceptance ratio, resulting in a collapsed Metropolis-within-Gibbs algorithm. 
Split-merge moves 
are used on the session-level topics $\bm t$ and command-level topics $\bm s$. 
A detailed description of the Gibbs sampling steps is reported in the Supplementary Material, 
which also contains details about Gibbs sampling in the PCNBC model (\textit{cf.} Section~\ref{sec:pcnbc}).

\subsection{Inference with unbounded number of topics and vocabulary} \label{sec:dp_model_inference}

If the number of topics and vocabulary is assumed to be unbounded, as in Section~\ref{sec:dp_model}, the posterior distribution takes a slightly different form, induced by the GEM prior.
To simplify the discussion on the GEM distribution, its link to the Dirichlet process, and its representation in the posterior distribution, consider $n$ objects allocated to $K_n$ non-empty groups, with labels $\bm x_n=(x_1,\dots,x_n)$, such that $x_i\in\mathds N$ and $K_n=\max(\bm x_n)$. Under a Dirichlet process with 
parameter $\beta$, the predictive distribution for the next label in the sequence is: 
\begin{equation}
p(x_{n+1}\mid\bm x_n) = \frac{\beta}{\beta+n} \mathds{I}_{\{K_n+1\}}(x_{n+1}) + \sum_{k=1}^{K_n} \frac{N_{k,n}}{\beta+n}\mathds{I}_{\{k\}}(x_{n+1}), \label{dp_pred}
\end{equation}
where $N_{k,n}=\sum_{i=1}^n \mathds{I}_{\{k\}}(x_i)$ is the number of the $n$ objects allocated to group $k$. The predictive equation \eqref{dp_pred} immediately provides a technique for Gibbs sampling: 
since the Dirichlet process assumes exchangeability of observations, any label can be considered as the last element of the sequence, and a new value resampled using \eqref{dp_pred}. This fact will be particularly useful when implementing the sampler. 
Using \eqref{dp_pred}, the joint distribution for the sequence is:
\begin{equation}
p(\bm x_n) = \prod_{j=1}^n p(x_j\mid\bm x_{j-1})= \frac{\alpha^{K_n}\Gamma(\alpha)}{\Gamma(\alpha+n)}\prod_{k=1}^{K_n} \Gamma(N_{k,n}).
\end{equation}
It follows that the components of the marginalised posterior distribution \eqref{eqn} take the 
form:
\begin{gather}
p(\bm t) = \frac{\gamma^{K(\bm t)}\Gamma(\gamma)}{\Gamma(\gamma+D)}\prod_{k=1}^{K(\bm t)} \Gamma(T_k), 
\hspace{.25cm}
p(\bm s\mid \bm t) = \prod_{k=1}^{K(\bm t)} \frac{\tau
^{\sum_{h=1}^{H(\bm s)} \mathds I_{\mathds N_{>0}}(S_{k,h})}\Gamma(\tau)}{\Gamma(\tau+\sum_{d}\mathds{I}_{\{k\}}(t_d) N_d)}
\prod_{h:S_{k,h}>0}\Gamma(S_{k,h}), \\
p(\bm z\mid \bm s) =\prod_{h=1}^{H(\bm s)}\frac{B(Z_h+\alpha_h,M_h^\ast-Z_h+\alpha_0)}{B(\alpha_h,\alpha_0)}, \label{dp_marginals} \\
p(\bm w\mid \bm z,\bm s) = \prod_{h=0}^{H(\bm s)} \frac{\eta^{\sum_{v=1}^{V(\bm w)} \mathds I_{\mathds N_{>0}} (W_{h,v})}\Gamma(\eta)}{\Gamma(\eta + \sum_{v=1}^{V(\bm w)} W_{h,v})}
\prod_{v:W_{h,v}>0}\Gamma(W_{h,v}),
\end{gather}
where $K(\bm t)=\sum_{k=1}^\infty \mathds{I}_{\mathds N_{>0}}(T_{k})$ and $H(\bm s)=\sum_{h=1}^\infty \mathds{I}_{\mathds N_{>0}}(\sum_{k=1}^\infty S_{k,h})$ are the number of unique session-level and command-level topics, 
and $V(\bm w)=\sum_{v=1}^\infty \mathds{I}_{\mathds N_{>0}}(\sum_{h=0}^\infty W_{h,v})$ is the observed number of unique words.
A detailed description about the Gibbs sampling steps for inference under the NCBC model with GEM priors is given in the Supplementary Material. 

\subsection{Initialisation schemes} \label{sec:initialisation}

In MCMC, setting good initial values could be helpful to achieve faster convergence, in particular for complex inferential tasks. In this work, two methods for initialisation are considered, based on spectral clustering and standard LDA. 

Spectral methods are commonly used for text analysis and topic modelling \citep{Ke22}. In order to initialise the algorithm via spectral clustering, a $(\sum_{d=1}^D N_d)\times V$ word occurrence matrix $\vec C=\{C_{sw}\}$ is constructed, where $C_{sw}$ counts the number of times word $w$ appears in command $s$. All commands are stacked in an individual matrix $\vec C$, initially disregarding information about the division into sessions. A truncated singular value decomposition of $\vec C$ is then calculated, considering only the largest $H $ singular values and corresponding left singular vectors. A clustering algorithm, like $k$-means, is then run on the resulting embedding, setting $H $ clusters. For initialisation of the session-level topics, a similar procedure is carried out, using the initial values of the command-level topics as words in a spectral clustering algorithm, obtaining a different form of the matrix of counts $\vec C$. First, the matrix $\vec C$, with dimension $D\times H $, is constructed, where each entry $C_{dh}$ counts the number of times a command assigned to the command-level topic $h$ appears in document $d$. Then, a $K $-dimensional truncated spectral decomposition of $\vec C$ is calculated, and $k$-means with $K $ clusters is run on the resulting embedding, obtaining initial values for the session-level topics. 

Alternatively, standard LDA could be used to initialise the MCMC sampler, via fast-performing software libraries such as \textit{Python}'s \textit{gensim} \citep{Rehurek10}. First, LDA with $H $ topics could be fitted, and subsequently used to predict a topic for all the words appearing in commands and sessions. Then the most common estimated topic within each command is selected as the initial command-level topic. If secondary topics are used, LDA is initially fitted with $H +1$ topics, and the most common estimated topic is selected as secondary topic. The command-level primary topic is then selected as the most common topic within each command, excluding the secondary topic. After command-level topics are estimated, session-level topics could be initialised by running LDA with $K $ topics, using the estimated command-level topics as words within the algorithm. For each command-level topic, now interpreted as a word, a topic can be estimated from the fitted LDA model, and the session-level topics are then initialised as the most common topic within each command.  

For initialisation of the secondary topic indicators, $z_{d,j,i}$ could be initially set to 1 if the proportion of sessions or commands where the word $w_{d,j,i}$ appears is less than a
threshold. This is because $\vec\phi_0$ should represent a distribution of common words, shared across topics. 

\section{Application to the Imperial College London honeypot data} \label{sec:honey}

The models described in Section~\ref{sec:models} are now applied to real data collected on a honeypot hosted within the Imperial College London (ICL) computer network. 
Within a time period between 21\textsuperscript{st} May, 2021, and 27\textsuperscript{th} January, 2022, the ICL honeypot collected approximately \numprint{40000} unique sessions, observed over 1.3 million times. This is a large corpus of attacks for a single machine.

\subsection{Data preprocessing} \label{sec:preprocessing}

As discussed in the introduction, such sessions and commands must be \textit{tokenised} to obtain the words and vocabulary. 
The tokenisation is performed with the \textit{Python} package NLTK \citep{Bird09}, setting the regular expression \verb|[a-zA-Z0-9_\.\-\*]+|. Also, commands observed in the ICL honeypot data often contain combinations of strings in hexadecimal form, preceded by the letter \mycode{x}. An example 
is:
\begin{minted}{shell}
bin busybox echo -e x6b x61 x6d x69 dev dev .nippon
\end{minted}
In these analyses, all such instances of hexadecimal strings (\mycode{x6b}, \mycode{x61}, \mycode{x6d} and \mycode{x69} in the above example) are replaced by the word \mycode{HEX}. Also, some commands display the word \mycode{GHILIMEA} appended to \mycode{HEX} strings. These are replaced with the word \verb|GHILIMEA_word|. Similarly to standard preprocessing techniques in natural language processing, extremely rare and extremely common words are removed from the dataset. For the ICL honeypot data, words appearing in less than 10 commands were removed, as were words appearing in over 10\% of commands. Such words are often denoted \textit{stopwords} in natural language processing and information retrieval \citep[see, for example,][]{Manning08}. After preprocessing, a vocabulary $\mathcal V$ of \numprint{1003} unique words is obtained, and \numprint{2617} uniquely observed sessions, for a total of \numprint{42640} commands and \numprint{261283} words.
Each session has an average of \numprint{16.29} commands (with median 15), whereas each command contains on average \numprint{6.12} words (with median 2). 
Given the malicious intent of the intruders, it is not uncommon to observe swear words and discriminatory language.  
Those terms have been redacted in 
the results. 
Additionally, a held-out corpus of \numprint{273} sessions was considered for testing. These documents preprocessed using the same procedure, resulting in a total of \numprint{5011} commands and \numprint{20177} words from the same vocabulary $\mathcal V$ constructed from the training set.

\subsection{Topic estimation} \label{sec:topic_estimation}

Before describing the results, some practical details about the 
estimation of topics from MCMC chains are discussed. 
In general, the number of session-level or command-level topics are unknown. The Dirichlet priors for $\vec\lambda$ and $\{\vec\phi_h\}$ in Section~\ref{sec:models} assume fixed, pre-specified values of $K$ and $H$. For inference with the Dirichlet prior, a maximum number of possible topics could be chosen, denoted $K_\text{max}$ and $H_\text{max}$, and the underlying number of topics could be estimated as the number of \textit{non-empty} topics at each iteration of the MCMC sampling procedure. 
Furthermore, estimates of topic allocations based on the MCMC sampler described in Section~\ref{sec:inference} could be affected by the issue of label switching \citep{Jasra05}. Therefore, session-level topic allocations are estimated in this work from the estimated posterior similarity between sessions $i$ and $j$, calculated as $\hat\pi _{ij}=\sum_{s=1}^{M} \mathds 1_{t^\star_{i,s}}\{t^\star_{j,s}\}/M$, where $M$ is the total number of posterior samples and $t^\star_{i,s}$ is the $s$-th sample for $t_i$. The posterior similarity matrix is invariant to permutations of the labels and therefore unaffected by label switching. After the posterior similarities are obtained for all pairs of sessions, hierarchical clustering with complete linkage is applied, with distance measure $1-\hat\pi_{ij}$ \citep{medve}. A similar procedure could be followed for the command-level topics, but the very large number of commands would make the size of the similarity matrix unfeasible to calculate and store in memory on a machine. Therefore, the last sample from the MCMC chain is considered as the estimate of the command-level topics. 
Furthermore, for comparisons between results of different models, topic-specific word distributions have been aligned via the Hungarian algorithm, commonly used in topic modelling \citep[see, for example,][]{Newman09}, using the Jensen-Shannon divergence as distance metric for comparing the estimated probability distributions. 

\subsection{Constrained Bayesian clustering (CBC)} \label{sec:mod1_fit}

\begin{figure}[!t]
\centering
\begin{subfigure}[t]{.55\textwidth}
\centering
\caption{Barplot of frequencies of estimated session-level topics}
\label{fig:mod1_1}
\includegraphics[height=4.25cm]{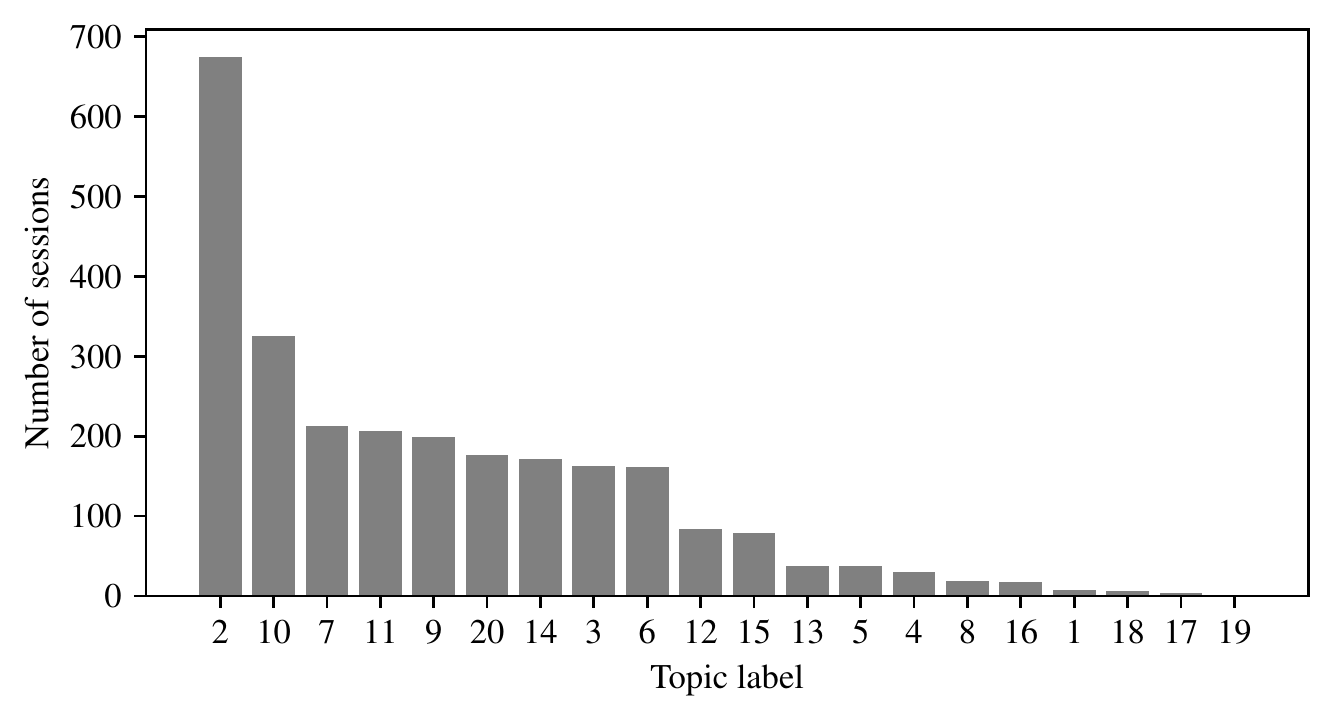}
\end{subfigure}
\begin{subfigure}[t]{.44\textwidth}
\centering
\caption{Barplot of estimated $K_\varnothing$}
\label{fig:mod1_2}
\includegraphics[height=4.25cm]{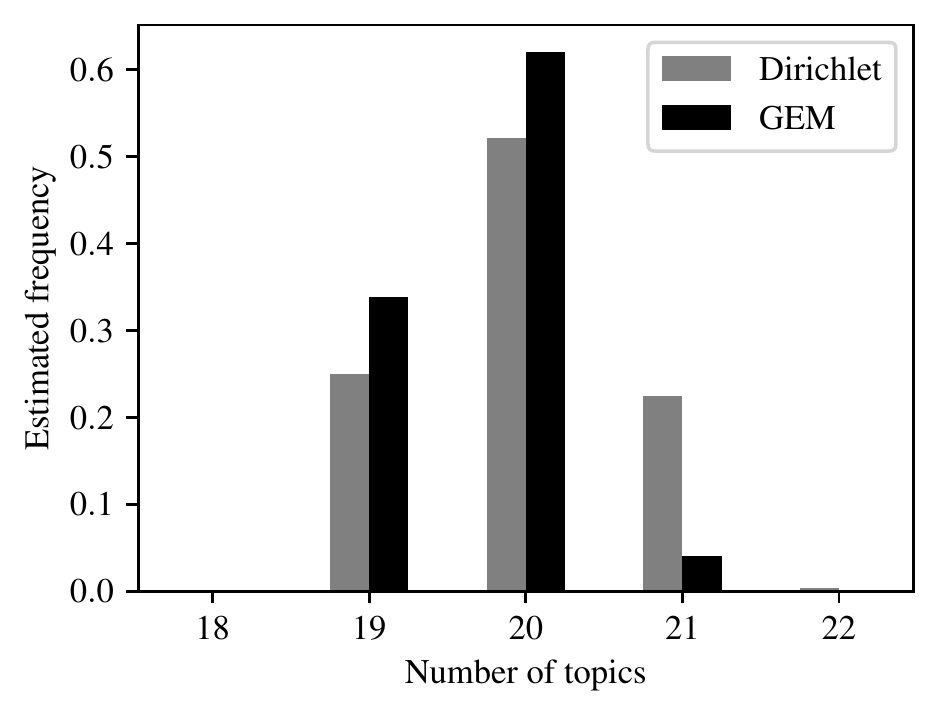}
\end{subfigure}
\caption{{Estimated topic frequencies and estimated distribution of the number of non-empty topics $K_\varnothing$ under the CBC model in \eqref{mod1}, fitted on the ICL honeypot data.}}
\label{fig:mod1_results}
\end{figure}

First, the CBC model in \eqref{mod1} is fitted on the postprocessed ICL honeypot data.
Under the Dirichlet prior for $\vec\lambda$, the hyperparameter $\vec\gamma$ is set to $\vec\gamma=0.1\cdot\vec 1_{K_\text{max}}$, with $K_\text{max}=30$. The hyperparameter of the Dirichlet prior for the topic-specific word distribution is set to $\vec\eta=\vec 1_{ V}$. 
The MCMC sampler is run for \numprint{250000} iterations with \numprint{50000} burn-in, initialising the topics via spectral clustering with $K_\text{max}$ clusters. The results are displayed in Figure~\ref{fig:mod1_results}. 
The session-level topics are estimated using the procedure described in Section~\ref{sec:topic_estimation}. Figure~\ref{fig:mod1_1} plots the resulting barplot of topic frequencies of estimated session-level topics, for a number of topics equal to the modal number of non-empty topics, $\hat K_\varnothing=20$. 
Furthermore, Figure~\ref{fig:mod1_2} displays the barplot of the estimated distribution for the number of non-empty topics. The barplot is also compared to the distribution obtained under a GEM prior for $\vec\lambda$, with hyperparameter $\gamma=3$, corresponding to $K_\text{max}\times 0.1$, using the same setup for the MCMC sampler. Additionally, the vocabulary is also assumed to be unbounded when GEM priors are used. The resulting distributions show agreement, demonstrating a similar performance of the Dirichlet and GEM priors in estimating the modal number of topics. In general, the interplay between the prior parameters $\vec\eta$ and $\vec\gamma$ appears to have an effect on the number of \textit{small} clusters that are estimated from the data: if $\vec\eta$ increases, the clusters in the right tail of Figure~\ref{fig:mod1_1} tend to be incorporated within the larger clusters. On the other hand, if $\vec\eta$ decreases towards zero, 
more topics are estimated.

Additional results are presented in Table~\ref{tab:mod_comparisons}, which reports the marginal log-likelihood per word on the training set and held-out sessions. Note that the table also contains the results for the remaining models fitted in this section, to ease comparisons. The table also contains results for the CBC model initialised at random and via the \textit{gensim} method described in Section~\ref{sec:initialisation}, showing that different initialisation methods lead to similar conclusions and results. 

\begin{table}[!t]
\caption{Average marginal log-likelihood per word on the training and held-out data from the Imperial College London honeypot for different configurations of the models fitted in Section~\ref{sec:honey}.}
\label{tab:mod_comparisons}
\scalebox{0.65}{
\begin{tabular}{ccccccccl}
& & & & & & Average marginal & Average marginal  \\
& & & & & &  log-likelihood per & log-likelihood per \\
Model & Prior & $K_\text{max}$ & $H_\text{max}$ & $\hat K_\varnothing$ & $\hat H_\varnothing$ & word (training data) & word (held-out data) & Initialisation \\ 
\hline\hline
CBC & Dirichlet & $30$& - & 20& - & $-2.30622$ & $-3.55584$ & Spectral\\
CBC & Dirichlet & $30$& - & 20& - & $-2.30766$ & $-3.55275$ & \textit{gensim} \\
CBC & Dirichlet & $30$& - & 20& - & $-2.31149$ & $-3.55624$ & Random \\
CBC & GEM & $\infty$ & - & 20& - & $-2.30846$ & $-3.55653$ & Spectral\\
CBC + secondary topic & Dirichlet & $30$& - & 20& - & $-2.56797$ & $-3.85121$ & CBC \\
CBC + secondary topic & Dirichlet & $20$& - & 20& - & $-2.57912$ & $-3.87293$ & CBC \\
CBC + secondary topic & Dirichlet & $30$& - & 20& - & $-2.57403$ & $-3.86829$ & Spectral + Random \\
NCBC & Dirichlet & 50 & 50 & 34 & 38 & $-2.11026$ & $-3.29999$ & Spectral \\
NCBC & Dirichlet & 50 & 50 & 36 & 37 & $-2.11128$ & $-3.29927$ & \textit{gensim} \\
NCBC & Dirichlet & 50 & 50 & 34 & 33 & $-2.11627$ & $-3.29252$ & Random \\
NCBC & GEM & $\infty$ & $\infty$ & 32 & 36 & $-2.12109$ & $-3.29956$ & Spectral \\
PCNBC & Dirichlet & 30 & 50 & 20 & 45 & $-2.18384$ & $-3.37315$ & Spectral \\
PCNBC & Dirichlet & 30 & 30 & 20 & 30 & $-2.18599$ & $-3.39523$ & Spectral \\
PCNBC & Dirichlet & 20 & 50 & 18 & 47 & $-2.17583$ & $-3.45624$ & Spectral \\
\hline
\end{tabular}
}
\end{table}

\subsubsection{Results: the model discovers a rare and unusual MIRAI variant}

The inferred meaning of each topic is summarised in Table~\ref{tab:mod1_interpretation}, according to the type of sessions that are assigned to each group. In addition, the 3 most representative words in the clusters for $p(w\mid t_d=k)$ and $p(t_d=k\mid w)$ are displayed in Table~\ref{tab:mod1_words}. The clusters appear to mostly contain botnets and different \textit{variants} of MIRAI malware. MIRAI is a type of botnet first emerged in 2016, which was specifically targeted towards compromising Internet of Things (IoT) devices, or launching Distributed Denial of Service (DDoS) attacks. Recently, it has been repurposed for Bitcoin mining on IoT devices compromised via brute-force attacks on protocols such as SSH and Telnet. Over the years, many different variants of MIRAI have emerged, and other bots with similar structure \citep{Lingenfelter20,Sadique21,Zhu22}, which appear to be assigned to different topics in Table~\ref{tab:mod1_interpretation}. Table~\ref{tab:mod1_words} shows that words within each topic are fairly heterogeneous, but a number of words appear frequently across multiple topics, such as \mycode{HEX}, \mycode{cd} or \mycode{sh}. 

Interestingly, careful examination of the sessions assigned to \textit{topic 5} estimated via the CBC model in \eqref{mod1} helped analysts to discover a rare and unusual variant of MIRAI, called MinerFinder \citep{Bevington21}. The objective of MinerFinder is to look for existing coin miner configurations, and try to gain root privileges to take control of the miner infrastructure, if found. This demonstrates that the CBC model with a single topic per session, even in its simplest form, could be extremely helpful for analysts to discover new attack patterns. Within \textit{topic 5}, MinerFinder is also mixed with other more common MIRAI variants, which share a common frequency distribution of words. Ideally, a clustering algorithm should be able to single-out MinerFinder from other MIRAI variants, despite their similarities. This might be possible when a nested structure is added to the topics, and the command structure is explicitly used, as demonstrated in Section~\ref{sec:mod3_fit}.   

\begin{table}[!t]
\caption{Estimated session-level topics and corresponding intent under the CBC model in \eqref{mod1}.}
\label{tab:mod1_interpretation}
\scalebox{0.65}{
\begin{tabular}{ccl}
Topic & Type of malware & Objective\\
\hline\hline
1 &	Shellbot & Install bot \\
2 &   (\mycode{ptmx}) unnamed botnet, MIRAI & Gather system information, change permissions, execute MIRAI variants \\
3 &	MIRAI & Download and execute MIRAI variants \mycode{kura} and \mycode{kurc}, fingerprint system \\
4 &	MIRAI & Download \mycode{sora} malware, write \mycode{upnp} and \mycode{updDl} malware  via echoing \mycode{HEX} strings \\ 
5 & MIRAI, \textbf{MinerFinder (new variant)} & Download malware, change permissions, gather system information, fingerprint system \\ 
6 &	Shellbot, SBIDIOT, coin miner & Download and execute coin miner and MIRAI malware, change SSH keys \\ 
7 &	(\mycode{s4y}, \mycode{LAYER}) unnamed botnet, MIRAI & Gather system information, change permissions, execute MIRAI variants \\
8 &	Coin miner & Download and execute coin mining malware \\
9 &	MIRAI & Download and execute MIRAI variants \mycode{PEDO}, \mycode{ECCHI}, \mycode{PEACH}...\\
10 &	MIRAI & Download and execute MIRAI variants \mycode{tftp1.sh}, \mycode{tftp2.sh}... \\
11 &	MIRAI & Determine shell executable, check \mycode{busybox} is present, print error message to console \\
12 &	(\mycode{misa}) unnamed botnets & Gather system information, change permissions, execute MIRAI variants \\
13 &	Shellbot, coin miner & Scan system, look for GPUs, look for coin miners, download malware \\ 
14 &	MIRAI & Download and execute MIRAI variants \mycode{sora}, \mycode{Pemex}...\\
15 &	MIRAI & Download and execute MIRAI variant \mycode{DNXFCOW} via echoing single \mycode{HEX} strings \\
16 &	MIRAI & Download and execute MIRAI variant \mycode{DNXFCOW} via echoing multiple \mycode{HEX} strings \\
17 &	MikroTik bot, coin miner & Gather system information, gather MikroTik router information, look for coin miners \\
18 &	GHILIMEA, PentaMiner coin miner script & Install coin miner, kill mining processes with high CPU usage in order to go undetected \\
19 &	MikroTik bot & Attempt to gain access to MikroTik router \\
20 &	Hive OS attack, coin miner & Download miner, attempt to take over configurations in Hive OS mining platform \\
\hline
\end{tabular}
}
\end{table}

\begin{table}[!t]
\caption{Top-3 words for $p(w\mid t_d=k)$ and $p(t_d=k\mid w)$ for the estimated session-level topics under CBC \eqref{mod1}.}
\label{tab:mod1_words}
\scalebox{0.65}{
\begin{tabular}{c | lll | lll}
Topic & \multicolumn{3}{c|}{Top-3 words for $p(w\mid t_d=k)$} & \multicolumn{3}{c}{Top-3 words for $p(t_d=k\mid w)$} \\[1pt]
\hline\hline
1 & \mycode{pkill} & \mycode{-rf} & \mycode{wget} & \mycode{sudo} & \mycode{hive-passwd} & \mycode{-L} \\
2 & \mycode{cd} & \mycode{.ptmx} & \mycode{var} & \mycode{.none} & \mycode{base64} & \mycode{con} \\
3 & \mycode{HEX} & \mycode{.nippon} & \mycode{cat} & \mycode{updDl} & \mycode{-ne} & \mycode{gsdfsdf424r24} \\
4 & \mycode{HEX} & \mycode{updDl} & \mycode{echo} & \mycode{ssh} & \mycode{sshd} & \mycode{bin.sh} \\
5 & \mycode{HEX} & \mycode{echo} & \mycode{tmp} & \mycode{.s4y} & \mycode{.LAYER} & \mycode{.lib} \\
6 & \mycode{cd} & \mycode{wget} & \mycode{tmp} & \mycode{kura} & \mycode{kurc} & \mycode{Uirusu} \\
7 & \mycode{cd} & \mycode{var} & \mycode{.s4y} & \mycode{.ptmx} & \mycode{.Switchblades} & \mycode{GSec} \\
8 & \mycode{sh} & \mycode{var} & \mycode{tsh} & \mycode{policy} & \mycode{7wmp0b4s.rsc} & \mycode{on-event} \\
9 & \mycode{HEX} & \mycode{PEDO} & \mycode{ECCHI} & \mycode{chroot} & \mycode{HEX} & \mycode{ftp} \\
10 & \mycode{cd} & \mycode{sh} & \mycode{chmod} & \mycode{GHILIMEA_word} & \texttt{then} & \texttt{fi} \\
11 & \mycode{shell} & \mycode{sh} & \mycode{enable} & \mycode{.cowbot.dropper} & \mycode{scanner.s.} & \mycode{rm7} \\
12 & \mycode{cd} & \mycode{.misa} & \mycode{var} & \mycode{sora2.sh} & \mycode{sora1.sh} & \mycode{sensi.sh} \\
13 & \mycode{grep} & \mycode{head} & \mycode{-c} & \mycode{retrieve} & \mycode{DNXFCOW} & \mycode{scanner.syn.} \\
14 & \mycode{cd} & \mycode{sh} & \mycode{chmod} & \mycode{cloud} & \mycode{-t} & \mycode{sms} \\
15 & \mycode{HEX} & \mycode{cd} & \mycode{.file} & \mycode{name} & \mycode{30} & \mycode{cpuinfo} \\
16 & \mycode{HEX} & \mycode{retrieve} & \mycode{DNXFCOW} & \mycode{.misa} & \mycode{buzz} & \mycode{GangWolf} \\
17 & \mycode{ip} & \mycode{find} & \mycode{remove} & \mycode{curl.sh} & \mycode{wget.sh} & \mycode{tftp.sh} \\
18 & \mycode{GHILIMEA_word} & \mycode{tmp} & \texttt{then} & \mycode{ECCHI} & \mycode{PEDO} & \mycode{mika} \\
19 & \mycode{policy} & \mycode{7wmp0b4s.rsc} & \mycode{None} & \mycode{tsh} & \mycode{1sh} & \mycode{-qO} \\
20 & \mycode{pkill} & \mycode{-s} & \mycode{sudo} & \mycode{LZRD} & \mycode{MIRAI} & \mycode{eb0t} \\
\hline
\end{tabular}
}
\end{table}

Overall, most of the activity on the honeypot seems to be related to attempts to install botnets or coin miners, but topics show remarkable separation between sessions and corresponding intents, as demonstrated in the list of malware and objectives in Table~\ref{tab:mod1_interpretation}. 
A potential solution to better visualise and further differentiate topic-specific word distributions would be to introduce a \textit{secondary topic}, which would capture the distribution of the most common words shared across multiple topics. 
This solution is explored in the next section. 

\subsection{Constrained Bayesian clustering with a secondary topic} \label{sec:mod2_fit}

As discussed in the previous section, a possible solution to aid topic interpretability and further discriminate topic-specific word distributions would be to add a secondary topic to the model. In this section, the CBC model with secondary topic in \eqref{mod2} is therefore fitted to the ICL honeypot data. The setup of the MCMC sampler is chosen to be identical to the previous section, and the additional hyperparameters are set to $\alpha_0=0.1$ and $\alpha_k=0.9$ for $k=1,\dots,K_\text{max}$, resulting in a prior probability of 90\% for a word to be allocated to a primary topic. The sampler is initialised using the topics estimated by the CBC model in \eqref{mod1}, fitted in Section~\ref{sec:mod1_fit}. 
The indicators $z_{d,j,i}$ are initialised from a Bernoulli distribution with probability equal to the proportion of documents in which the word $w_{d,j,i}$ occurred. For additional comparisons, the model was also fitted setting $K_\textit{max}=20$ corresponding to the value $\hat K_\varnothing$ estimated via CBC in Section~\ref{sec:mod1_fit}. Additionally, the Gibbs sampler was also run with random initial values for the primary-secondary topic indicators, and spectral clustering for the session-level groups. Results are displayed in Table~\ref{tab:mod_comparisons}, demonstrating agreement across different initialisation procedures, but inferior performance in terms of average marginal log-likelihood compared to standard CBC. 

\begin{figure}[!t]
\centering
\begin{minipage}{.475\textwidth}
\begin{subfigure}[t]{\textwidth}
\centering
\caption{Jensen-Shannon distances between estimated topics}
\label{fig:hm2_1}
\includegraphics[width=\textwidth]{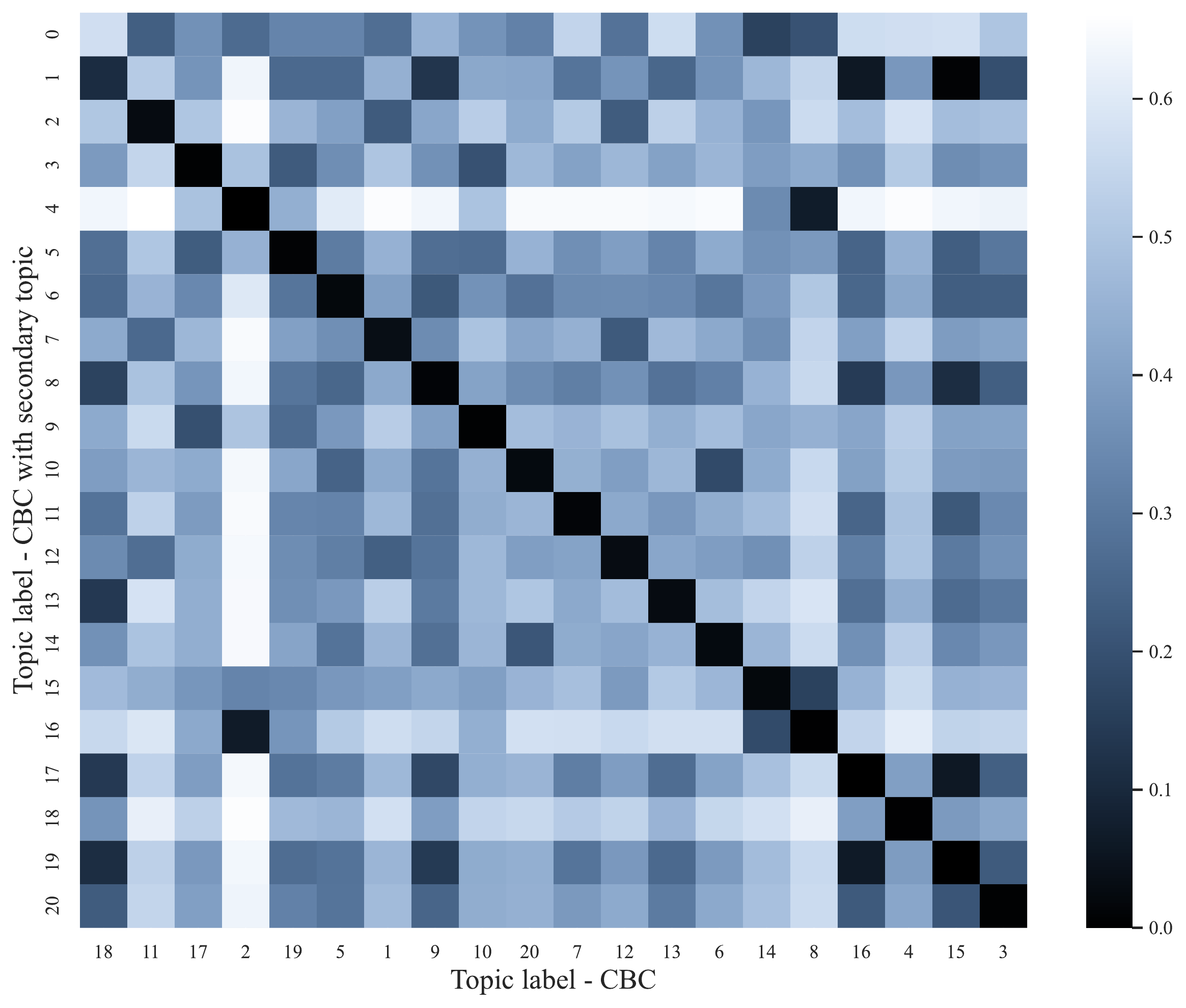}
\end{subfigure}
\end{minipage}
\hspace*{0.025\textwidth}
\begin{minipage}{.475\textwidth}
\begin{subfigure}[t]{\textwidth}
\centering
\caption{Jaccard coefficients between estimated groups}
\label{fig:hm2_2}
\includegraphics[width=0.7\textwidth]{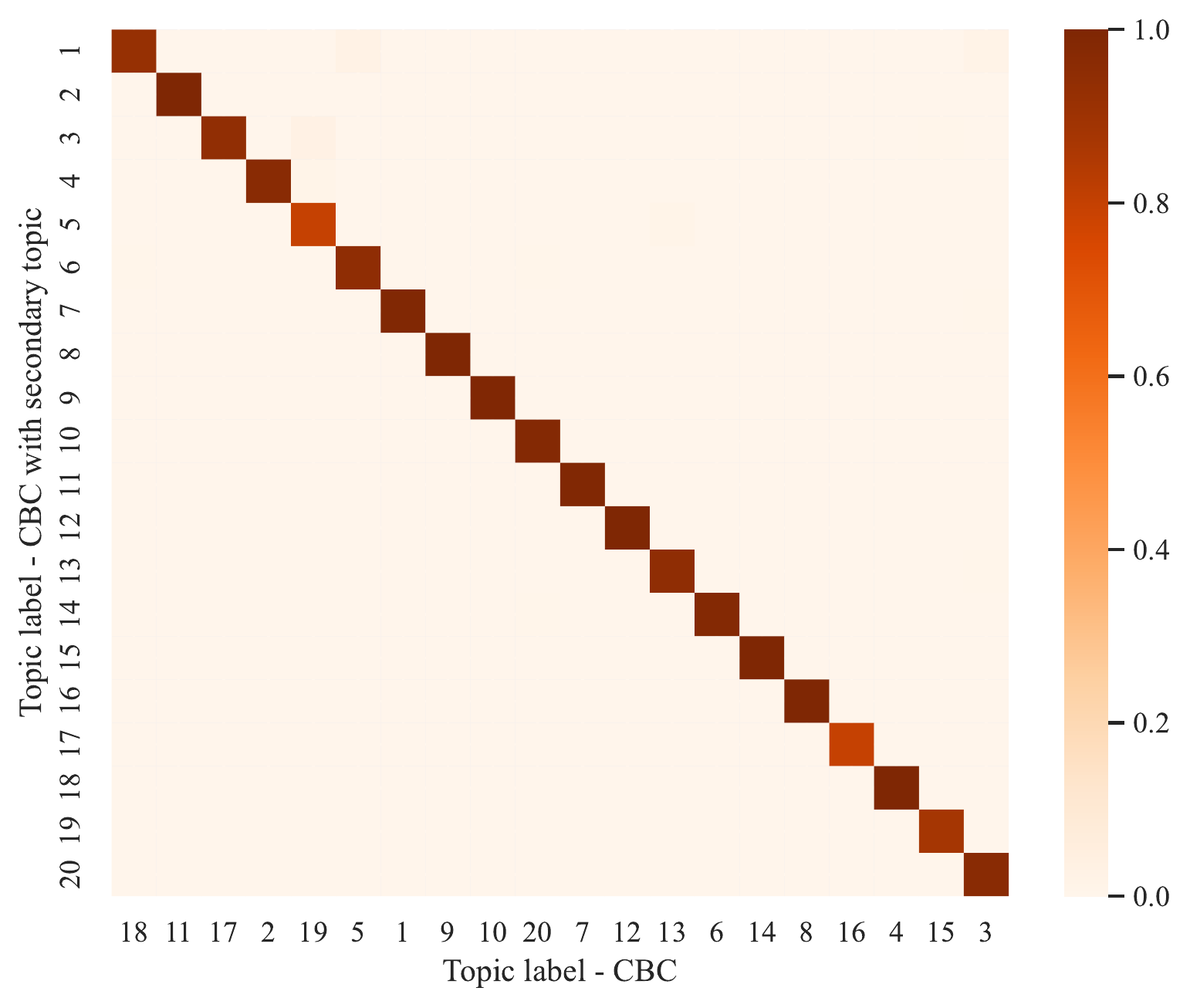}
\end{subfigure}
\begin{subfigure}[t]{\textwidth}
\centering
\caption{Entropy difference between estimated word distributions}
\label{fig:hm2_3}
\includegraphics[width=\textwidth]{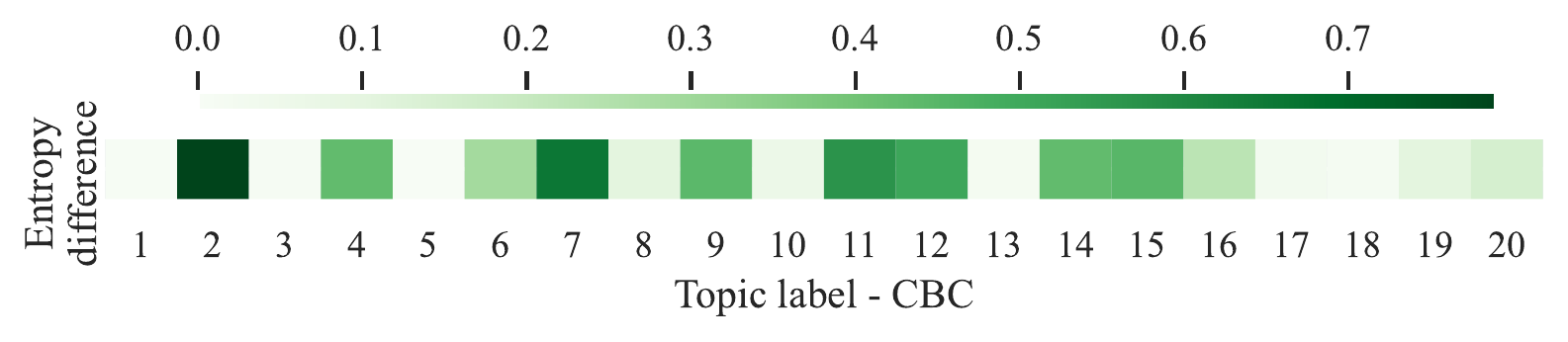}
\end{subfigure}
\end{minipage}
\caption{Heatmaps of Jensen-Shannon divergences between the session-level word distributions, Jaccard coefficients for the estimated session-level groups, and difference in entropy of the estimated topic distributions under CBC \underline{without} secondary topic and CBC \underline{with} secondary topic, after alignment via the Hungarian algorithm. } 
\label{fig:heatmap_secondary}
\end{figure}

Figure~\ref{fig:hm2_1} displays the Jensen-Shannon divergences between the estimated session-level topics and secondary topics, aligned to the topics discovered via CBC in Section~\ref{sec:mod1_fit} via the Hungarian algorithm. The plot shows that the topic distributions estimated via the two models are in agreement, whereas the distribution of \textit{topic 0} is not closely related to any of the other distributions, implying that the estimated values of $\theta_k$, representing the topic-specific secondary topic proportions, are relatively low. 
The multivariate Jensen-Shannon divergence between the topics obtained via CBC with secondary topic is $3.57989$, whereas the value decreases to $3.23813$ for the CBC model without secondary topic, fitted in Section~\ref{sec:mod1_fit}. This demonstrates that the topic distributions estimated from the model with secondary topics are \textit{more heterogeneous}, which could explain why cyber analysts found them more informative for estimating the intents associated with each topic. 
Additionally, Figure~\ref{fig:hm2_2} shows the Jaccard similarity scores between the groups estimated from the model with and without secondary topics, demonstrating agreement in the clusters obtained from the two methodologies. 
Figure~\ref{fig:hm2_3} demonstrates that when the secondary topic is \textit{not} used, a more even usage of the vocabulary occurs, resulting in more overlap between the estimated word distributions and higher entropy. On the other hand, when the secondary topic is used, the topic distributions appear to have smaller entropy, corresponding to less usage of the vocabulary and overlap.  

With a secondary topic, words are implicitly sampled from the mixture distribution $\tilde{\vec\phi}_{t_d} = \theta_{t_d}\vec\phi_{t_d} + (1-\theta_{t_d})\vec\phi_0$. This could aid interpretability of the latent intent of the session-level topic $t_d$, since the common words shared across most topics are filtered out and included in $\vec\phi_0$ instead; this comes at the expense of introducing a large number of additional parameters via the indicators $z_{d,j,i}$, which could cause a decrease in the marginal likelihood. This is confirmed when comparing Table~\ref{tab:mod2_words} with Table~\ref{tab:mod1_words} (obtained from a model that \textit{does not include} a secondary topic). 
Overall, Table~\ref{tab:mod2_words} shows that the secondary topic captures the most common words across documents, such as the string \mycode{HEX} and simple shell commands, for example \mycode{cd}, \mycode{tmp}, \mycode{var} (and other common commands such as \mycode{chmod}, \mycode{echo} and \mycode{shell}, excluded from the table). When comparing with Table~\ref{tab:mod2_words}, the relevant topics appear to be less dominated by common words. For example, for topic 1, the words \mycode{tmp} and \mycode{cd} appear to be among the most representative words in Table~\ref{tab:mod1_words}, but they have a much less prominent role in Table~\ref{tab:mod2_words}, where words such as \mycode{x86_64}, \mycode{uname} and \mycode{Xorg} gain importance (see also additional figures in the Supplementary Material). 
Qualitative analysis of these results from cyber-security analysts confirmed that these words appear to be more representative of the actual intents of the session, which is particularly helpful when communicating results. 

Overall, the assumption of having only one topic per session might be limiting, even if an additional secondary shared topic is added. This is mainly because most sessions could be considered as mixtures of commands, where each command has its own intent. Therefore, a more precise clustering of topics and commands might be provided by the Nested Constrained Bayesian clustering Model (NCBC) in Section~\ref{sec:hctm}, considered in the next section.

\begin{table}[!t]
\caption{Top-4 words 
for a subset of the estimated session-level topics under CBC with secondary topic \eqref{mod2}, after alignment with the results of CBC in Section~\ref{sec:mod1_fit} via the Hungarian algorithm with Jensen-Shannon divergence.}
\label{tab:mod2_words}
\scalebox{0.65}{
\begin{tabular}{c | llll | llll}
Topic & \multicolumn{4}{c|}{Top-4 words for $p(w\mid t_dz_{d,j,i}=k)$} & \multicolumn{4}{c}{Top-4 words for $p(t_dz_{d,j,i}=k\mid w)$} \\[1pt]
\hline\hline
0 & \mycode{HEX} & \mycode{cd} & \mycode{tmp} & \mycode{var} & \mycode{LZRD} & \mycode{MIRAI} & \mycode{eb0t} & \mycode{BOTNET} \\
1 & \mycode{pkill} & \mycode{tmp} & \mycode{x86_64} & \mycode{history} & \mycode{sudo} & \mycode{Xorg} & \mycode{LC_ALL} & \mycode{x11vnc} \\
7 & \mycode{.s4y} & \mycode{cd} & \mycode{var} & \mycode{.LAYER} & \mycode{beastmode} & \mycode{.Switchblades} & \mycode{dark} & \mycode{19ju3d} \\
10 & \mycode{tftp1.sh} & \mycode{sh} & \mycode{tftp2.sh} & \mycode{cd} & \mycode{chroot} & \mycode{x86_64} & \mycode{ftp} & \mycode{a.sh} \\
14 & \mycode{sh} & \mycode{tftp} & \mycode{anonymous} & \mycode{777} & \mycode{retrieve} & \mycode{DNXFCOW} & \mycode{scanner.syn.} & \mycode{-en} \\
\hline
\end{tabular}
}
\end{table}
 
\subsection{Nested constrained Bayesian clustering (NCBC)} \label{sec:mod3_fit}

\begin{figure}[!t]
\centering
\begin{subfigure}[t]{.55\textwidth}
\centering
\caption{Barplot of frequencies of estimated session-level topics}
\label{fig:mod3_1}
\includegraphics[height=3.85cm]{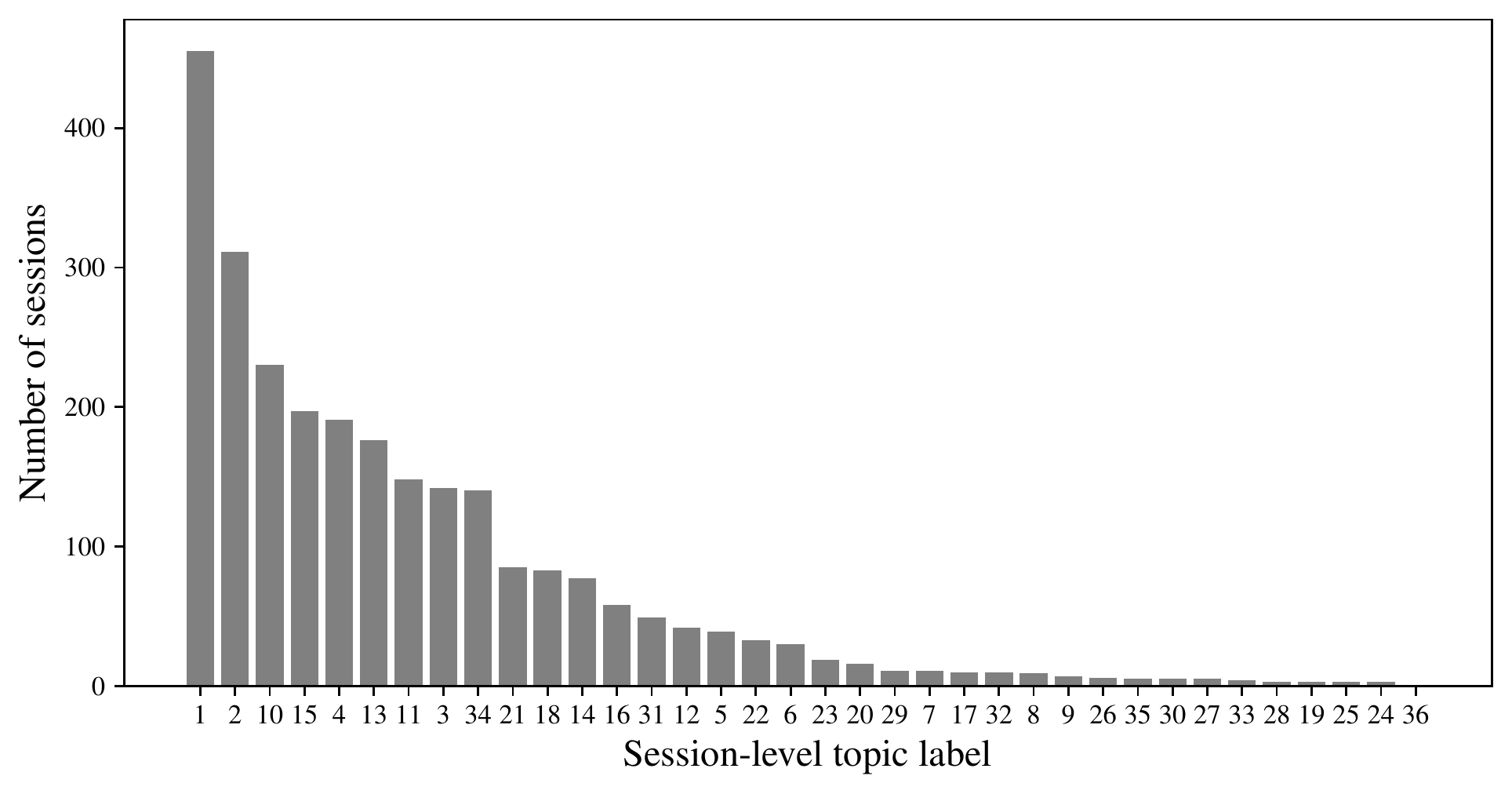}
\end{subfigure}
\begin{subfigure}[t]{.44\textwidth}
\centering
\caption{Barplot of estimated $K_\varnothing$}
\label{fig:mod3_2}
\includegraphics[height=4cm]{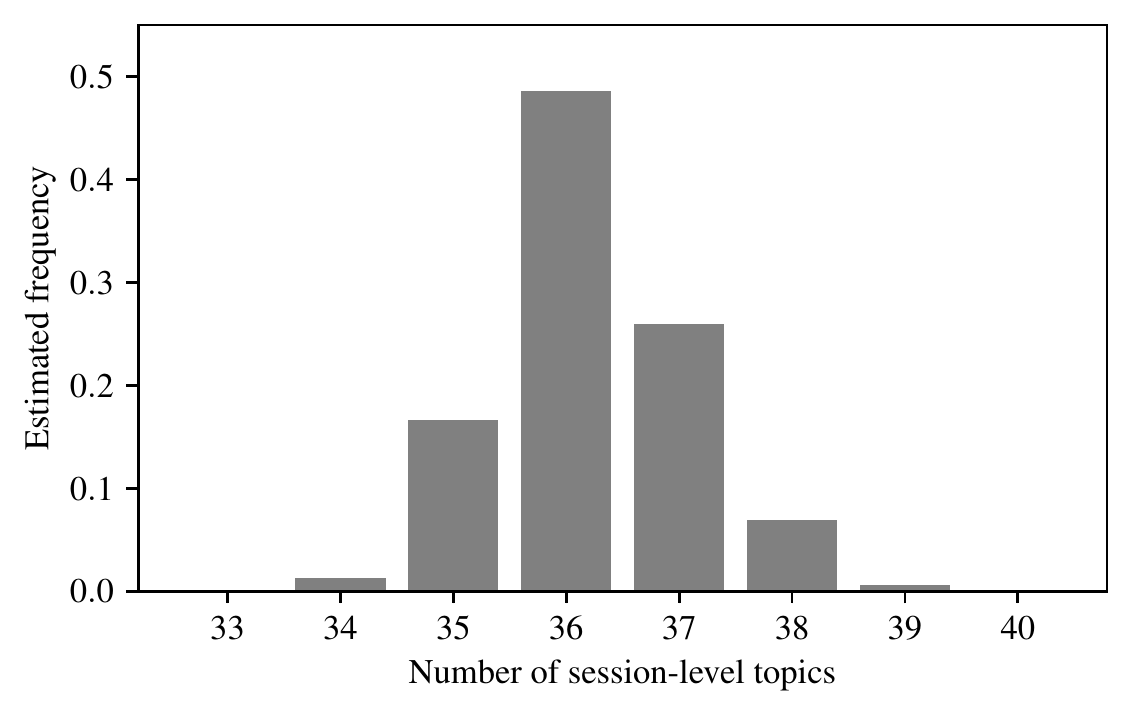}
\end{subfigure}
\begin{subfigure}[t]{.55\textwidth}
\centering
\caption{Barplot of frequencies of estimated command-level topics}
\label{fig:mod3_3}
\includegraphics[height=3.85cm]{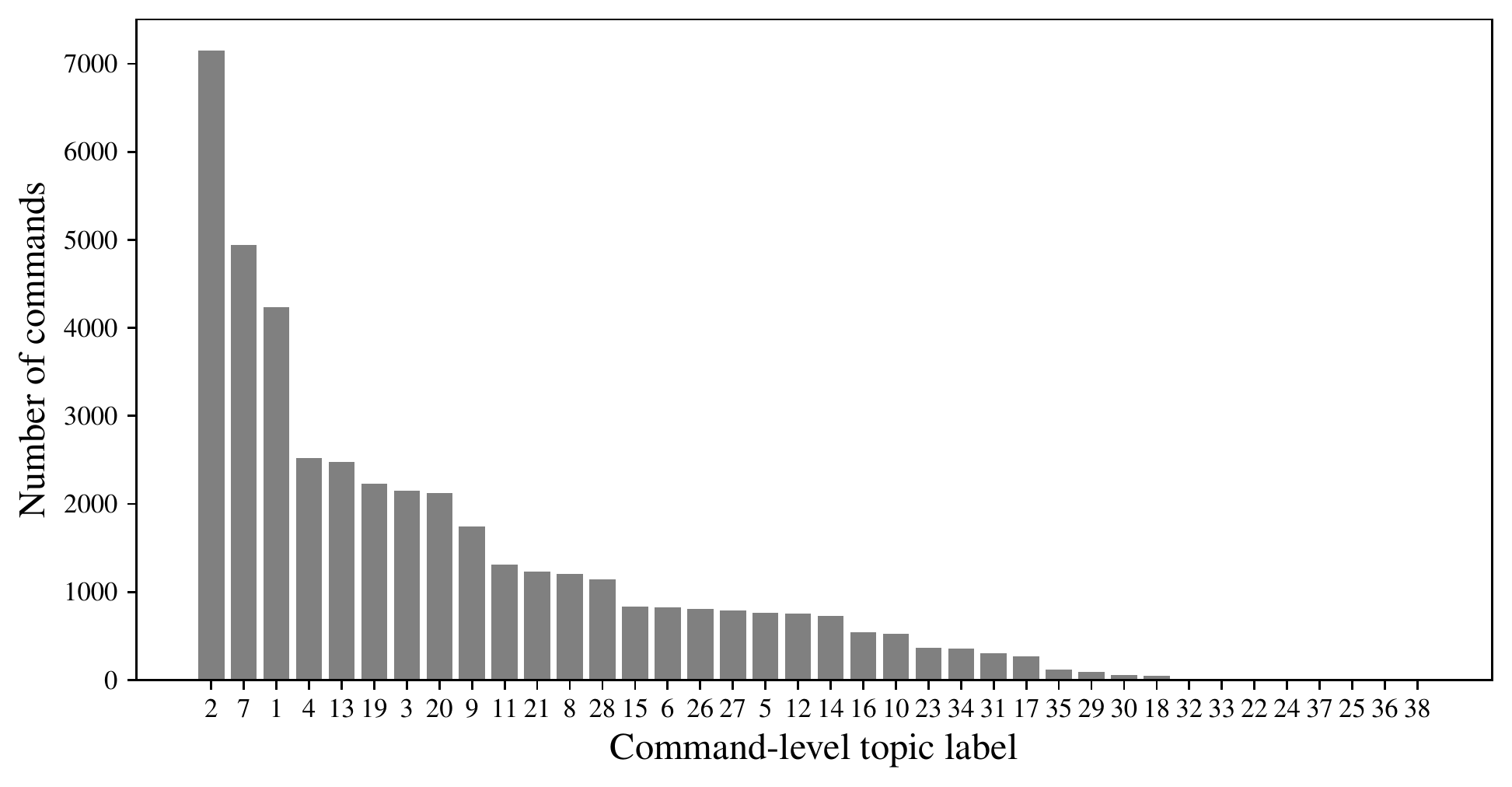}
\end{subfigure}
\begin{subfigure}[t]{.44\textwidth}
\centering
\caption{Barplot of estimated $H_\varnothing$}
\label{fig:mod3_4}
\includegraphics[height=4cm]{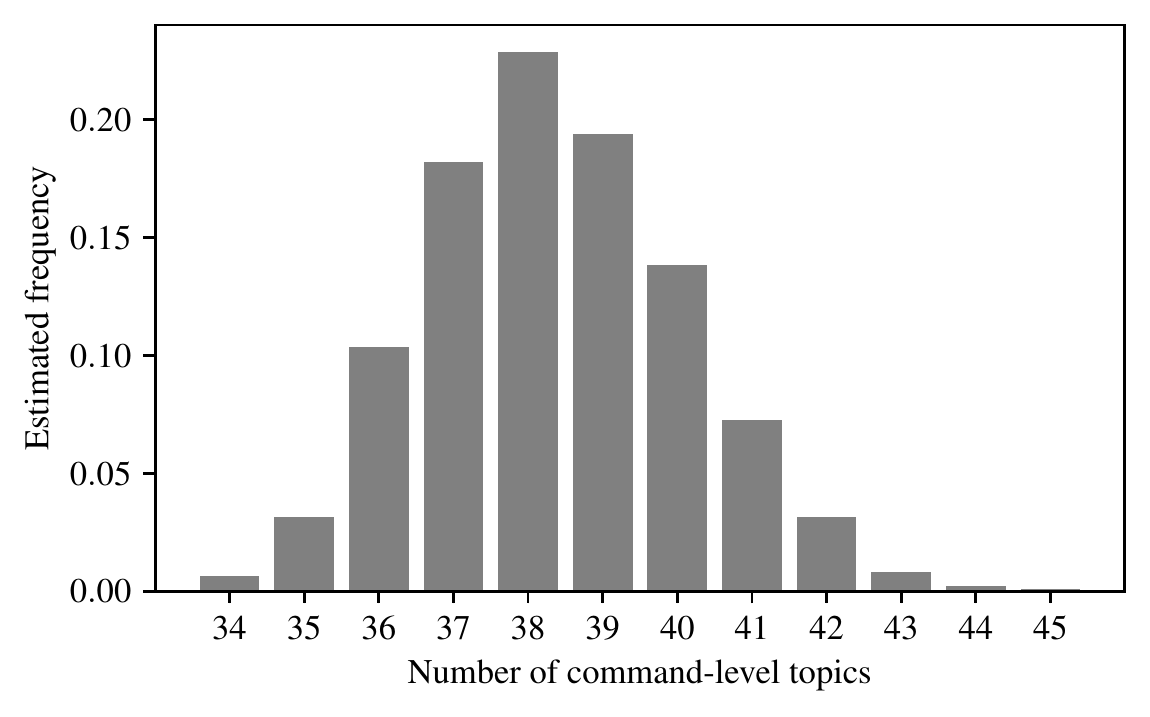}
\end{subfigure}
\caption{{Frequency distributions of the estimated session-level and command-level topics, and estimated distribution of the number of non-empty session-level topics $K_\varnothing$ and command-level topics $H_\varnothing$, under the nested constrained model in \eqref{mod3}, fitted on the ICL honeypot data.}}
\label{fig:mod3_results}
\end{figure}

As discussed in the previous section, the NCBC in Section~\ref{sec:hctm} could help to further elucidate the underlying group structure within the ICL honeypot data. Similarly to Section~\ref{sec:mod1_fit}, the MCMC is run for \numprint{250000} iterations with \numprint{50000} burn-in. The command-level and session-level topics are initialised via the spectral clustering algorithm described in Section~\ref{sec:initialisation}, setting Dirichlet priors of dimension $K_\text{max}=50$ and $H_\text{max}=50$, with hyperparameters $\vec\eta=\vec 1_{ V},\ \vec\tau=0.1\cdot\vec 1_{H_\text{max}}, \vec\gamma=0.1\cdot\vec 1_{K_\text{max}}$. Table~\ref{tab:mod_comparisons} also shows additional comparisons. The session-level and command-level topics are estimated following the procedure described in Section~\ref{sec:topic_estimation}, setting $\hat K_\varnothing = 36$ and $\hat H_\varnothing=38$, corresponding to the modal number of non-empty topics. Figure~\ref{fig:mod3_results} displays the frequency distribution of the estimated session-level (Figure~\ref{fig:mod3_1}) and command-level topics (Figure~\ref{fig:mod3_3}), followed by the estimated distributions of the number of non-empty session-level (Figure~\ref{fig:mod3_2}) and command-level topics (Figure~\ref{fig:mod3_4}). 

Table~\ref{tab:mod_comparisons} shows that NCBC achieves the best performance in terms of average marginal log-likelihood per word across all models considered in this work. Similar results are obtained when utilising different initialisation schemes, such as \textit{gensim} or random intialisation, or when using GEM priors over the Dirichlet distribution. This is not surprising: the CBC model in \eqref{mod1} gives the \textit{same} primary topic to all words in a session, whereas the NCBC admits command-specific topics, providing more flexibility. In the ICL honeypot data, and more generally in session data, individual commands tend to have a specific intent identified by specific words in the command (for example, \mycode{wget} for downloading files from a web server under HTTP, HTTPS and FTP). Therefore, having command-specific topics helps in identifying the intents of individual commands, making the command-level topic-specific word distributions highly interpretable. 

Similarly to the previous section, the intents for the estimated session-level and command-level topics are summarised in Table~\ref{tab:mod3_interpretation} and \ref{tab:mod3_interpretation2}. In general, the two tables show that some topics correspond to the same intent, achieved through different words or commands. In particular, Table~\ref{tab:mod3_interpretation} shows the same malware types as Table~\ref{tab:mod1_interpretation}, with similar objectives. 
Similarly to the results in the previous section, different MIRAI variants are observed, and malware types such as shellbots, coin miners, the Hive OS attack and the MikroTik bot are all allocated to separate topics. 
In addition to the session-level objectives in Table~\ref{tab:mod1_interpretation}, Table~\ref{tab:mod3_interpretation} also shows topics representing reconnaissance sessions, where intruders attempted to gather system information, for example by checking directories and determining shell executables. 
MinerFinder is again discovered and relevant sessions are 
singled-out in the topic with label 24. In particular, under the nested constrained model, MinerFinder is explicitly split from the similar MIRAI variants that were present in \textit{topic 5} under the CBC model (\textit{cf.} Table~\ref{tab:mod1_interpretation}), which are instead allocated to \textit{topic 22} under the NCBC (\textit{cf.} Table~\ref{tab:mod3_interpretation}). If the topics are estimated using the initialisation procedures described in Section~\ref{sec:initialisation}, MinerFinder is usually allocated to large clusters, with a number of sessions ranging between $80$ and $\numprint{1000}$. 

\begin{table}[!t]
\caption{Estimated session-level topics and corresponding intent under the nested constrained model in \eqref{mod3}.}
\label{tab:mod3_interpretation}
\scalebox{0.65}{
\begin{tabular}{ccl}
Topic & Type of malware & Objective\\
\hline\hline
1 &	MIRAI & Check shell and directories, download and execute MIRAI malware, delete files \\
2 &   (\mycode{ptmx}) unnamed botnet, MIRAI & Gather system information, change permissions, execute MIRAI variants \\
3 &	Shellbot, coin miner, MIRAI & Download and execute coin miner and MIRAI malware \\ 
4 &   MIRAI & Determine shell executable, check \mycode{busybox} is present, print error message to console \\
5 &	Shellbot, coin miner & SSH backdoor botnet, download malware, change SSH keys, check CPU/GPU information \\
6 &   MIRAI & Download and execute MIRAI malware (for example, \mycode{garm} or \mycode{gmips}), delete files \\
7 & Shellbot, coin miner & Gather CPU/GPU information, download coin miners, kill existing coin miners \\
8 & Reconnaissance & Check mounted file system, gather system information, fingerprint system  \\
9 & MIRAI & Write malware (for example \mycode{dvrHelpwer} and \mycode{updDl}) via echoing \mycode{HEX} strings \\ 
10 & Reconnaissance & Check shell and directories, delete files (\mycode{.ptmx}, \mycode{Switchblades}) \\
11 & Hive OS attack, coin miner & Download miner, attempt to take over configurations in Hive OS mining platform \\ 
12 & MIRAI & Execute MIRAI variant \mycode{PEDO}, fingerprint system \\ 
13 & MIRAI & Execute MIRAI variants \mycode{kura} and \mycode{kurc}, fingerprint system \\ 
14 & MIRAI & Determine shell, download and execute MIRAI variant \mycode{DNXFCOW} via echoing \mycode{HEX} strings \\
15 & Reconnaissance & Check shell and directories, delete files (\mycode{.ptmx}, \mycode{.s4y}) \\
16 & MIRAI & Download and execute MIRAI variant \mycode{PEDO} and \mycode{mika}, fingerprint system \\ 
17 & Coin miner & Download coin miner (\mycode{c3pool}) \\
18 & MIRAI & Download, execute MIRAI variant \mycode{ECCHI}, delete files, fingerprint system \\
19 & MIRAI & Download malware, write \mycode{updDl} malware via echoing \mycode{HEX} strings \\ 
20 & MIRAI & Download \mycode{sora} malware, write \mycode{upnp} and \mycode{updDl} malware via echoing \mycode{HEX} strings \\
21 & MIRAI & Gather system information, change permissions, execute MIRAI variants \\
22 & MIRAI & Download malware, change permissions, gather system information, fingerprint system \\
23 & Coin miner, SBIDIOT & Download and execute coin mining malware \\
24 & \textbf{MinerFinder (new MIRAI variant)} & Changes SSH keys, looks for coin miners, attempt to take over miners \\
25 & MIRAI & Check shell and directories, execute MIRAI malware (\mycode{Skyline} and \mycode{Akim}), delete files \\
26 & GHILIMEA, PentaMiner coin miner script & Install coin miner, kill mining processes with high CPU usage in order to go undetected \\
27 & Reconnaissance & Attempt to read and change SSH keys \\
28 & MIRAI & Write \mycode{RONALD} malware via echoing \mycode{HEX} strings \\ 
29 & MIRAI & Gather system information, print error message (component of \mycode{DNXFCOW} MIRAI variant) \\
30 & MIRAI & Gather system information, change permissions, execute MIRAI variant \mycode{cowffxxna} \\ 
31 &	MikroTik bot, coin miner & Gather system information, gather MikroTik router information, kill existing coin miners \\
32 &	MIRAI & Download and execute MIRAI variant \mycode{DNXFCOW} via echoing multiple \mycode{HEX} strings \\
33 & Shellbot, coin miner & Scan system, look for GPUs, look for coin miners, download malware \\
34 & MIRAI & Gather system information, change permissions, execute MIRAI variants \\
35 & Coin miner & Download coin miner (\mycode{TeamTNT}) \\
36 & MikroTik bot & Remove firewall NAT rules on MikroTik router \\
\hline
\end{tabular}
}
\end{table}

\begin{figure}[!t]
\centering
\begin{subfigure}[t]{.475\textwidth}
\centering
\caption{NCBC vs. Spectral clustering}
\label{fig:spectral_ncbc}
\includegraphics[width=.9\textwidth]{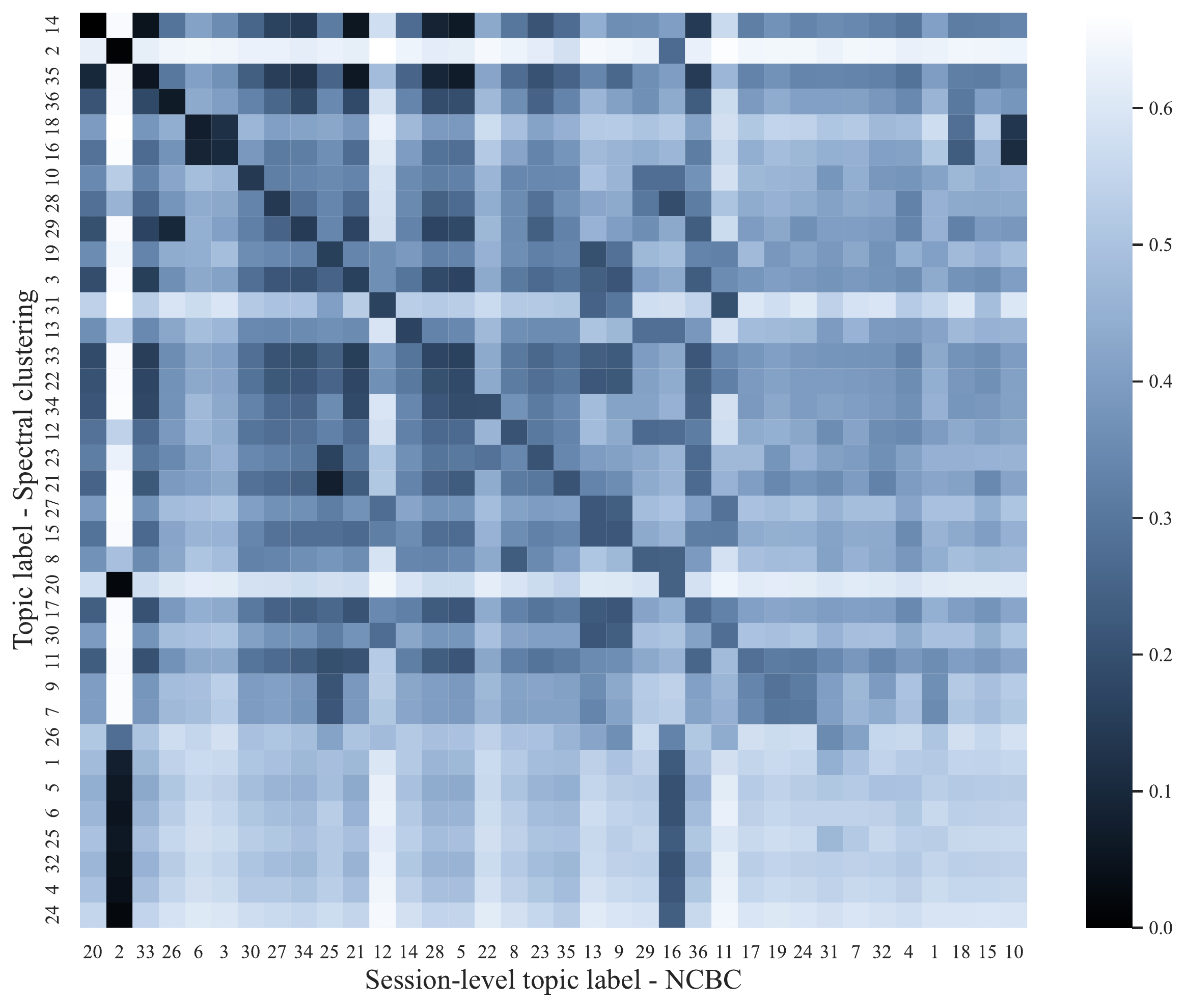}
\end{subfigure}
\hspace*{0.025\textwidth}
\begin{subfigure}[t]{.475\textwidth}
\centering
\caption{NCBC vs. LDA}
\label{fig:spectral_lda}
\includegraphics[width=.9\textwidth]{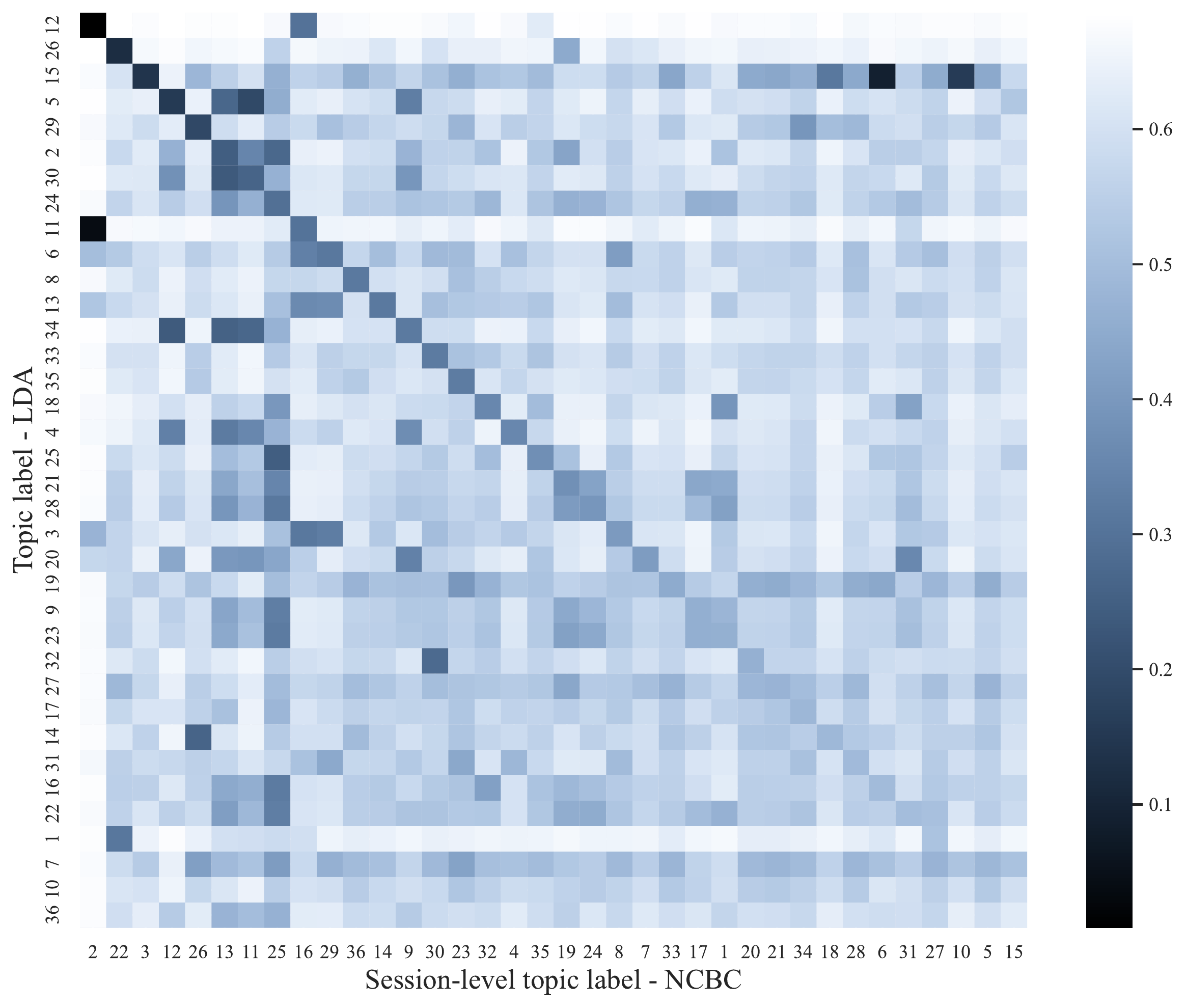}
\end{subfigure}
\caption{Heatmap of Jensen-Shannon divergences between the session-level word distributions under NCBC, compared to LDA and spectral clustering. Topics are aligned via the Hungarian algorithm and the resulting distance matrix is sorted by the diagonal entries in increasing order, after alignment. } 
\label{fig:heatmap_js_ncbc}
\end{figure}

It must be remarked that MinerFinder is \textit{not detected} using alternative clustering approaches based on spectral clustering or standard LDA fitted via \textit{gensim} (\textit{cf.} Section~\ref{sec:initialisation}). This is confirmed by Figure~\ref{fig:heatmap_js_ncbc}, which shows a comparison between the topics discovered via NCBC and the results of spectral clustering (Figure~\ref{fig:spectral_ncbc}) and Latent Dirichlet Allocation fitted via \textit{gensim} (Figure~\ref{fig:spectral_lda}) with $K_{\text{max}}=36$. In both plots, no distribution appears to be close to \textit{topic 24} obtained via NCBC, which exclusively contains MinerFinder. 

\begin{table}[!t]
\caption{Estimated command-level topics and corresponding intent under the nested constrained model in \eqref{mod3}.}
\label{tab:mod3_interpretation2}
\scalebox{0.65}{
\begin{tabular}{cl}
Topic & Objective\\
\hline
\hline
1 &	Attempt to write file \mycode{.ptmx} or \mycode{LAYER} to directory \mycode{var}, \mycode{run} or \mycode{tmp} and change directory if writeable \\ 
2 &	Attempt to write file \mycode{.ptmx} to directory \mycode{netslink}, \mycode{mnt} or \mycode{shm} and change directory if writeable \\
3 &	Attempt to copy, write and delete files \\ 
4 &	Attempt to grant full permissions to all users on malware files \\
5 &	Gather system information about CPU architectures \\
6 &	Kill processes, gather system information, Hive OS logon attempt \\
7 &	Check available commands, determine shell executable \\
8 &	Check if \mycode{busybox} exists, print error message to console (component of \mycode{PEDO} MIRAI variant) \\
9 &	Gather system information about mounted file systems, copy files \\
10 &	Fingerprint readable and writeable directories to hidden file \mycode{.nippon} \\
11 &	Download malware from internet using \mycode{wget} and \mycode{curl} \\
12 &	Attempt to read and delete files \\
13 &	Attempt to execute malware, delete files after execution \\
14 &	Check if directories such as \mycode{var}, \mycode{run} or \mycode{tmp} exist, by attempting to change directory \\
15 &	Download malware using \mycode{tftp} \\
16 &	Download, execute and delete malware files \\
17 &	Download malware using \mycode{ftpget} \\
18 &	Download and install GHILIMEA coin mining malware, kill own processes if CPU usage is high \\
19 &	Check if \mycode{busybox} exists, check available commands, determine shell executable \\
20 & Check if \mycode{busybox} exists, print error message to console (component of \mycode{KURA} and \mycode{OWARI} MIRAI variants) \\
21 & Check if \mycode{busybox} exists, print error message to console (component of \mycode{ECCHI} MIRAI variant) \\
22 & Download malware to attempt to compromise MikroTik router \\
23 & Look for existing coin miners, gather GPU and CPU information \\ 
24 & Attempt to gather information about MikroTik router \\
25 & Determine shell executable \\
26 & Attempt to write file \mycode{.file} to directory \mycode{netslink}, \mycode{mnt} or \mycode{boot} and change directory if writeable \\
27 & Copy, grant full permission, execute and delete \mycode{.cowbot} malware (MIRAI variant) \\
28 & Write malware binary to disk via echoing \mycode{HEX} bits \\
29 & Download coin miner malware from \mycode{c3pool} \\
30 & Read SSH authorised keys, attempt to delete and replace authorised keys \\ 
31 & Exit shell (part of GHILIMEA coin mining malware) \\
32 & Install GHILIMEA coin mining malware, kill own processes if CPU usage is high \\
33 & Check internet connectivity, check if GHILIMEA is already installed, kill own processes if CPU usage is high \\ 
34 & Read and delete files (for example, \mycode{.none} and \mycode{.human}) \\ 
35 & Check if \mycode{busybox} exists, print error message to console (component of \mycode{RONALD} and \mycode{Akim} MIRAI variant) \\
36 & Remove firewall NAT rules on MikroTik router \\
37 & Remove temporary directory \mycode{p} (singleton cluster) \\
38 & Exit shell (singleton cluster) \\
\hline
\end{tabular}
}
\end{table}

\begin{table}[!t]
\caption{Top 3 words for $p(w\mid s_{d,j}=\ell)$ and $p(s_{d,j}=\ell \mid w)$ for the estimated command-level topics under NCBC \eqref{mod3}.}
\label{tab:mod3_words}
\scalebox{0.55}{
\begin{tabular}{c | lll | lll}
Topic & \multicolumn{3}{c|}{Top-3 words for $p(w\mid s_{d,j}=\ell)$} & \multicolumn{3}{c}{Top-3 words for $p(s_{d,j}=\ell \mid w)$} \\[1pt]
\hline\hline
1 & \mycode{var} & \mycode{cd} & \mycode{tmp} & \mycode{mounts} & \mycode{cat} & \mycode{proc} \\
2 & \mycode{cd} & \mycode{.ptmx} & \mycode{netslink} & \mycode{name} & \mycode{30} & \mycode{cpuinfo} \\
3 & \mycode{-rf} & \mycode{cp} & \mycode{GSec} & \mycode{local} & \mycode{U6} & \mycode{scheduler} \\
4 & \mycode{chmod} & \mycode{777} & \mycode{mika} & \mycode{find} & \mycode{-t} & \mycode{add} \\
5 & \mycode{i} & \mycode{echo} & \texttt{do} & \mycode{7wmp0b4s.rsc} & \mycode{policy} & \mycode{scheduler} \\
6 & \mycode{ls} & \mycode{pkill} & \mycode{sudo} & \mycode{PEDO} & \mycode{1.} & \mycode{nc} \\
7 & \mycode{shell} & \mycode{sh} & \mycode{enable} & \mycode{vOqkOc77} & \mycode{.none} & \mycode{cR37KUaG} \\
8 & \mycode{PEDO} & \mycode{1.} & \mycode{nc} & \mycode{kura} & \mycode{Uirusu} & \mycode{LOLKEK} \\
9 & \mycode{cat} & \mycode{echo} & \mycode{.nippon} & \mycode{ECCHI} & \mycode{IHCCE} & \mycode{telnet.wget.x86_64} \\
10 & \mycode{HEX} & \mycode{echo} & \mycode{-e} & \mycode{Akim} & \mycode{RONALD} & \mycode{6KafCk3x0a} \\
11 & \mycode{wget} & \mycode{-O} & \mycode{curl} & \mycode{10} & \mycode{pids.txt} & \mycode{g} \\
12 & \mycode{-rf} & \mycode{updDl} & \mycode{cd} & \mycode{interval} & \mycode{10m} & \mycode{on-event} \\
13 & \mycode{sh} & \mycode{-rf} & \mycode{tftp1.sh} & \mycode{DNXFCOW} & \mycode{start-shell} & \mycode{config} \\
14 & \mycode{cd} & \mycode{tmp} & \mycode{run} & \mycode{-h} & \mycode{disown} & \mycode{-y} \\
15 & \mycode{tftp} & \mycode{-r} & \mycode{-g} & \mycode{sms} & \mycode{smsd.conf*} & \mycode{modem*} \\
16 & \mycode{-rf} & \mycode{wget} & \mycode{cd} & \mycode{authorized_keys} & \mycode{.ssh} & \mycode{ssh-rsa} \\
17 & \mycode{anonymous} & \mycode{-u} & \mycode{-P} & \mycode{Skyline} & \mycode{.file} & \mycode{ips} \\
18 & \mycode{GHILIMEA_word} & \mycode{tmp} & \mycode{lib} & \mycode{cpr} & \mycode{REMOVED} & \mycode{PROCESS} \\
19 & \mycode{DNXFCOW} & \mycode{start-shell} & \mycode{config} & \mycode{already} & \mycode{is} & \mycode{miner_pid.txt} \\
20 & \mycode{kura} & \mycode{ps} & \mycode{cd} & \mycode{cowffxxna} & \mycode{.cowbot.bin} & \mycode{scanner.s.} \\
21 & \mycode{ECCHI} & \mycode{ps} & \mycode{cd} & \mycode{curl.sh} & \mycode{1sh} & \mycode{tftp.sh} \\
22 & \mycode{None} & \mycode{a} & \mycode{name} & \mycode{anonymous} & \mycode{21} & \mycode{ftpget} \\
23 & \mycode{grep} & \mycode{head} & \mycode{-c} & \mycode{Sofia.file} & \mycode{Sofia} & \mycode{var.s4y} \\
24 & \mycode{var} & \mycode{config} & \mycode{etc} & \mycode{GnoSec} & \mycode{z0x3n} & \mycode{v} \\
25 & \mycode{sh} & \mycode{0xft6426467.sh} & \mycode{sshd} & \mycode{-L} & \mycode{download.c3pool.com} & \mycode{LC_ALL} \\
26 & \mycode{cd} & \mycode{.file} & \mycode{mnt} & \mycode{get} & \mycode{-g} & \mycode{-r} \\
27 & \mycode{retrieve} & \mycode{chmod} & \mycode{777} & \mycode{root} & \mycode{USER} & \mycode{home} \\
28 & \mycode{HEX} & \mycode{echo} & \mycode{-ne} & \mycode{HEX} & \mycode{-en} & \mycode{HEX} \\
29 & \mycode{-s} & \mycode{bash} & \mycode{curl} & \mycode{netslink} & \mycode{shm} & \mycode{boot} \\
30 & \mycode{.ssh} & \mycode{cd} & \mycode{authorized_keys} & \mycode{sudo} & \mycode{hive-passwd} & \mycode{Xorg} \\
31 & \mycode{GHILIMEA_word} & \mycode{exit} & \mycode{awk} & \mycode{-e} & \mycode{base64} & \mycode{openssl} \\
32 & \mycode{GHILIMEA_word} & \mycode{cpr} & \mycode{0} & \mycode{interval} & \mycode{10m} & \mycode{on-event} \\
33 & \mycode{GHILIMEA_word} & \mycode{echo} & \texttt{fi} & \mycode{-O} & \mycode{-qO} & \mycode{ssh} \\
34 & \mycode{tmp} & \mycode{.none} & \mycode{cat} & \mycode{shell} & \mycode{enable} & \mycode{exec} \\
35 & \mycode{RONALD} & \mycode{Akim} & \mycode{Skyline} & \texttt{do} & \mycode{read} & \texttt{while} \\
36 & \mycode{ip} & \mycode{find} & \mycode{remove} & \mycode{*.sh} & \mycode{help} & \mycode{ftp1.sh} \\
37 & \mycode{p} & \mycode{0xt984767.sh} & \mycode{xvf} & \mycode{..file} & \mycode{selfrep} & \mycode{qjmiguza} \\
38 & \mycode{exit} & \mycode{0xft6426467.sh} & \mycode{sshd} & \mycode{mika} & \mycode{kurc} & \mycode{exad} \\
\hline
\end{tabular}
}
\end{table}

\begin{figure}[!t]
\centering
\begin{subfigure}[t]{.475\textwidth}
\centering
\caption{CBC vs. NCBC}
\label{fig:cbc_ncbc}
\includegraphics[width=.9\textwidth]{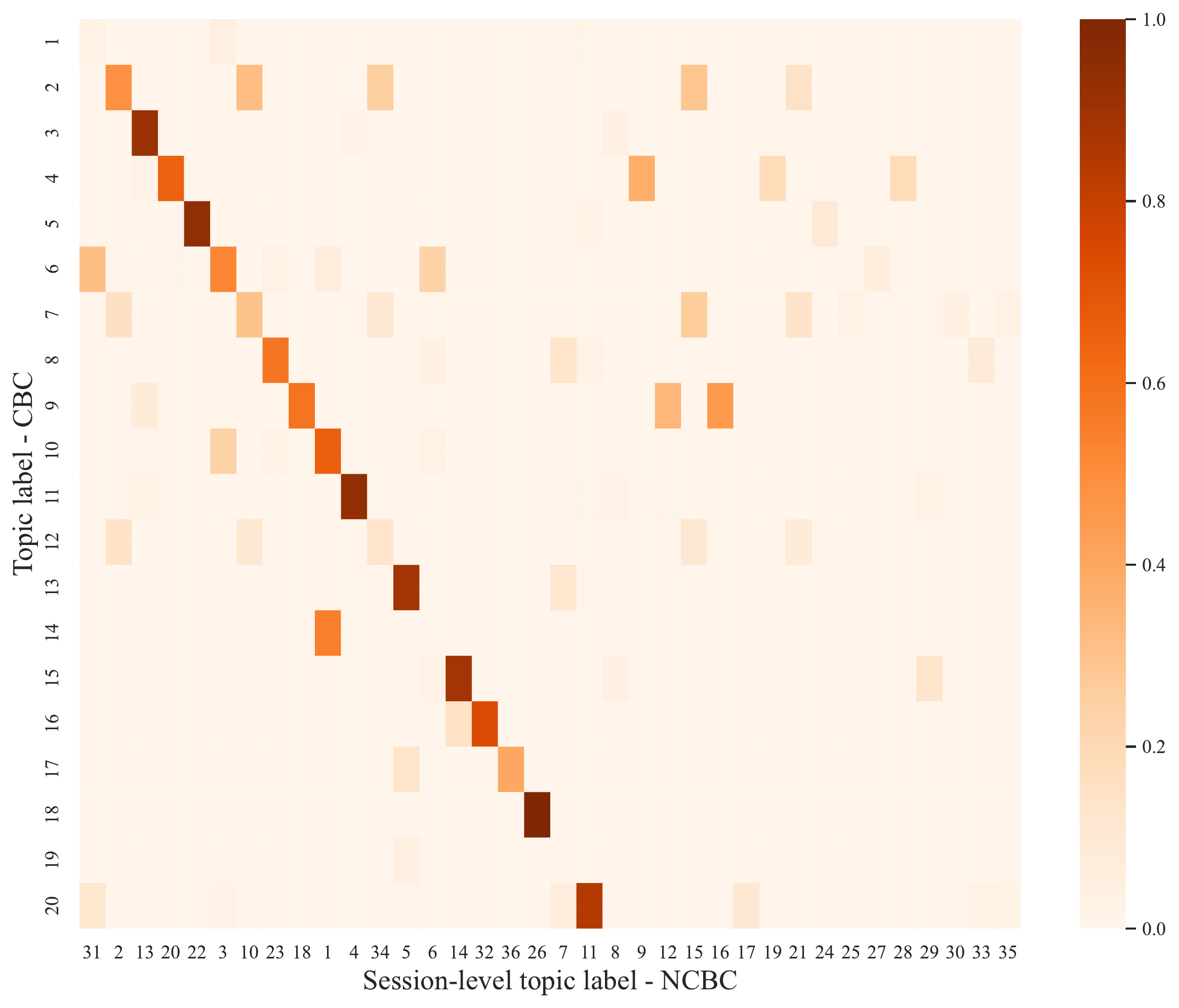}
\end{subfigure}
\hspace*{0.025\textwidth}
\begin{subfigure}[t]{.475\textwidth}
\centering
\caption{PCNBC vs. NCBC}
\label{fig:pcnbc_ncbc}
\includegraphics[width=.9\textwidth]{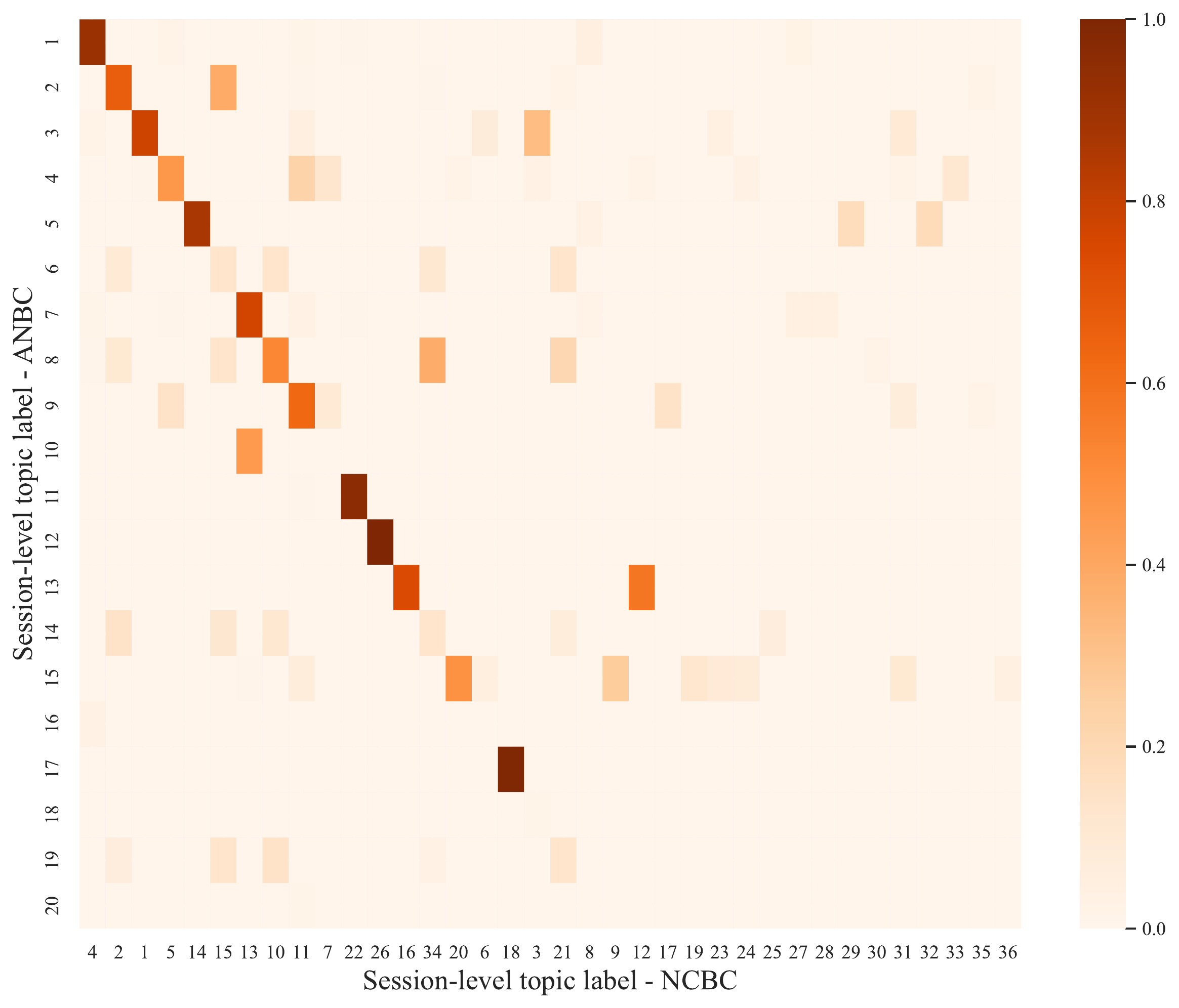}
\end{subfigure}
\caption{Heatmap of Jaccard scores between session groups under CBC (Section~\ref{sec:mod1_fit}) and PCNBC (Section~\ref{sec:pcnbc_fit}), compared to the groups fitted via NCBC (Section~\ref{sec:mod3_fit}). Topics are aligned via the Hungarian algorithm. } 
\label{fig:heatmap}
\end{figure}

Furthermore, Figure~\ref{fig:cbc_ncbc} displays the heatmap of Jaccard similarity scores comparing the results of CBC (\textit{cf.} Section~\ref{sec:mod1_fit}) and NCBC, fitted in this section, demonstrating significant agreement. The level of agreement is remarkable, considering that the session-level topics are obtained under two different modelling assumptions: in Section~\ref{sec:mod1_fit}, the CBC model in \eqref{mod1} directly uses the session-level topic to obtain the word distribution for the entire session. On the other hand, the session-level topics in NCBC are only estimated from sequences of command-level topics, which are themselves unknown and therefore estimated.

More details about the commands appearing in the sessions are given in Tables~\ref{tab:mod3_interpretation2} and~\ref{tab:mod3_words}, and such objectives can be associated with the command-level topic-specific wordclouds in the Supplementary Material. 
Overall, analysing the results of NCBC confirms the conclusions in the previous section, with the added benefit of having estimated command-level topics. 

\subsection{Parent-Child Nested Bayesian Clustering} \label{sec:pcnbc_fit}

In conclusion, the results are compared with the output of PCNBC (Section~\ref{sec:pcnbc}), a model specifically tailored to the characteristics of honeypot data. 
Similarly to previous sections, the collapsed Gibbs sampler is run for \numprint{250000} iterations with \numprint{50000} burn-in. The session-level and parent topics are initialised via the spectral clustering algorithm described in Section~\ref{sec:initialisation}, adapted for PCNBC, setting Dirichlet priors of dimension $K_\text{max}=30$ and $H_\text{max}=50$, with hyperparameters $\vec\eta=\vec 1_{ V},\ \vec\tau=0.1\cdot\vec 1_{V},\ \vec\gamma=0.1\cdot\vec 1_{K_\text{max}},\ \vec\chi=0.1\cdot\vec 1_{H_\text{max}}$. Additional comparisons are also presented in Table~\ref{tab:mod_comparisons}. The topics are estimated following a procedure similar to the description in Section~\ref{sec:topic_estimation}, adapted to PCNBC, setting $\hat K_\varnothing=20$ and $\hat H_\varnothing=45$, corresponding to the modal number of non-empty topics. 

The results in Table~\ref{tab:mod_comparisons} demonstrate that PCNBC has a good performance on the ICL honeypot data, second only to NCBC in terms of average marginal log-likelihood per word. Figure~\ref{fig:pcnbc_ncbc} shows the heatmap of Jaccard similarity scores between the session-level groups obtained from PCNBC and NCBC (\textit{cf.} Section~\ref{sec:mod3_fit}). Table~\ref{tab:pcnbc_interpretation} provides an interpretation of the session-level groups estimated by PCNBC. It appears that PCNBC is unable to single-out MinerFinder, which is isolated in a single cluster only by NCBC.  

It must be remarked that PCNBC provides less flexibility compared to NCBC. Under NCBC, sub-commands starting with the same parent word can have different intents across different sessions depending on their session-level or command-level topic, whereas in PCNBC, segments starting with the same parent will necessarily have the same topic, even if the session-level topics differ. Additionally, the role of session-level topics in PCNBC is different compared to CBC and NCBC: in PCNBC, the session-level topics only determine the distribution of parent words, and they do not strongly influence the distribution of the child words. Therefore, session-level topics are only expected to capture general behaviours, such as installation of malware, but they are not necessarily suitable to distinguish different types of malware installed using similar sequences of instructions, starting with the same parent words. Despite this limitation the topic descriptions in Table~\ref{tab:pcnbc_interpretation} appear to be closely linked with the session-level topics discussed in Tables~\ref{tab:mod1_interpretation} and \ref{tab:mod3_interpretation}.

\begin{table}[!t]
\caption{Estimated session-level topics and corresponding intent under the PCNBC model in \eqref{eq:pcnbc}.}
\label{tab:pcnbc_interpretation}
\scalebox{0.65}{
\begin{tabular}{ccl}
Topic & Type of malware & Objective\\
\hline\hline
1 &	MIRAI & Determine shell executable, check \mycode{busybox} is present, print error message to console \\
2 &   (\mycode{ptmx}) unnamed botnet, MIRAI & Reconnaissance, change permissions, execute MIRAI variants \\
3 &   MIRAI & Download and execute MIRAI variants \mycode{tftp.sh}, \mycode{tftp1.sh}, \mycode{tftp2.sh}, \mycode{ftp.sh}, \mycode{}... \\
4 &	Hive OS attack, shellbot, coin miner & Scan system, look for GPUs, download and look for coin miners, download malware \\ 
5 &	MIRAI & Download and execute MIRAI variant \mycode{DNXFCOW} via echoing single and multiple \mycode{HEX} strings \\
6 &	\mycode{s4y} unnamed botnet, MIRAI & Gather system information, change permissions, execute MIRAI variants \\
7 &   MIRAI & Download and execute MIRAI variants \mycode{nippon}, \mycode{PEACH}, \mycode{Uirusu} and others \\ 
8 &   (\mycode{ptmx}) unnamed botnet, MIRAI & Reconnaissance, change permissions, execute MIRAI variants \\
9 &	Hive OS attack, coin miner & Download miner, attempt to take over configurations in Hive OS mining platform \\
10 &	MIRAI & Download and execute MIRAI variants \mycode{kura} and \mycode{kurc}, fingerprint system \\
11 & MIRAI & Download malware, change permissions, gather system information, fingerprint system \\ 
12 &	GHILIMEA, PentaMiner script & Install coin miner, kill mining processes with high CPU usage in order to go undetected \\ 
13 & MIRAI & Download and execute MIRAI variants \mycode{PEDO}, \mycode{mika}, \mycode{nippon}...\\
14 &	(\mycode{misa}) unnamed botnets & Gather system information, change permissions, execute MIRAI variants \\
15 & MIRAI + \textbf{MinerFinder} & Download malware and coin miner, write \mycode{upnp} and \mycode{updDl} malware via echoing \mycode{HEX} strings \\
16 & Shell & Determine executable \\
17 & 	MIRAI & Download and execute MIRAI variants \mycode{ECCHI}, \mycode{nippon}, \mycode{dvrHelper}...\\
18 & Update & Attempt to update installers \\
19 &	(\mycode{LAYER}) unnamed botnet, MIRAI & Gather system information, change permissions, execute MIRAI variants \\
20 &	Reconnaissance & Check for downloaded files \\
\hline
\end{tabular}
}
\end{table}

\section{Conclusion}

Models for clustering session data collected on honeypots were proposed, with the objective of finding groups of similar attacks which could aid cyber analysts in detecting emerging intrusion attempts and vulnerabilities. The proposed models are based on modifications of Latent Dirichlet Allocation, aimed at improving topic interpretability and convergence properties. In particular, the concepts of primary and secondary topics were introduced, along with session-level and command-level topics. Secondary topics are used to represent common, high-frequency words, whereas the primary topic is used for the words which define the latent intent of the session. Furthermore, two nested layers of topics were introduced: a command-level topic which determines the word distribution, and a session-level topic which controls the distribution of the command-level intents. The proposed models are broadly denoted Nested Constrained Bayesian Clustering (NCBC) models with a secondary topic. The methodologies are also extended to a Bayesian nonparametric framework, admitting an unbounded (and unknown) number of latent intents and words in the vocabulary. Additionally, Parent-Child Nested Bayesian Clustering (PCNBC) is proposed, a model that clusters initial words of sub-commands separately from the rest of the session. PCNBC is specifically tailored to the unique characteristics of honeypot data.  

The proposed models could be further extended by introducing dependencies between command-level topics. In particular, a Markov model with $H$ states could be devised, whereby each session-level topic corresponds to different transition probabilities between command-level topics. 
Also, a possible extension of this work could consider dynamically-evolving topics \citep[dynamic topic modelling,][]{Blei06}, or explicitly encode correlation between topics \citep[correlated topic models,][]{Lafferty05}, or a combination of the two approaches \citep[see, for example,][]{Tomasi20}. Additionally, a current limitation of the proposed model is the largely serial nature of the proposed sampling schemes, which do not provide computational scalability. Therefore, scalable alternatives could be devised \citep[see, for example,][]{Yan09}, relying on variational approximations to the posterior \citep[see, for example,][]{Teh062}, or utilising uncollapsed Gibbs samplers, which are easily parallelisable \citep{Tristan15}. 

Sections~\ref{sec:datadesc} and~\ref{sec:preprocessing} briefly discussed 
challenges related to tokenisation in cyber-security. 
Depending on the 
tokenisation method, important context about the commands might be lost, which might prevent cyber-security analysts to make rapid decisions 
on the content of the session. Therefore, it must be carefully decided which representation into words is most accurately representative of the actual content. 
One possibility 
would be to explore deep neural network methods \citep[see, for example,][]{Hanif22} for parsing cyber-security logs, inspired by their use for large language models. 

Overall, session data collected on honeypots are a valuable resource for cyber analysts, and principled statistical modelling is required to obtain actionable insights from these data. The proposed methodologies have provided useful groupings of the attacks observed on a university network, including the discovery of an unusual MIRAI variant which attempts to take over existing coin miner infrastructure. This demonstrates the potential of topic models to elucidate hidden structure within text data, obtaining valuable insights for cyber-defence. 

\section*{Acknowledgements}
The authors thank Andy Thomas and the ICT team at Imperial College London for their support on data collection. FSP and NAH acknowledge funding from Microsoft, through the research grant \textit{``Understanding the enterprise: Host-based event prediction for automatic defence in cyber-security''}. 
AM and NAH acknowledge funding from the Data-Centric Engineering programme at the Alan Turing Institute, for the Grand Challenge \textit{``Monitoring Complex Systems''.} 
Part of this work was carried out when AM was a Postdoctoral Research Associate at Imperial College London and the Alan Turing Institute. 

\section*{Code}

A \textit{Python} library that implements the methodologies proposed in this article is available in the Github repository \href{https://www.github.com/fraspass/ncbc}{\texttt{fraspass/ncbc}}.

\bibliographystyle{imsart-nameyear} 
\bibliography{biblio}       


\newpage

\setcounter{figure}{0}
\renewcommand{\thefigure}{S.\arabic{figure}}

\setcounter{table}{0}
\renewcommand{\thetable}{S.\arabic{table}}

\begin{appendix}

\section{Gibbs sampling in NCBC with secondary topic} \label{sec:gibbs_sampling_steps}

\subsection{Resampling the session-level topic allocations} \label{sec:session_res}

The Gibbs sampler requires to sample from the conditional distribution of a subset of the parameters, conditional on the observed data and remaining parameters. Therefore, for resampling the session-level topic allocation $t_d$ for a given document, it is required to sample from 
$p(t_d \mid \bm t^{-d}, \bm w, \bm z,\bm s)$, 
where the superscript $-d$ denotes that the calculations of the corresponding quantity \textit{exclude} the $d$-th document. For the $r$-th session-level topic, the probability can be written as:
\begin{equation}
p(t_d=r \mid \bm t^{-d}, \bm w, \bm z,\bm s) 
\propto p(t_d=r\mid\bm t^{-d}) 
 p(\bm s\mid t_d=r, \bm t^{-d})
\propto \frac{B(\bm\gamma + \bm T)}{B(\bm\gamma + \bm T^{-d})} \prod_{k=1}^{K }B(\bm\tau + \bm S_k). 
\end{equation}
where the quantities $\bm T$ and $\bm S_k$ in the expression are calculated assuming that $t_d=r$. The ratio of beta functions in the conditional distribution can be simplified using the properties of the gamma function, yielding $B(\bm\gamma + \bm T)/B(\bm\gamma + \bm T^{-d})\propto(\gamma_r+T_r^{-d})$. Similarly, the product of beta functions in the final part of the expression could be further simplified using the fact that $S_{r,h}=S_{r,h}^{-d} + S_h^d$, where $S_{k,h}^{-d}=\sum_{u\neq d}\mathds{I}_{\{k\}}(t_u)\sum_{j}\mathds{I}_{\{h\}}(s_{j,u})$ and $S_h^d=\sum_{j}\mathds{I}_{\{h\}}(s_{d,j})$. From the properties of the gamma function, the probability can be then expressed as:
\begin{equation}
p(t_d=r \mid \bm t^{-d}, \bm w, \bm z,\bm s)
\propto (\gamma_r+T_r^{-d})\frac{\prod_{h=1}^{H } \prod_{\ell=1}^{S^d_h} (\tau_h + S_{r,h}^{-d} + \ell - 1)}{\prod_{\ell=1}^{N_d} \{\sum_{h=1}^{H } (\tau_h + S_{r,h}^{-d}) + \ell - 1\}}. \label{resamp_t}
\end{equation}

\subsection{Resampling the command-level topic allocations} \label{sec:command_res}

The Gibbs sampler also requires to sample the command-level topic allocations from the distribution $p(s_{d,j}=\ell \mid \bm s^{-d,j}, \bm w, \bm t, \bm z)$, where the superscript denotes that the quantities have been calculated excluding the $(d,j)$-th terms. For the $\ell$-th command-level topic, the probability factorises as:
\begin{multline}
p(s_{d,j}=\ell \mid \bm s^{-d,j}, \bm w, \bm t, \bm z) \propto \\
 p(s_{d,j}=\ell, \bm s^{-d,j} \mid \bm t) \times p(\bm z\mid s_{d,j}=\ell, \bm s^{-d,j}) \times p(\bm w\mid s_{d,j}=\ell, \bm s^{-d,j}, \bm z). \label{fact_s}
\end{multline}
The first term in the factorisation \eqref{fact_s} 
can be simplified 
using \eqref{marg_s} and $S_{t_d,\ell}=S_{t_d,\ell}^{-d,j}+1$:
\begin{equation}
p(s_{d,j}=\ell, \bm s^{-d,j} \mid \bm t) 
\propto (\tau_\ell + S_{t_d,\ell}^{-d,j}). \label{prob_s1}
\end{equation}
A similar reasoning could be used to simplify the second term in the factorisation \eqref{fact_s}: 
$Z_\ell = Z_\ell^{-d,j}+Z^{d,j}$, where $Z_h^{-d,j}=\sum_{(u,q)\neq (d,j)} \mathds I_{\{h\}}(s_{u,q})\sum_{i=1}^{M_{d,j}}z_{u,q,i}$ and $Z^{d,j}=\sum_{i=1}^{M_{d,j}}z_{d,j,i}$. 
Simplifying the gamma functions, the corresponding probability is expressed as:
\begin{multline}
p(\bm z\mid s_{d,j}=\ell, \bm s^{-d,j}) \propto\\ 
 \frac{\prod_{q=1}^{Z^{d,j}} (\alpha_\ell+Z_\ell^{-d,j}+q-1)\prod_{u=1}^{M_{d,j} - Z^{d,j}}(\alpha_0+M_\ell^{\ast-d,j} - Z_\ell^{-d,j}+ u-1)}{\prod_{q=1}^{M_{d,j}} (\alpha_0+\alpha_\ell+M_\ell^{\ast-d,j} + q-1)}.
\label{prob_s2}
\end{multline}
The last term in 
\eqref{fact_s} admits a similar simplification to \eqref{resamp_t}, using 
$W_{\ell,v} = W_{\ell,v}^{-d,j} + W_{v}^{d,j}$, where $W_{h,v}^{-d,j} = \sum_{u,q,i:(u,q)\neq (d,j)} \mathds I_{\{h\}}(z_{u,q,i} ~s_{u,q})\mathds I_{\{v\}}(w_{u,q,i})$, 
and $W_{v}^{d,j}=\sum_{i}[1-\mathds I_{\{0\}}(z_{d,j,i} ~s_{d,j})]\mathds I_{\{v\}}(w_{d,j,i})$. The resulting probability is: 
\begin{equation}
p(\bm w\mid s_{d,j}=\ell, \bm s^{-d,j}, \bm z) 
\propto \frac{\prod_{v=1}^{ V} \prod_{q=1}^{W_v^{d,j}} (\eta_v + W_{\ell,v}^{-d,j} + q - 1)}{\prod_{q=1}^{\sum_v W_v^{d,j}} \{\sum_{v=1}^{ V}(\eta_v + W_{\ell,v}^{-d,j}) + q - 1\}}.
\label{prob_s3}
\end{equation}
The probability \eqref{fact_s} is obtained 
by normalising the product of terms \eqref{prob_s1}, \eqref{prob_s2}, \eqref{prob_s3}.

\subsection{Resampling the primary-secondary topic indicators} 

In the models with primary-secondary topics, the Gibbs sampler also requires to resample the binary indicators $z_{d,j,i}$, conditional on $\bm w, \bm s,\bm t$ and $\bm z^{-d,j,i}$, denoting all the indicators except $z_{d,j,i}$. 
Each binary indicator is drawn from a Bernoulli distribution with unnormalised probabilities:
\begin{equation}
p(z_{d,j,i}=b \mid \bm z^{-d,j,i}, \bm w,\bm s,\bm t) \propto p(z_{d,j,i}=b \mid \bm z^{-d,j,i}, \bm s) \times p(\bm w \mid z_{d,j,i}=b,\bm z^{-d,j,i}, \bm s), \label{fact_z}
\end{equation}
where $b\in\{0,1\}$.
Noting that $Z_{s_{d,j}} = Z_{s_{d,j}}^{-d,j,i} + b$, where the term $Z_{s_{d,j}}^{-d,j,i} $ is defined as $Z_h^{-d,j,i}=\sum_{(u,q)\neq(d,j)}\mathds{I}_{\{h\}}(s_{u,q}) \sum_{i=1}^{M_{u,q}}z_{u,q,i}$, the first term in the factorisation becomes:
\begin{equation}
p(z_{d,j,i}=b \mid \bm z^{-d,j,i}, \bm s) \propto (\alpha_{s_{d,j}} + Z_{s_{d,j}}^{-d,j,i})^b (\alpha_0 + M_{s_{d,j}}^\ast - Z_{s_{d,j}}^{-d,j,i} - 1)^{1-b}.
\end{equation}
For the marginal likelihood of observed words, the only terms affected by a change in the binary indicator $z_{d,j,i}$ are $W_{0,w_{d,j,i}}=W_{0,w_{d,j,i}}^{-d,j,i}+1-b$ and $W_{s_{d,j},w_{d,j,i}}=W_{s_{d,j},w_{d,j,i}}^{-d,j,i}+b$, where $W_{h,v}^{-d,j,i}=\sum_{(u,q,r)\neq(d,j,i)}\mathds{I}_{\{h\}}(z_{u,q,r} ~s_{u,q}) \mathds{I}_{\{v\}}(w_{u,q,r})$, giving:
\begin{equation}
p(\bm w \mid z_{d,j,i}=b,\bm z^{-d,j,i}, \bm s)
\propto \left\{\frac{\eta_{w_{d,j,i}} + W_{0,w_{d,j,i}}^{-d,j,i}}{\sum_{v=1}^{V} (\eta_v+W_{0,v}^{-d,j,i})}\right\}^{1-b}\left\{\frac{\eta_{w_{d,j,i}} + W_{s_{d,j},w_{d,j,i}}^{-d,j,i}}{\sum_{v=1}^{ V} (\eta_v + W_{s_{d,j},v}^{-d,j,i})}\right\}^b. 
\end{equation}

\subsection{Split-merge topic allocations} \label{sec:split_merge}

There are two types of split-merge moves that can be proposed: (i) split-merge session-level topics, and (ii) split-merge command-level topics.
For the split-merge move on session-level topics, two sessions $d$ and $d^\prime$ are sampled at random from the $D$ observed sessions. 
If $t_d=t_{d^\prime}=t^\ast$, the proposal for the session-level topics splits the sessions assigned to $t^\ast$ in two different clusters using the following iterative procedure: (i) assign  topic $t^\ast$ to document $d$, and topic $\tilde t$ 
to document $d^\prime$ (where $\tilde t$ corresponds to the number of non-empty clusters, plus one -- note that if $\tilde t>K $ the split move should be immediately rejected); (ii) documents previously assigned to topic $t^\ast$ are sequentially allocated to topics $t^\ast$ or $\tilde t$ in random order, with probabilities proportional to the predictive distribution \eqref{resamp_t}, restricted to the session already reallocated to topics $t^\ast$ and $\tilde t$. 
This allocation procedure is adapted from common split-merge MCMC moves in related clustering problems \citepSM[see, for example,][]{Dahl03,SannaPassino20}.
The final proposal is denoted as $\bm t^\ast$, with probability $q(\bm t^\ast\mid\bm t)$, corresponding to the product of sequential probabilities obtained in the splitting procedure. 
The resulting acceptance probability for the move from $\bm t$ to $\bm t^\ast$ is:
\begin{equation}
\min\left\{1,\frac{p(\bm t^\ast) p(\bm s \mid \bm t^\ast)}
{p(\bm t) p(\bm s \mid \bm t) q(\bm t^\ast \mid \bm t)}\right\}.
\label{acc_ratio}
\end{equation} 
On the other hand, if $t_d\neq t_{d^\prime}$, the proposal $\bm t^\ast$ assigns topic $t_d$ to all documents previously given topic $t_{d^\prime}$, corresponding to a merge move. 
In this case, the acceptance ratio in \eqref{acc_ratio} must be further multiplied by the proposal probability $q(\bm t \mid \bm t^\ast)$, calculated by simulating a split move from $\bm t^\ast$ to $\bm t$. Since there is only one way to merge two topics, the proposal probability at the denominator of \eqref{acc_ratio} is $q(\bm t^\ast \mid \bm t)=1$ for a merge move.

A similar split-merge move can be constructed for the command-level topics: two commands $j$ and $j^\prime$ are randomly sampled from two random documents $d$ and $d^\prime$ respectively. If $s_{d,j}\neq s_{j^\prime,d^\prime}$, a merge move is proposed. Alternatively, if $s_{d,j}=s_{j^\prime,d^\prime}=s^\ast$, the split move proceeds similarly to the procedure described for the topic-level sessions, and the command previously assigned topic $s^\ast$ are sequentially allocated to $s^\ast$ or $\tilde s$ (corresponding to the number of non-empty command-level topics, plus one) 
with probabilities proportional to the predictive distribution \eqref{fact_s}, limited to the commands already reassigned to $s^\ast$ and $\tilde s$. As before, if $\tilde s>H $, the move is rejected. In summary, the acceptance probability for a vector of command-level topics $\bm s^\ast$ obtained via the split-merge procedure is:
\begin{equation}
\min\left\{1,\frac{p(\bm s^\ast \mid \bm t)p(\bm z \mid \bm s^\ast)p(\bm w \mid \bm z,\bm s^\ast)q(\bm s\mid \bm s^\ast)}
{p(\bm s \mid \bm t)p(\bm z \mid \bm s)p(\bm w \mid \bm z,\bm s)q(\bm s^\ast\mid \bm s)}\right\},
\end{equation}
where the probabilities $q(\bm s^\ast\mid \bm s)$ and $q(\bm s\mid \bm s^\ast)$ are either $1$, or the product of allocation probabilities calculated from the sequential splitting procedure.

The acceptance probability of the split-merge moves varied between 0.05\% and 2\% across different models on the ICL honeypot data, with higher values usually reached for the first 500 proposed split-merge moves (up to 5\%). Normally, these moves appear to be useful before convergence to achieve fast jumps in the marginal log-likelihood, but their contribution is not substantial after convergence is reached.

\subsection{Bayesian inference with GEM priors} \label{sec:inf_gem}

Inference in the model with unbounded number of topics and vocabulary size can be carried out using a similar algorithm to the Gibbs sampling described in Section~\ref{sec:inference}. Only minor modifications are required, since the marginals in \eqref{dp_marginals} take a different form under the GEM priors. 
For example, for resampling the session-level topics, the probabilities in \eqref{resamp_t} are modified as follows:
\begin{multline}
p(t_d=r \mid \bm t^{-d}, \bm w, \bm z,\bm s)
\propto \ \gamma^{\mathds I_{\{K(\bm t^{-d})+1\}}(r)}(T_r^{-d})^{1-\mathds I_{\{K(\bm t^{-d})+1\}}(r)} \\ \times\frac{\prod_{h:S_{h}^d > 0} \tau^{\mathds I_{\{0\}}(S_{r,h}^{-d})} (S_{r,h}^{-d})^{\mathds I_{\mathds N_{>0}}(S_{r,h}^{-d})}\left[\prod_{\ell=2}^{S^d_h} (S_{r,h}^{-d} + \ell - 1)\right]^{\mathds I_{\mathds N_{>1}}(S_h^d)}}{\prod_{\ell=1}^{N_d} \{\tau + (\sum_{h=1}^{H(\bm s)} S_{r,h}^{-d}) + \ell - 1\}}, \label{resamp_t_dp}
\end{multline}
where $r\in\{1,\dots,K(\bm t^{-d})+1\}$, and the convention $0^0=1$ is adopted.
Similarly, the probabilities in \eqref{fact_s} for resampling command-level topics become:
\begin{align}
p(&s_{d,j} = \ell\ \mid\ \bm s^{-d,j}, \bm w, \bm t, \bm z) \propto \
\tau^{\mathds I_{\{0\}}(S_{t_d,\ell})} (S_{t_d,\ell}^{-d,j})^{\mathds I_{\mathds{N}_{>0}}(S_{t_d,\ell})}\\ &\times
\frac{\prod_{v:W_v^{d,j}>0} \eta^{\mathds I_{\{0\}}(W_{\ell,v}^{-d,j})} (W_{\ell,v}^{-d,j})^{\mathds I_{\mathds N_{>0}}(W_{\ell,v}^{-d,j})} \left[\prod_{q=2}^{W_v^{d,j}} (W_{\ell,v}^{-d,j} + q - 1)\right]^{\mathds I_{\mathds N_{>1}}(W_v^{d,j})}}{\prod_{q=1}^{\sum_v W_v^{d,j}} \{\eta + (\sum_{v=1}^{V(\bm w)} W_{\ell,v}^{-d,j}) + q - 1\}} \\
&\times \frac{\prod_{q=1}^{Z^{d,j}} (\alpha_\ell+Z_\ell^{-d,j}+q-1)\prod_{u=1}^{M_{d,j} - Z^{d,j}}(\alpha_0+M_\ell^{\ast-d,j} - Z_\ell^{-d,j}+ u -1)}{\prod_{q=1}^{M_{d,j}} (\alpha_0+\alpha_\ell+M_\ell^{\ast-d,j} + q - 1)},
\end{align}
where $\ell\in\{1,\dots,H(\bm s^{-d,j})+1\}$.
A modification is required also for the conditional probabilities of resampling the indicator $z_{d,j,i}$ in \eqref{fact_z}, resulting in:
\begin{align}
p(z_{d,j,i}=b \mid \bm z^{-d,j,i}, \bm w,\bm s,\bm t) \propto\ &
(\alpha_{s_{d,j}} + Z_{s_{d,j}}^{-d,j,i})^b (\alpha_0 + M_{s_{d,j}}^\ast - Z_{s_{d,j}}^{-d,j,i} - 1)^{1-b} \\
&\times \left\{\frac{\eta^{\mathds I_{\{0\}}(W_{0,w_{d,j,i}}^{-d,j,i})}(W_{0,w_{d,j,i}}^{-d,j,i})^{\mathds I_{\mathds N_{>0}}(W_{0,w_{d,j,i}}^{-d,j,i})}}{\eta+\sum_{v=1}^{V(\bm w^{-d,j,i})} W_{0,v}^{-d,j,i}}\right\}^{1-b} \\ &\times
\left\{\frac{\eta^{\mathds I_{\{0\}}(W_{s_{d,j},w_{d,j,i}}^{-d,j,i})}(W_{s_{d,j},w_{d,j,i}}^{-d,j,i})^{\mathds I_{\mathds N_{>0}}(W_{s_{d,j},w_{d,j,i}}^{-d,j,i})}}{\eta + \sum_{v=1}^{V(\bm w^{-d,j,i})}W_{s_{j,d},v}^{-d,j,i}}\right\}^b, \notag
\end{align}
where $b\in\{0,1\}$. 
The split-merge move in Section~\ref{sec:split_merge} can be equivalently extended to the model with GEM priors, using the same ideas presented in this section. Split-merge moves are expected to improve convergence of Gibbs sampling 
in models based on Dirichlet processes \citepSM{Jain04}.
All the probabilities described in this section are extremely similar to the equations in Section~\ref{sec:inference}, with the added complexity of handling previously unseen topics. 
Alternative MCMC algorithms for models based on the Dirichlet processes are 
also discussed in the literature 
\citepSM[for example,][]{Neal00,Ishwaran01,Teh06}. 

\section{Gibbs sampling in the PCNBC model} \label{sec:inf_pcnbc}

The parent-child nested Bayesian clustering (PCNBC) model introduced in Section~\ref{sec:pcnbc} has a slightly different structure from NCBC. The marginalised posterior distribution for PCNBC in \eqref{eq:pcnbc} factorises as follows:
\begin{equation}
p(\bm t, \bm u \mid \bm w)
\propto p(\bm w, \bm t,\bm u) 
= p(\bm t) 
\times p(\bm u)
\times p(\bm w \mid \bm t,\bm u). 
\label{eq:eqn_pcnbc}
\end{equation} 
The marginal distribution for the session-level topics $p(\bm t)$ is identical to \eqref{marg_t}, whereas $p(\bm u)$ and $p(\bm w \mid \bm t,\bm u)$ take the following form: 
\begin{align}
p(\bm u) = \frac{B(\bm\chi + \bm U)}{B(\bm\chi)},
& & 
p(\bm w \mid \bm t,\bm u) = \prod_{k=1}^K \frac{B(\bm\tau + \bm W_k^p)}{B(\bm\tau)} \times \prod_{h=1}^H \frac{B(\bm\eta + \bm W_h^c)}{B(\bm\eta)},
\label{eq:pcnbc_marginals}
\end{align}
where $U_h=\sum_{v=1}^V \mathds I_{\{h\}}(u_v)$ denotes the number of words allocated to the group $h$, $\bm U=(U_1,\dots,U_H)$, $W_{k,v}^p = \sum_{i,j,d} \mathds I_{\{k\}}(t_d)\mathds I_{\mathcal A_{d,j}}(i)\mathds I_{\{v\}}(w_{d,j,i})$ is the number of times word $v$ was an parent word in a document assigned to topic $k$, $\bm W_k^p=(W_{k,1}^p,\dots,W_{k,V}^p)$, and $W_{h,v}^c = \sum_{i,j,d} \mathds I_{\{h\}}(u_{w_{d,j,a_{d,j,i}^\ast}})[1-\mathds I_{\mathcal A_{d,j}}(i)]\mathds I_{\{v\}}(w_{d,j,i})$ is the number of times a non-parent (or child) word $v$ followed an parent word assigned to topic $h$, $\bm W_h^c=(W_{h,1}^c,\dots,W_{h,V}^c)$.

Following similar guidelines to Sections~\ref{sec:session_res} and \ref{sec:command_res}, the Gibbs sampler for PCNBC is based on the conditional distributions $p(t_d=r \mid \bm t^{-d}, \bm u, \bm w)$ and $p(u_v=\ell \mid \bm u^{-v}, \bm t, \bm w)$. To derive their analytic expressions from \eqref{eq:pcnbc_marginals}, it should be noted that $W_{k,v}^p = W_{k,v}^{a,-d} + W_v^{a,d}$ conditional on $t_d=k$, and $W_{h,\varpi}^c = W_{h,\varpi}^{c,-v} + W_\varpi^{c,v}$ conditional on $u_v=h$, where $W_v^{a,d}=\sum_{i,j}\mathds I_{\mathcal A_{d,j}}(i)\mathds I_{\{v\}}(w_{d,j,i})$ is the number of times the word $v$ appears as an parent word in the document $d$, and $W_\varpi^{c,v} = \sum_{i,j,d} \mathds I_{\{v\}}(w_{d,j,a_{d,j,i}^\ast})[1-\mathds I_{\mathcal A_{d,j}}(i)]\mathds I_{\{\varpi\}}(w_{d,j,i})$ is the number of times a word $\varpi$ is a non-parent word following an parent word $v$. Using a similar machinery as the derivations described in Sections~\ref{sec:session_res} and \ref{sec:command_res}, the resulting expressions are: 
\begin{align}
p(t_d=r \mid \bm t^{-d}, \bm u, \bm w) &\propto p(t_d=r \mid \bm t^{-d})\times p(\bm w\mid t_d=r, \bm t^{-d}, \bm u) \\
&\propto 
(\gamma_r + T_r^{-d})
\frac{\prod_{\varpi=1}^{ V} \prod_{q=1}^{W_\varpi^{a,d}} (\tau_\varpi + W_{r,\varpi}^{a,-d} + q - 1)}
{\prod_{q=1}^{\sum_\varpi W_\varpi^{a,d}} \{\sum_{\varpi=1}^{V}(\tau_\varpi + W_{r,\varpi}^{a,-d}) + q - 1\}}, \\
p(u_v=\ell \mid \bm u^{-v}, \bm t, \bm w) &\propto p(u_v=\ell \mid \bm u^{-v})\times p(\bm w\mid u_v=\ell, \bm u^{-v}, \bm t) \\
&\propto 
(\chi_\ell + U_\ell^{-v})
\frac{\prod_{\varpi=1}^{ V} \prod_{q=1}^{W_\varpi^{c,v}} (\eta_\varpi + W_{\ell,\varpi}^{c,-v} + q - 1)}
{\prod_{q=1}^{\sum_\varpi W_\varpi^{c,v}} \{\sum_{\varpi=1}^{V}(\eta_\varpi + W_{\ell,\varpi}^{c,-v}) + q - 1\}}.
\end{align}

\section{Simulations and results on synthetic data} \label{sec:results}

In this section, the proposed models are compared and contrasted on synthetic datasets, in order to assess their performance in recovering the session-level topics, 
the main quantity of inferential interest. Note that the true topics are \textit{known} when data are simulated. 
 If synthetic data are used, the true underlying session-level topics are available, and the true allocations could be compared to the estimated topics via the Adjusted Rand Index \citepSM[ARI,][]{Hubert85}. 
Each simulation is repeated for 50 datasets using different seeds, setting $ V=50,\ D=100,\ N_d=10,\ M_{d,j}=10,\ \vec\lambda=1/K\cdot\vec 1_K$ for each simulated dataset. The MCMC sampler is run for \numprint{10000} iterations with \numprint{1000} burn-in, initialising the topics via spectral clustering. Unless otherwise specified, the hyperparameter $\vec\gamma$ is set to $\vec\gamma=0.1\cdot\vec 1_{K_\text{max}}$.

\begin{figure}[!t]
\centering
\begin{subfigure}[t]{.495\textwidth}
\centering
\caption{Boxplot of ARIs for estimated session-level topics}
\label{sim1_1}
\includegraphics[width=\textwidth]{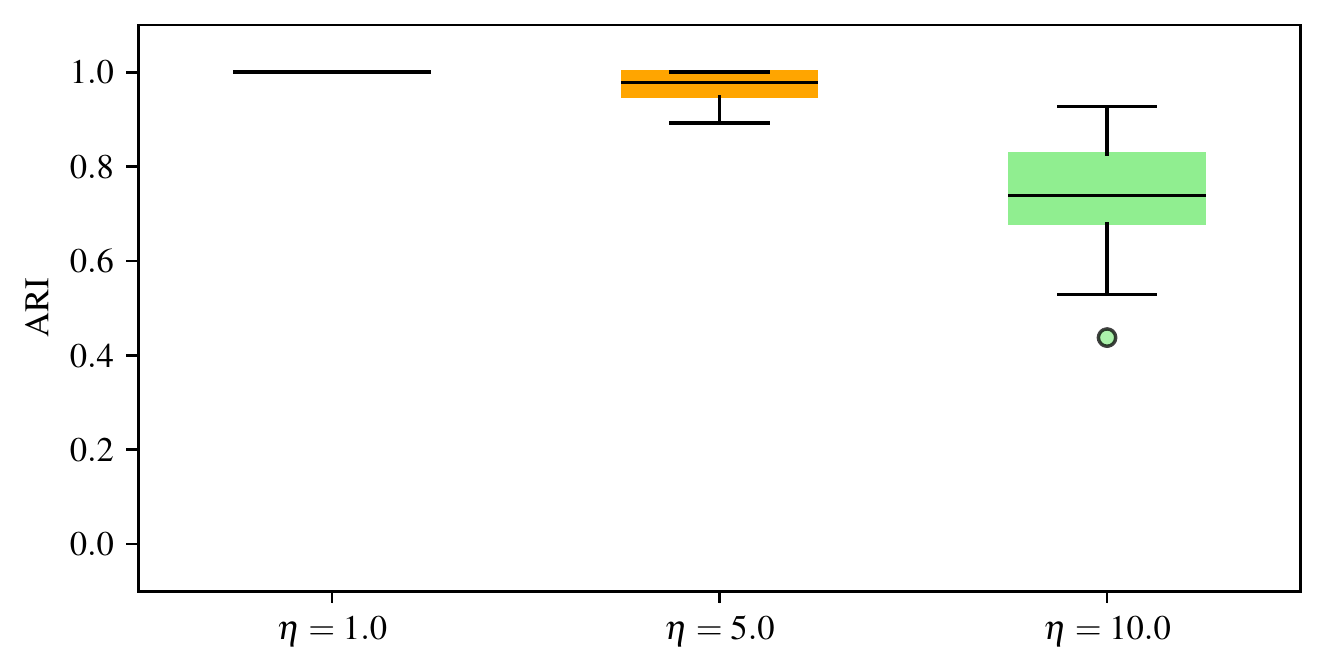}
\end{subfigure}
\begin{subfigure}[t]{.495\textwidth}
\centering
\caption{Barplot of estimated number of non-empty topics $K_\varnothing$}
\label{sim1_2}
\includegraphics[width=\textwidth]{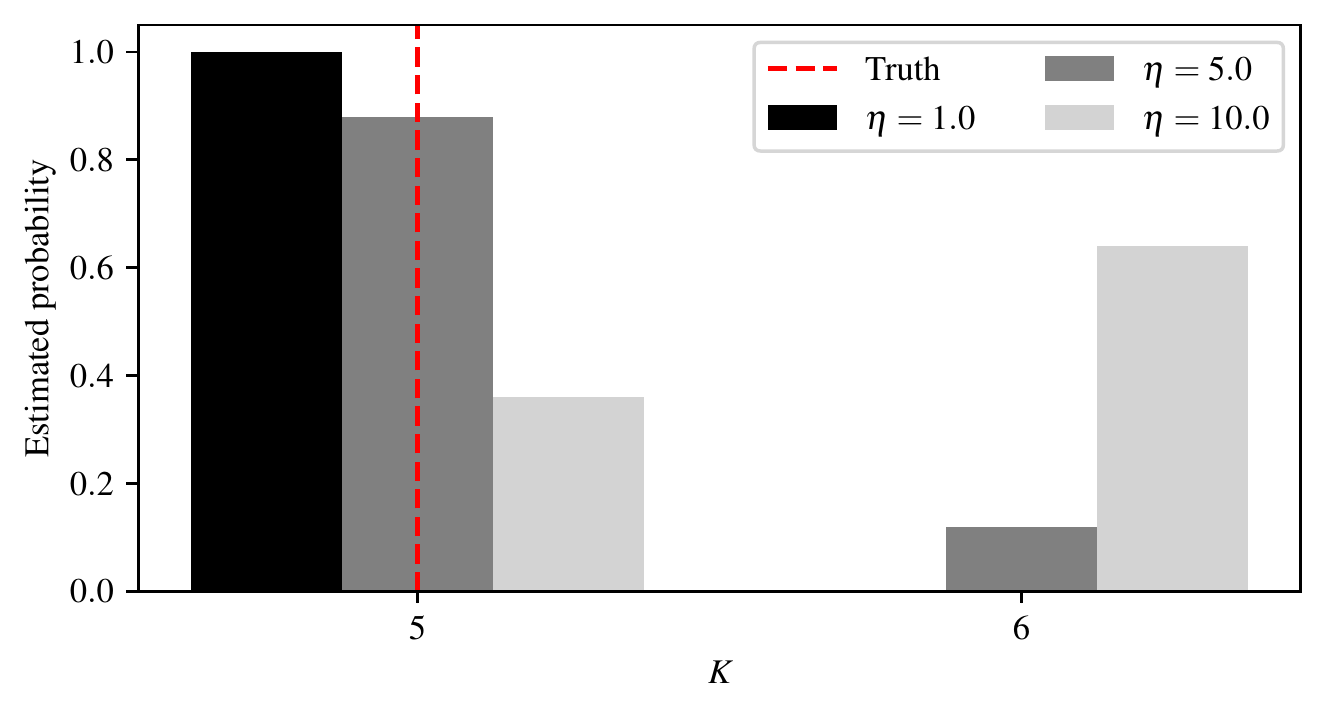}
\end{subfigure}
\caption{Summary plots obtained from 50 simulated datasets from model \eqref{mod1}, with $ V=50,\ K=5,\ D=100,\ N_d=10,\ M_{d,j}=10,\ \vec\lambda=1/K\cdot\vec 1_K$, using different values for $\vec\eta$, such that $\vec\eta=\eta\cdot\vec 1_K$. The MCMC sampler is run for \numprint{10000} iterations with \numprint{1000} burn-in, setting $K_\text{max}=10$.}
\label{sim1}
\end{figure}

First, datasets are simulated from model \eqref{mod1}, setting $K=5$ and using three different values of the parameter $\vec\eta$. If $\vec\eta=\eta\cdot\vec 1_K$, the parameter $\eta$ controls the \textit{spikiness} of the topic-specific word distributions: low values of $\eta$ correspond to distributions which assign most of the probability mass to a small number of words, whereas larger values of $\eta$ correspond to distributions where the probability mass is distributed more uniformly across words. In the MCMC sampler, $K_\text{max}$ is set to $10$, implying that the sampler should identify $5$ empty topics. The results are reported in Figure~\ref{sim1}: Figure~\ref{sim1_1} reports the ARIs for the estimated session-level topics, whereas Figure~\ref{sim1_2} shows the barplot of the estimated number of non-empty topics, denoted $K_\varnothing$. As expected, the ARI decreases when the value of $\eta$ increases, since the topic-specific distributions become increasingly uniform and therefore similar. Increasing the value of $\eta$ also provides worse estimates of the true number of topics used in the simulation. Ideally, $K_\varnothing$ should correspond to the value of $K$ used in the simulation, provided that $K_\text{max}$ in the MCMC algorithm is chosen to be larger than the true underlying $K$. 

\begin{figure}[!t]
\centering
\begin{subfigure}[t]{.475\textwidth}
\centering
\caption{Boxplot of ARIs for estimated session-level topics obtained via different initialisation schemes}
\label{sim11_1}
\includegraphics[width=\textwidth]{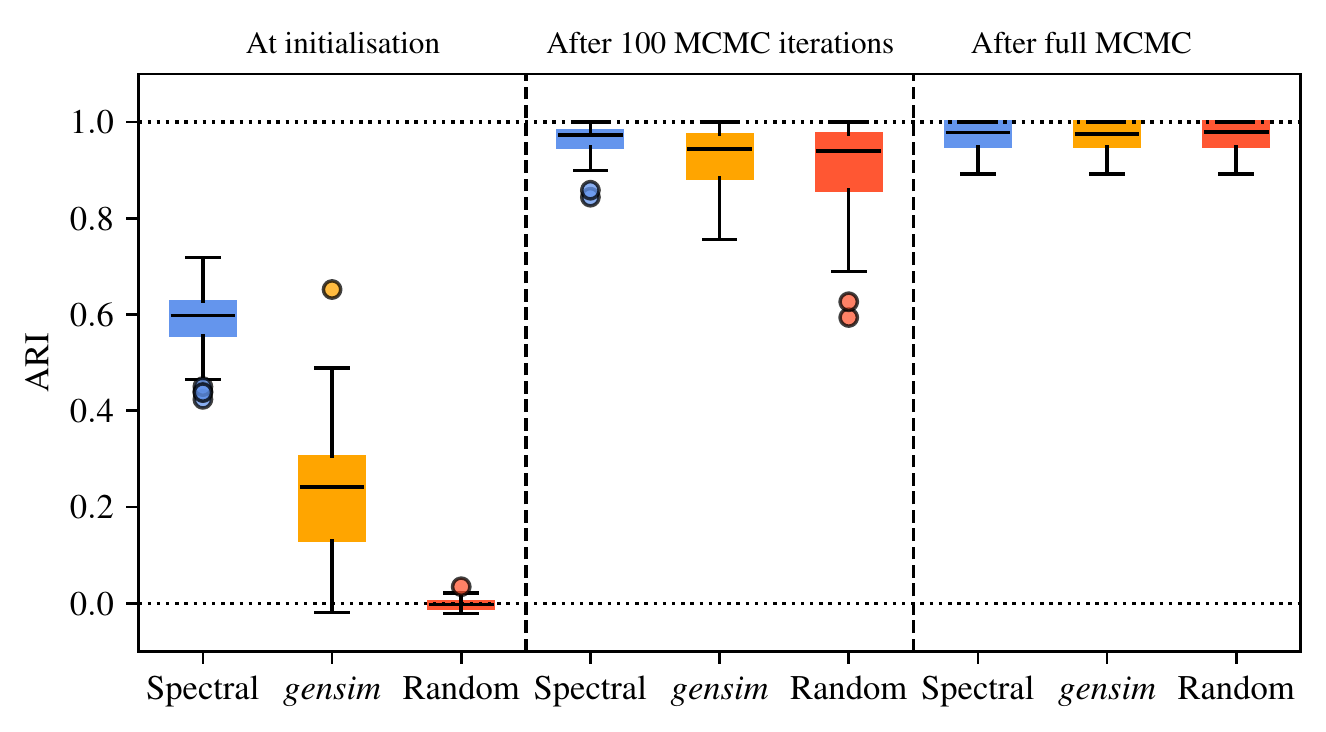}
\end{subfigure}
\hspace*{.02\textwidth}
\begin{subfigure}[t]{.475\textwidth}
\centering
\caption{Barplot of estimated number of non-empty topics $K_\varnothing$ for different priors on $\vec\lambda$ ($K_\text{max}$-dimensional Dirichlet or GEM)}
\label{sim11_2}
\includegraphics[width=\textwidth]{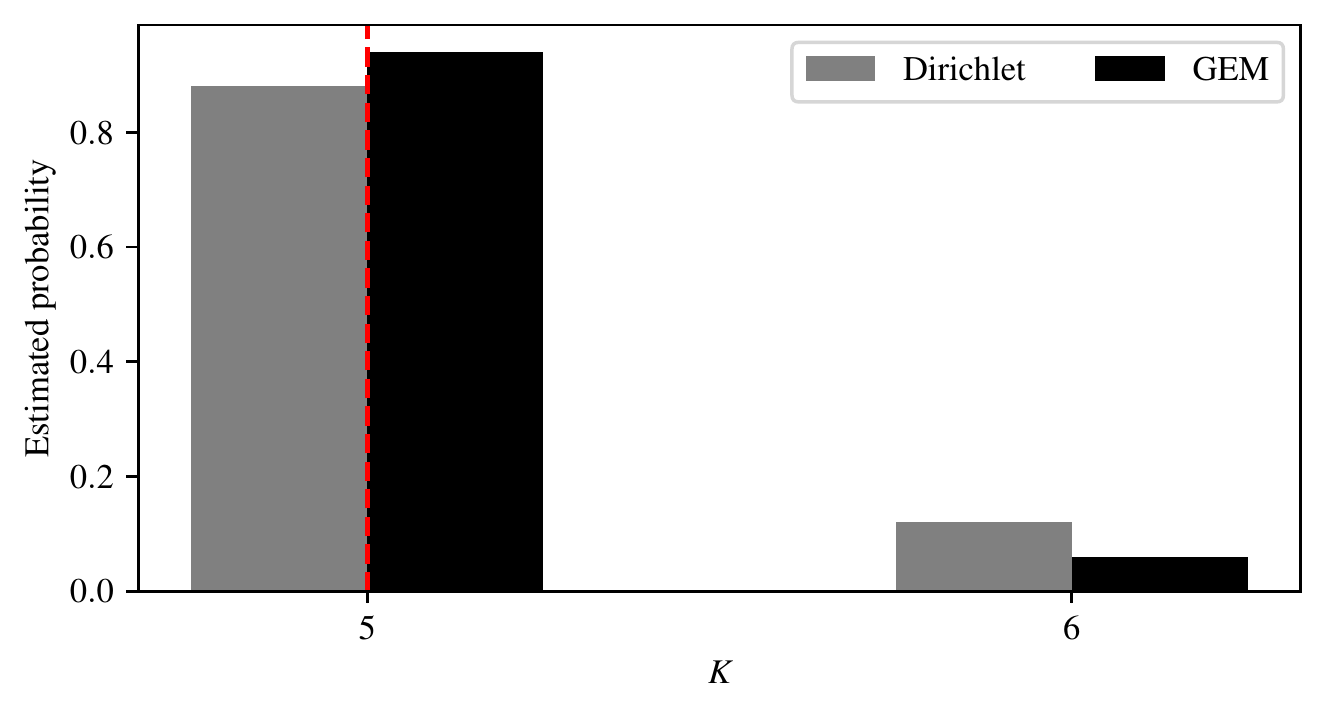}
\end{subfigure}
\caption{Summary plots obtained from 50 simulated datasets from model \eqref{mod1}, with $ V=50,\ K=5,\ D=100,\ N_d=10,\ M_{d,j}=10,\ \vec\lambda=1/K\cdot\vec 1_K$, and $\vec\eta=5\cdot\vec 1_K$. The MCMC sampler is run for \numprint{10000} iterations with \numprint{1000} burn-in, setting $K_\text{max}=10$ or a GEM prior for $\vec\lambda$ with $\gamma=0.1$.}
\label{sim11}
\end{figure}

Next, inference is repeated on the 50 datasets simulated as in Figure~\ref{sim1} with $\vec\eta=5\cdot\vec 1_K$, using different initialisation methods and priors for $\bm\lambda$. In particular, results are compared between the spectral and \textit{gensim} initialisation schemes described in Section~\ref{sec:initialisation}. Figure~\ref{sim11_1} displays the boxplots of ARIs for the session-level topics across the different datasets, after initialisation, before and after running the MCMC sampling scheme. The plot shows that the spectral initialisation scheme appears to have a better performance initially, but both methods lead to equivalent results after MCMC sampling. Furthermore, results on estimation of the number of topics are compared between two different priors: a $K_\text{max}$-dimensional Dirichlet distribution with $K_\text{max}=10$ and $\vec\gamma=0.1\cdot\vec 1_{K_\text{max}}$ (also used in Figure~\ref{sim1_2}) or a GEM prior on $\vec\lambda$ with $\gamma=0.1$. Figure~\ref{sim11_2} reports the resulting barplot for the estimated modal number of non-empty communities, 
suggesting that the two priors have a similar performance. 

\begin{figure}[!t]
\centering
\begin{subfigure}[t]{.475\textwidth}
\centering
\caption{Boxplot of ARIs for estimated session-level topics}
\label{sim2_1}
\includegraphics[width=\textwidth]{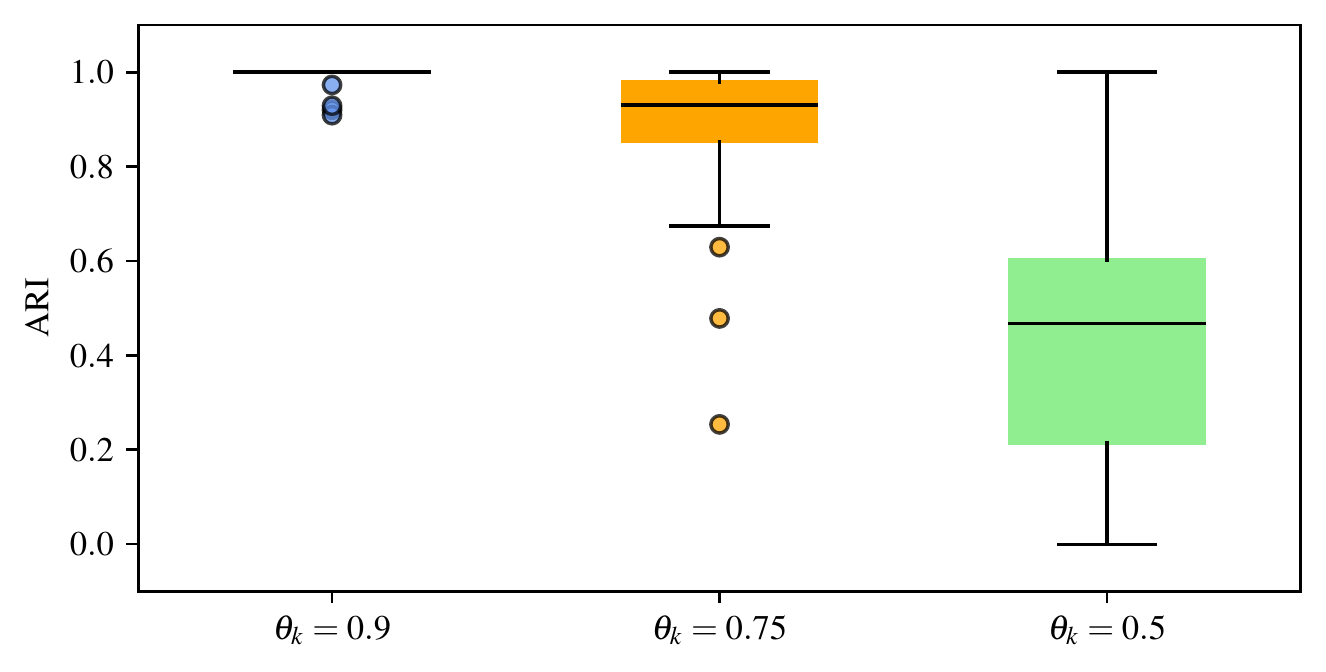}
\end{subfigure}
\hspace*{.02\textwidth}
\begin{subfigure}[t]{.475\textwidth}
\centering
\caption{Barplot of estimated number of non-empty topics $K_\varnothing$}
\label{sim2_2}
\includegraphics[width=\textwidth]{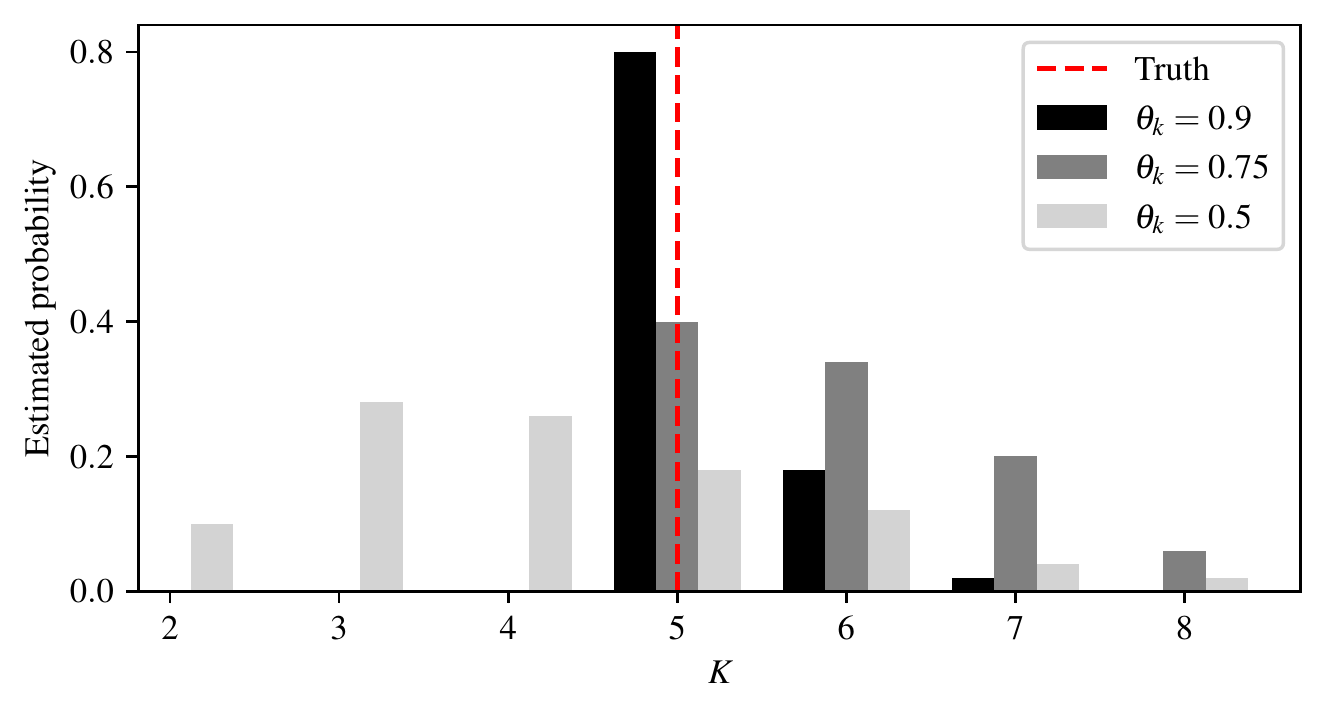}
\end{subfigure}
\caption{Summary plots obtained from 50 simulated datasets from model \eqref{mod2}, with $ V=50,\ K=5,\ D=100,\ N_d=10,\ M_{d,j}=10,\ \vec\lambda=1/K\cdot\vec 1_K,\ \vec\eta=\vec 1_K$, using different values for $\theta_k$. The MCMC sampler is run for \numprint{10000} iterations with \numprint{1000} burn-in, setting $K_\text{max}=10,\ \alpha_h=0.9, \alpha_0=0.1$.}
\label{sim2}
\end{figure}

Second, datasets are simulated from model \eqref{mod2}, using $\vec\eta=\vec 1_K$, $K=5$ and different values of $\theta_k$. The value of $\theta_k$ corresponds to the expected proportion of words sampled from the primary topic. Therefore, small values of $\theta_k$ are expected to make inferential procedures more difficult, since less observations are available from the primary topics, and words are sampled instead from a secondary topic, shared across sessions and commands. Furthermore, the secondary topic makes the primary topics more difficult to estimate, since the words are implicitly sampled from the mixture distribution $\tilde{\vec\phi}_{t_d} = \theta_{t_d}\vec\phi_{t_d} + (1-\theta_{t_d})\vec\phi_0$. If $\theta_k$ decreases for all $k$, the distributions $\tilde{\bm\phi}_1,\dots,\tilde{\bm\phi}_K$ become increasingly similar, and a drop in the ARI similar to Figure~\ref{sim1_1} is expected. In the MCMC sampler, the required parameters are set to $K_\text{max}=10$, $\alpha_h=0.9$ and $\alpha_0=0.1$. The results are presented in Figure~\ref{sim2}. As expected, Figure~\ref{sim2_1} shows that the ARI for the estimated session-level topics decreases when $\theta_k$ decreases. Furthermore, Figure~\ref{sim2_2} shows that estimation of the number of primary topics becomes increasingly imprecise when $\theta_k$ decreases, causing the drop in ARI in Figure~\ref{sim2_1}.  

\begin{figure}[!t]
\centering
\begin{subfigure}[t]{.475\textwidth}
\centering
\caption{Boxplot of ARIs for estimated session-level topics}
\label{sim3_1}
\includegraphics[width=\textwidth]{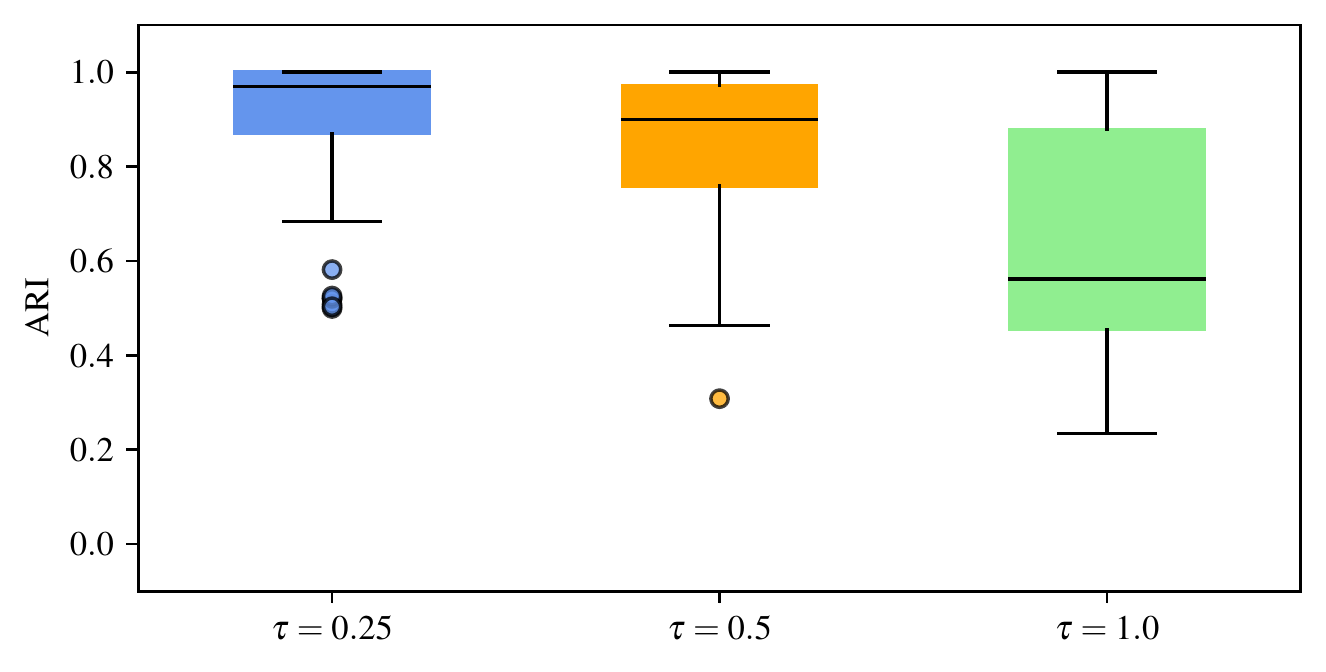}
\end{subfigure}
\hspace*{.02\textwidth}
\begin{subfigure}[t]{.475\textwidth}
\centering
\caption{Barplot of estimated number of non-empty topics $K_\varnothing$}
\label{sim3_2}
\includegraphics[width=\textwidth]{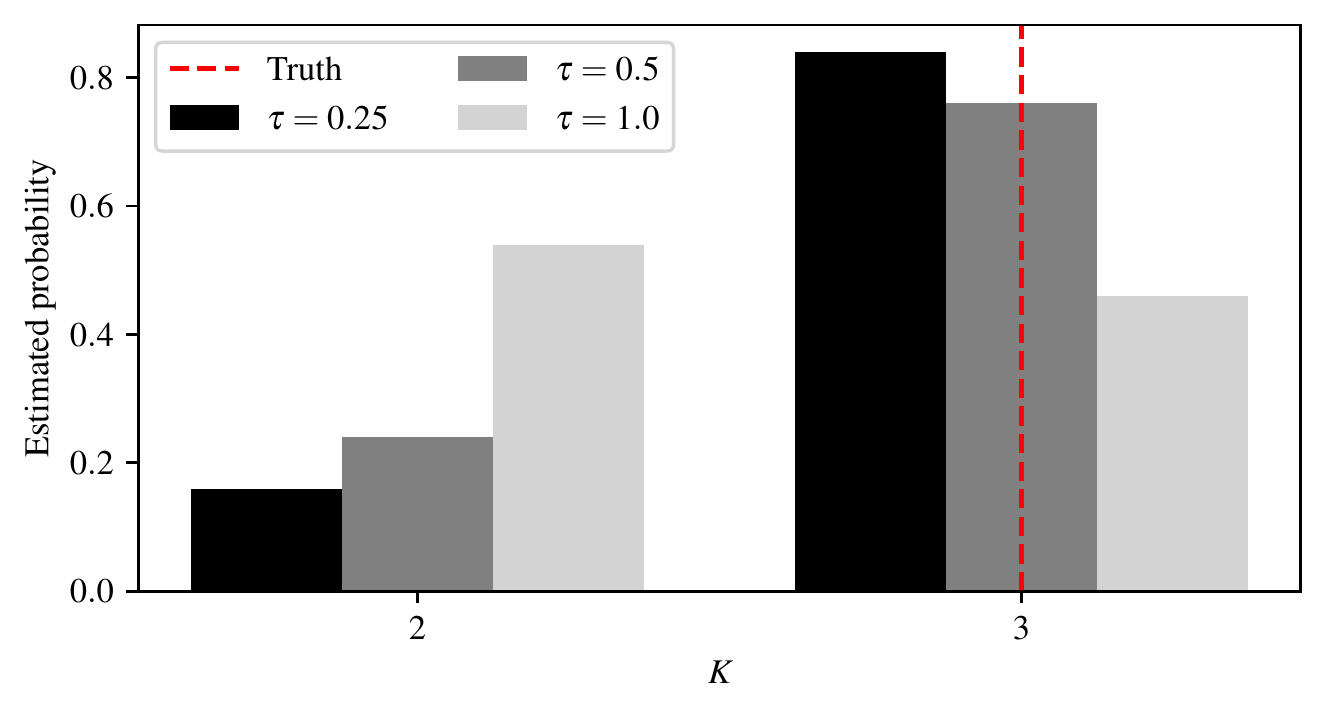}
\end{subfigure}
\caption{Summary plots obtained from 50 simulated datasets from model \eqref{mod3}, with $ V=50,\ K=3,\ H=5,\ D=100,\ N_d=10,\ M_{d,j}=10,\ \vec\lambda=1/K\cdot\vec 1_K,\ \vec\eta=\vec 1_K$, using different values for $\vec\tau=\tau\cdot\vec 1_H$. The MCMC sampler is run for \numprint{10000} iterations with \numprint{1000} burn-in, setting $K_\text{max}=10$, $H_\text{max}=10$.}
\label{sim3}
\end{figure}

Third, datasets are simulated from model \eqref{mod3}, using $\vec\eta=0.1\cdot\vec 1_K$, $K=3$, $H=5$ and $\vec\tau=\tau\cdot\vec 1_H$, for different values of $\tau$. In this case, $\tau$ expresses how concentrated around a peak the command-level topic distributions are. Small values of $\tau$ imply that the probability mass is mostly concentrated around one topic, whereas larger values correspond to more evenly distributed probability mass functions. In the MCMC sampler, the values $K_\text{max}=10$ and $H_\text{max}=10$ are used. The task of recovering the session-level topics is much more complex than previous simulations, since for model \eqref{mod3}, the session-level topic only controls the distribution of the command-level topics. Hence, the session-level topic must be estimated from the $N_d$ command-level topics within each document, which are themselves estimated. This makes the inferential task substantially harder. Figure~\ref{sim3} displays the results: Figure~\ref{sim3_1} shows that the ARI for the estimated session-level topics tends to decrease when $\tau$ increases, and Figure~\ref{sim3_2} shows that estimates of the number of non-empty session-level topics are more precise when the value of $\tau$ used in the simulation is smaller. 
Notice that, since $\vec\eta=0.1\cdot\vec 1_K$, the topic-specific word distributions have most of their mass concentrated around a small number of words, and the command-level topics are therefore easy to recover, with ARIs averaging above $0.99$ across the three settings for $\tau$ shown in Figure~\ref{sim3}. 

\section{Additional plots from the application to the Imperial College London honeypot data} \label{sec:plots}

\begin{figure}[!t]
\centering
\includegraphics[width=.9\textwidth]{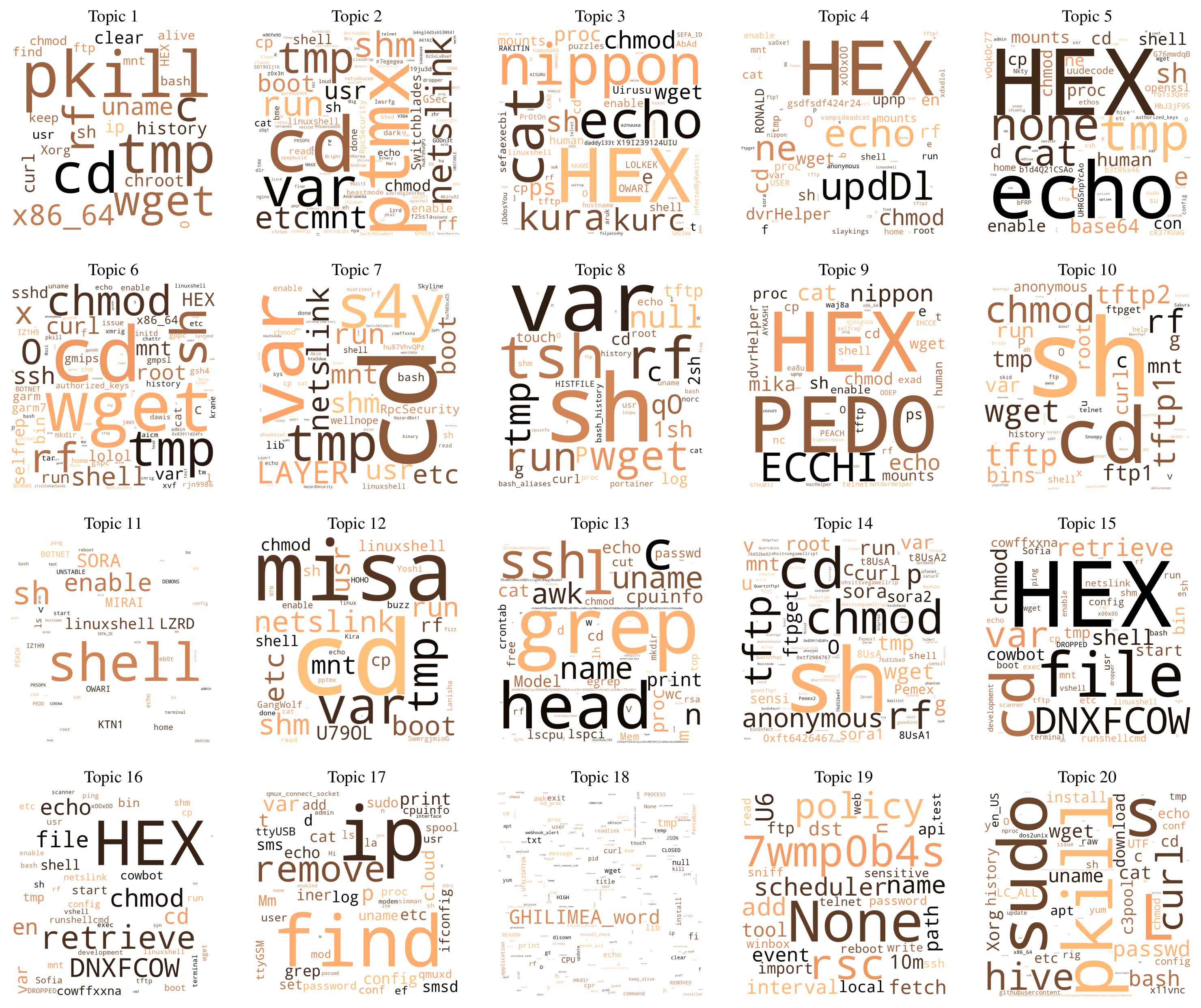}
\caption{Wordclouds of estimated topic-specific distributions under CBC \eqref{mod1}, fitted on the ICL honeypot data.}
\label{fig:mod1_wordcloud}
\end{figure}

\begin{figure}[!t]
\centering
\includegraphics[width=.9\textwidth]{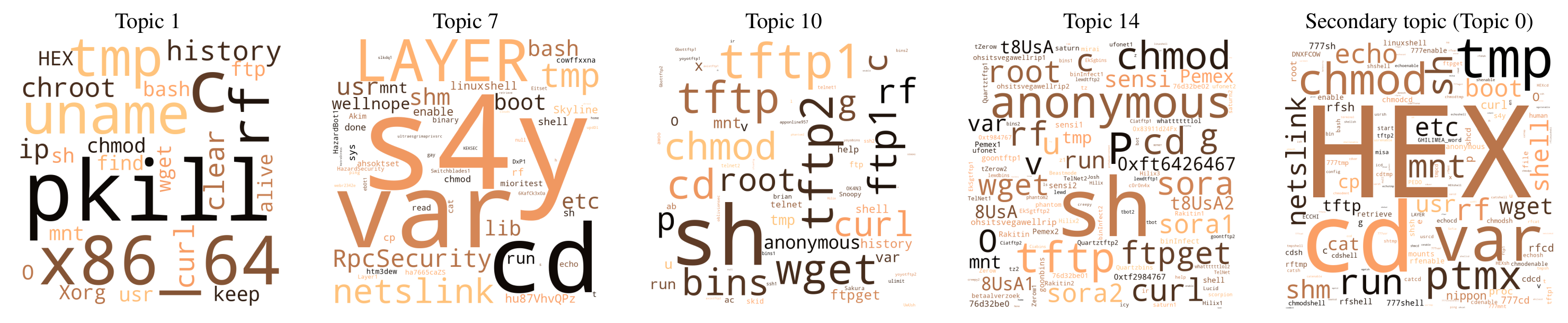}
\caption{Wordclouds of estimated topic-specific distributions under CBC with secondary topic \eqref{mod2}, after alignment of indices to the topics estimated via CBC.}
\label{fig:mod2_wordcloud}
\end{figure}

\begin{figure}[!t]
\centering
\begin{subfigure}[t]{.55\textwidth}
\centering
\caption{Barplot of frequencies of estimated session-level topics}
\label{fig:pcnbc_1}
\includegraphics[height=4.25cm]{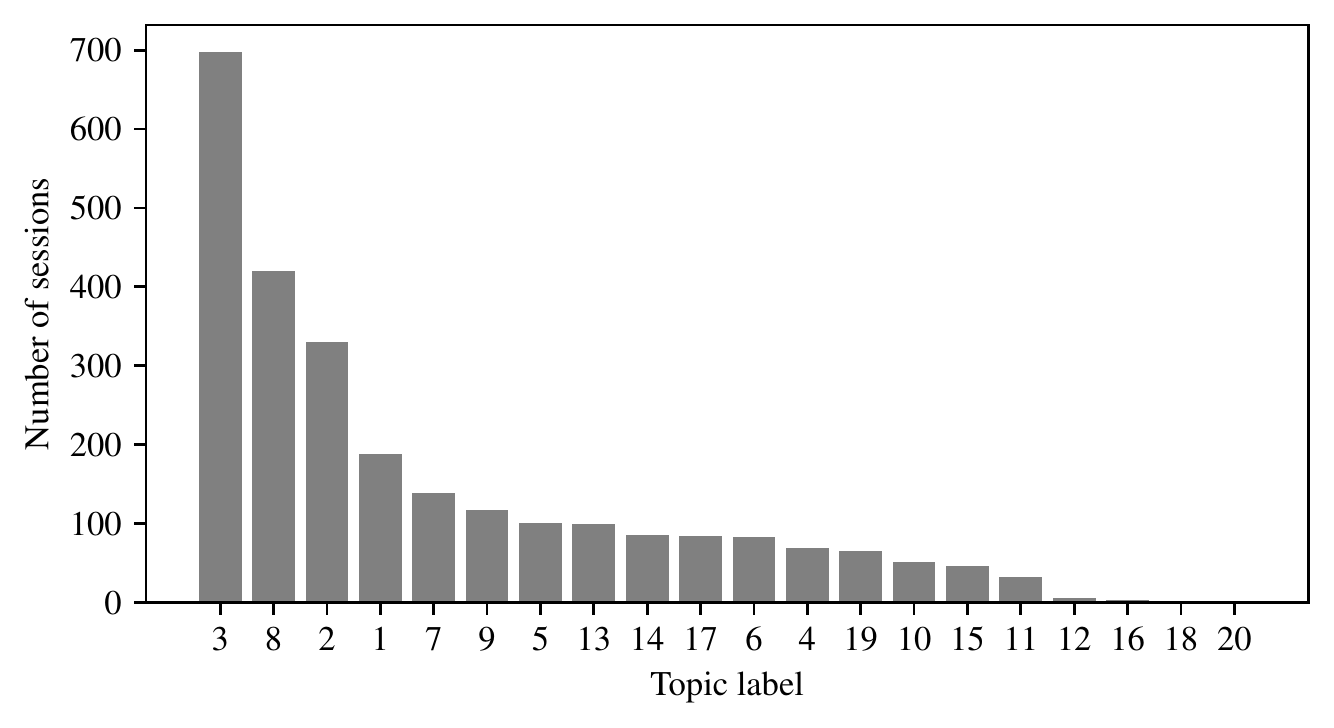}
\end{subfigure}
\begin{subfigure}[t]{.44\textwidth}
\centering
\caption{Barplot of estimated $K_\varnothing$}
\label{fig:pcnbc_2}
\includegraphics[height=4.25cm]{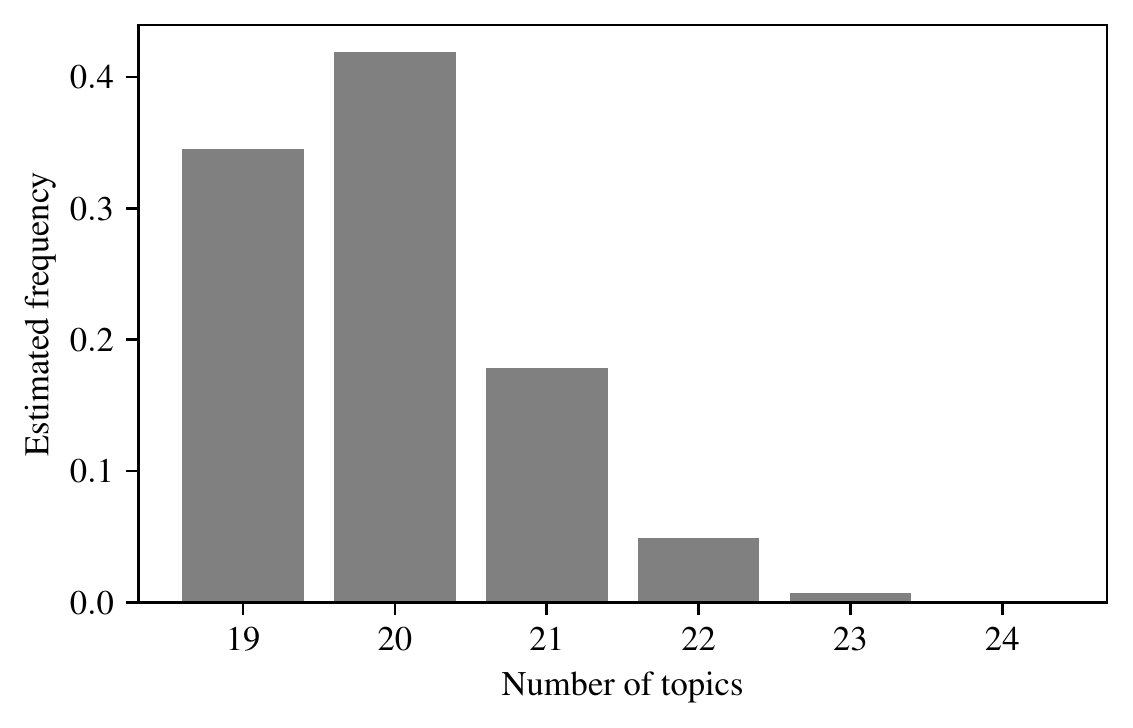}
\end{subfigure}
\caption{{Estimated topic frequencies and estimated distribution of the number of non-empty topics $K_\varnothing$ under the PCNBC model in \eqref{eq:pcnbc}, fitted on the ICL honeypot data.}}
\label{fig:pcnbc_results}
\end{figure}

\begin{figure}[!t]
\centering
\includegraphics[width=.9\textwidth]{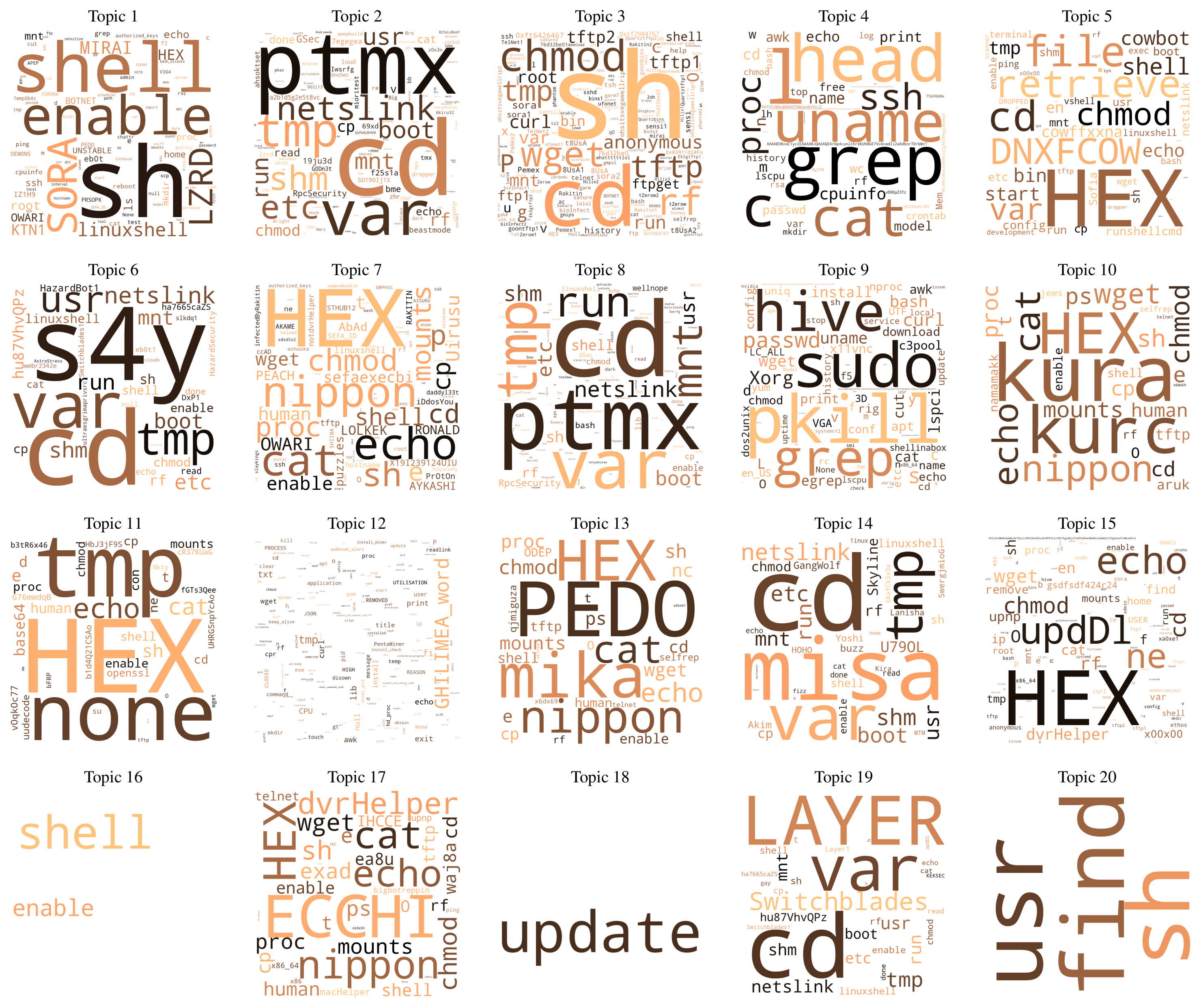}
\caption{Wordclouds of estimated topic-specific distributions under PCNBC \eqref{eq:pcnbc}.}
\label{fig:pcnbc_wordcloud}
\end{figure}

\begin{figure}[p]
\centering
\begin{subfigure}[t]{.99\textwidth}
\centering
\caption{Session-level topic-specific word distributions}
\includegraphics[width=\textwidth]{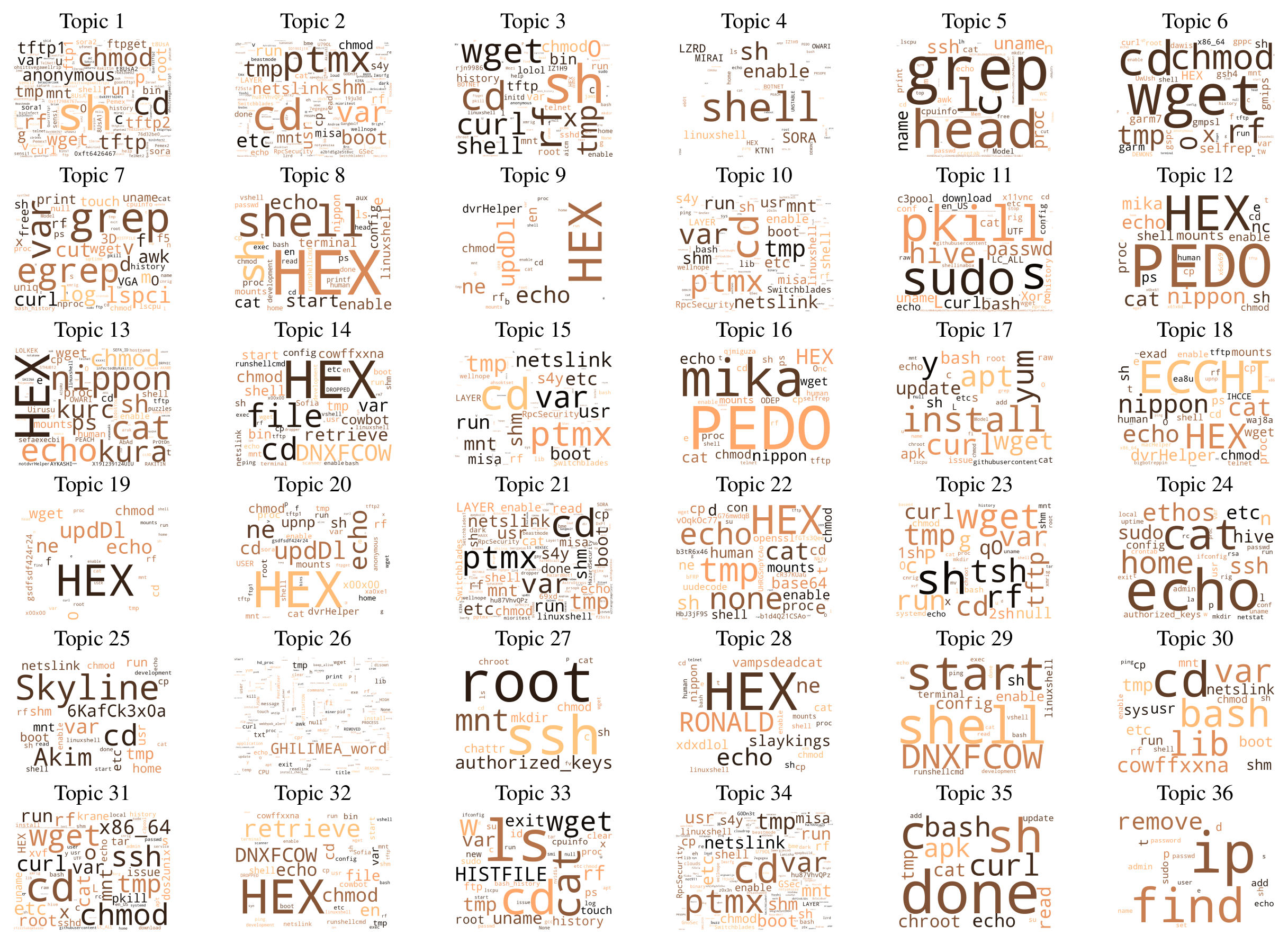}
\label{fig:mod3_wordcloud_sessions}
\end{subfigure}
\begin{subfigure}[t]{.99\textwidth}
\centering
\caption{Command-level topic-specific word distributions}
\includegraphics[width=\textwidth]{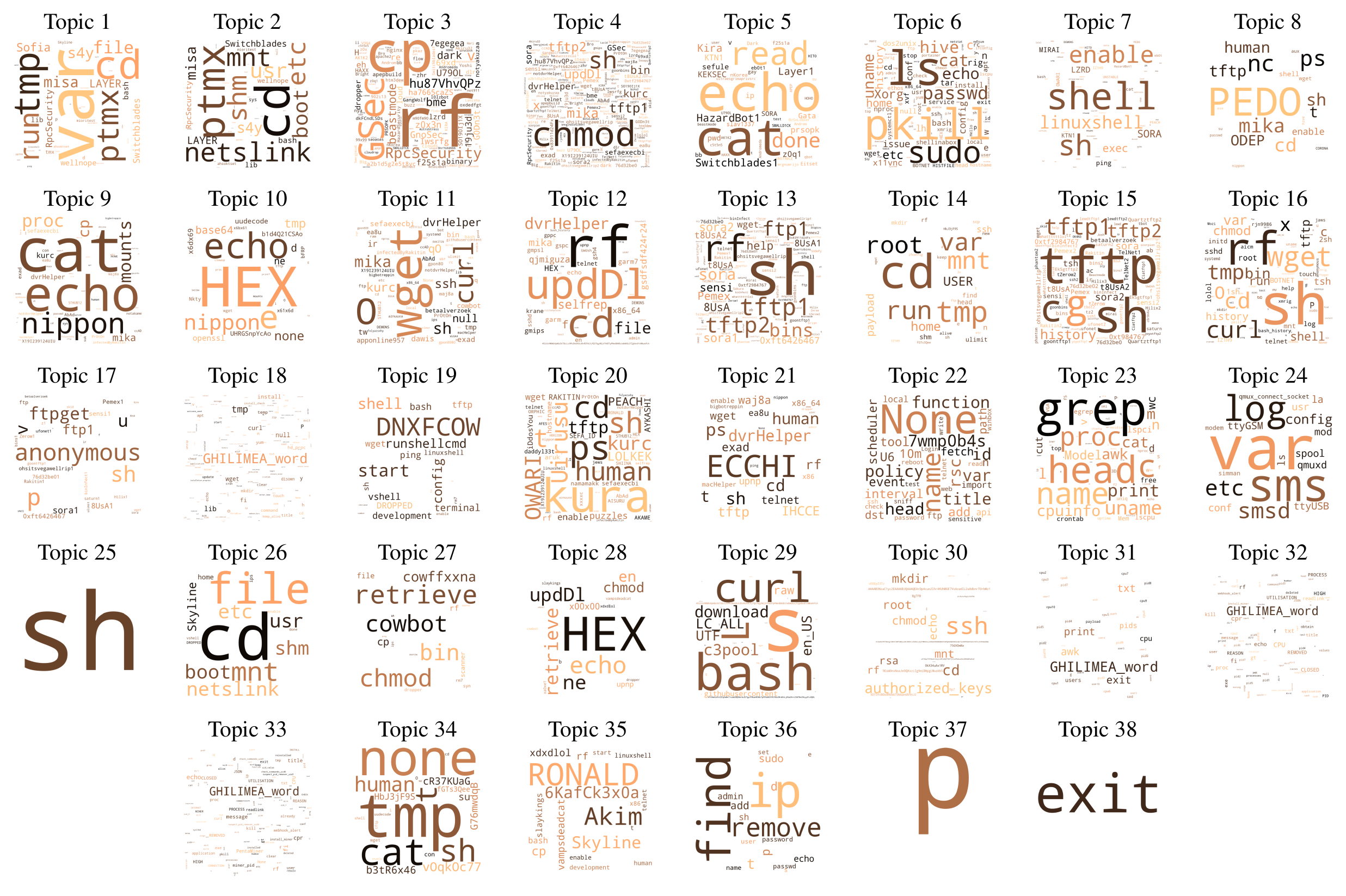}
\label{fig:mod3_wordcloud_commands}
\end{subfigure}
\caption{Wordclouds of estimated session-level and command-level topic-specific distributions under NCBC \eqref{mod3}.}
\label{fig:mod3_wordcloud}
\end{figure}

Further intuition about the differences between topics could be provided by examining the topic-specific word distributions, which can be displayed via wordclouds, plotted in Figure~\ref{fig:mod1_wordcloud} for the CBC model in Section~\ref{sec:mod1_fit}. 
For the CBC model with secondary topic fitted in Section~\ref{sec:mod2_fit}, the resulting wordclouds for some of the topic-specific word distributions are plotted in Figure~\ref{fig:mod2_wordcloud}, including the distribution of the secondary topic, labelled \textit{topic 0}. 
For the NCBC model fitted in Section~\ref{sec:mod3_fit}, the wordclouds 
are displayed in Figure~\ref{fig:mod3_wordcloud}. Figure~\ref{fig:mod3_wordcloud_sessions} displays the topic-specific word distributions obtained from the estimated session-level topics, whereas Figure~\ref{fig:mod3_wordcloud_commands} plots the wordclouds corresponding to the estimated command-level topics. Finally, Figure~\ref{fig:pcnbc_results} shows the estimated posterior distribution for $K_\varnothing$ and the distribution of session-level topics under the PCNBC model fitted in Section~\ref{sec:pcnbc_fit}, and Figure~\ref{fig:pcnbc_wordcloud} displays the wordclouds for these topics. 

\clearpage
\newpage

\setcounter{figure}{0}
\renewcommand{\thefigure}{S.\arabic{figure}}

\setcounter{table}{0}
\renewcommand{\thetable}{S.\arabic{table}}

\bibliographystyleSM{imsart-nameyear} 
\bibliographySM{biblio}

\end{appendix}

\end{document}